\documentclass[aps,prd,nofootinbib,superscriptaddress,reprint,onecolumn,notitlepage,11pt,twoside]{revtex4-1}

\usepackage[includeheadfoot,paperheight = 269mm,paperwidth = 190mm,top=18mm, left = 28mm, right=18mm, bottom=22mm]{geometry}
\usepackage{graphicx}
\usepackage{amsfonts,amssymb,amsmath}
\usepackage{hyperref}
\usepackage{color}
\usepackage{eimodels}
\usepackage{eivars}
\usepackage{xspace}
\usepackage{fancyhdr}
\usepackage{bm}

\hypersetup{
    colorlinks=true,       
    linkcolor=red,          
    citecolor=cyan,        
    filecolor=magenta,      
    urlcolor=blue,           
}

\newcommand{\BibitemShut}[1]{}

\newcommand{\ie}{i.e.\xspace}

\newlength{\wsingfig}
\newlength{\wdblefig}
\newlength{\wfull}
\newlength{\hfull}
\newcommand{\sss}[1]{{\scriptscriptstyle{#1}}}

\newcommand{\usssREF}{\sss{\mathrm{REF}}}

\newcommand{\calH}{\mathcal{H}}
\newcommand{\calP}{\mathcal{P}}

\newcommand{\calO}{\mathcal{O}}

\newcommand{\calL}{\mathcal{L}}

\newcommand{\calR}{\mathcal{R}}
\newcommand{\calE}{\mathcal{E}}
\newcommand{\calM}{\mathcal{M}}

\newcommand{\dd}{\mathrm{d}} 

\newcommand{\MeV}{\mathrm{MeV}}
\newcommand{\GeV}{\mathrm{GeV}}

\newcommand{\Mpc}{\mathrm{Mpc}}

\newcommand{\Mp}{M_\usssPl}

\newcommand{\COSMOMC}{\texttt{COSMOMC}\xspace}

\def\vx{{\bm x}}
\def\vk{{\bm k}}
\def\vka{{\bm k}_{1}}
\def\vkb{{\bm k}_{2}}
\def\vkc{{\bm k}_{3}}
\def\ka{k_{1}}
\def\kb{k_{2}}
\def\kc{k_{3}}
\def\cB{{\cal B}}
\def\cG{{\cal G}}
\def\Mpl{M_{_{\rm Pl}}}
\def\beq{\begin{equation}}
\def\eeq{\end{equation}} 
\def\bea{\begin{eqnarray}}
\def\eea{\end{eqnarray}}
\def\benu{\begin{enumerate}}
\def\eenu{\end{enumerate}}
\def\nn{\nonumber} 
\def\pa{{\partial}}
\def\f{\frac}
\def\l{\left}
\def\r{\right}
\def\d{{\rm d}}
\def\cR{{\mathcal{R}}}

\newcommand{\Pstar}{P_*}

\newcommand{\Hstar}{H_*}

\newcommand{\fnl}{f_{_\uNL}}
\newcommand{\gnl}{g_{_\uNL}}
\newcommand{\fnlloc}{f_{_\uNL}^\uloc}
\newcommand{\fnleq}{f_{_\uNL}^\ueq}
\newcommand{\fnlortho}{f_{_\uNL}^\uortho}

\newcommand{\epsstar}[1]{\epsilon_{#1}}

\newcommand{\BayesFactor}[2]{B^{#1}_{#2}}
\newcommand{\Bref}[1]{\BayesFactor{#1}{\usssREF}}

\newcommand{\EI}{\textit{Encyclopaedia Inflationaris}}

\newcounter{contribution}
\numberwithin{equation}{contribution}

\begin{document}
\setcounter{page}{3}


\title{The Observational Status of Cosmic Inflation after Planck}

\author{J\'er\^ome Martin\footnote{E-mail: jmartin@iap.fr}}
\affiliation{Institut d'Astrophysique de Paris, UMR 7095-CNRS, 98bis
  boulevard Arago, 75014 Paris, France}

\begin{abstract}
The observational status of inflation after the Planck $2013$ and
$2015$ results and the BICEP2/Keck Array and Planck joint analysis is
discussed. These pedagogical lecture notes are intended to serve as a
technical guide filling the gap between the theoretical articles on
inflation and the experimental works on astrophysical and cosmological
data. After a short discussion of the central tenets at the basis of
inflation (negative self-gravitating pressure) and its experimental
verifications, it reviews how the most recent Cosmic Microwave
Background (CMB) anisotropy measurements constrain cosmic
inflation. The fact that vanilla inflationary models are, so far,
preferred by the observations is discussed and the reason why
plateau-like potential versions of inflation are favored within this
subclass of scenarios is explained. Finally, how well the future
measurements, in particular of $B$-Mode CMB polarization or primordial
gravity waves, will help to improve our knowledge about inflation is
also investigated.
\end{abstract}

\maketitle


\tableofcontents

\section{Introduction}
\label{sec:intro}

With the release of the Planck data
$2013$~\cite{Ade:2013zuv,Ade:2013uln,Ade:2013ydc} and
$2015$~\cite{Planck:2015xua,Ade:2015oja,Ade:2015ava}, and the recent
BICEP2/Keck Array and Planck joint
analysis~\cite{BICEP2/Keck:2015tva}, the theory of cosmic
inflation~\cite{Starobinsky:1979ty,Starobinsky:1980te,
  Guth:1980zm,Linde:1981mu,Mukhanov:1981xt,Mukhanov:1982nu,
  Starobinsky:1982ee} has acquired a new status. Several of its
predictions such as spatial flatness of our Universe, the presence of
Doppler peaks in the Cosmic Microwave Background (CMB) multipole
moments, almost scale invariant power spectrum for density
perturbations have been definitively confirmed by the recent CMB
anisotropy measurements. That makes inflation a predictive and
verified theory of the early Universe.

\par

In fact, another remarkable outcome of the Planck data is that they
also allow us to identify which version of inflation is most likely to
have been realized in
Nature~\cite{Martin:2014vha,Martin:2013nzq,Martin:2013gra}. As is
well-known, inflation comes in different flavors but these different
scenarios make different predictions and, thus, one can, at least in
principle, distinguish among them. The fact that the primordial
fluctuations are adiabatic and Gaussian to a relatively high degree of
accuracy~\cite{Ade:2013uln,Ade:2013ydc} is an important indication
that we probably deal with single-field slow-roll inflation (with
standard kinetic term), the simplest but non-trivial model of
inflation. Of course, the final word has not yet been spoken since
many non-vanilla inflationary scenarios are still compatible with the
data. But, presently, they are just not needed in order to explain CMB
measurements even if this situation could change in the future.

\par

The fact that we now have high accuracy CMB data at our disposal also
allows us to detect the ``fine structure'' of inflation and to
constrain the shape of the inflaton potential. Here again, the Planck
data have provided precious information. We now know that the
potential is of the plateau type and that simple monomials are
disfavored~\cite{Ijjas:2013vea,Martin:2014vha,Martin:2013nzq,
  Martin:2013gra}. Moreover, we now start probing the reheating
epoch~\cite{Martin:2010kz,Martin:2014nya}. Reheating is the epoch,
after inflation and before the radiation dominated era of the standard
hot Big bang phase, where the inflaton field decays and where all
matter we see around us was
produced~\cite{Turner:1983he,Traschen:1990sw,Kofman:1997yn,Amin:2014eta}. It
is therefore of major conceptual importance. And Planck $2013$ and
$2015$ data put non trivial constraints on the physical processes that
took place at that
time~\cite{Martin:2006rs,Lorenz:2007ze,Martin:2010kz,Martin:2014nya,Munoz:2014eqa,Dai:2014jja,Gong:2015qha}.

\par

The goal of these lectures, given at the second Jose Plinio Baptista
school on Cosmology held in Pedra Azul (Brazil) in March $2014$, is to
review how the above conclusions can be established. Many reviews on
inflation can be found in the
literature~\cite{Martin:2003bt,Martin:2004um,Martin:2007bw,Sriramkumar:2009kg}
and there are technical papers reporting the astrophysical and
cosmological observations, such as the Planck
papers~\cite{Ade:2013zuv,Ade:2013uln,Ade:2013ydc}. But, in between,
few things can be found and the present article aims at filling this
gap. In some sense, it can be viewed as a technical guide which, from
a reasonable prior knowledge of inflation, permits a detailed
understanding of the implications for inflation of the recent high
accuracy CMB data.

\par

These lecture notes are also written at a special time: the Planck
$2013$ and $2015$
data~\cite{Ade:2013zuv,Ade:2013uln,Ade:2013ydc,Planck:2015xua,Ade:2015oja,Ade:2015ava}
have been released and their consequences (in fact, mainly the
consequences of Planck $2013$) already analyzed in several
works. Moreover, very recently, a joint analysis made by the
BICEP2/Keck Array team and the Planck
collaboration~\cite{BICEP2/Keck:2015tva} has been published showing
that the BICEP2 detection of $B$-mode CMB polarization announced in
Ref.~\cite{Ade:2014xna} is mainly due to dust and cannot be attributed
to primordial gravity waves produced during inflation. At the time of
writing, the Planck $2015$ scientific products (in particular, the
likelihood) are expected to be delivered in June $2015$ only. This
means that reproducing or extending the Planck $2015$ analysis is not
yet possible. However, from what is already known, the Planck $2015$
results are in good agreement with Planck $2013$. Therefore, the
conclusions discussed in the present article (model comparison,
constraints on reheating etc \dots ) will most likely remain valid for
the second release of the Planck data. Whenever available, we quote
the values obtained by Planck $2015$.

\par

These lectures are also related to the lectures given by C.~Byrnes on
Non-Gaussianities~\cite{Byrnes:2014pja} and by D.~Wands on CMB
physics. Hopefully, these three reviews should provide the reader with
a rather complete overview of modern primordial Cosmology and its
observational implications. In particular, Ref.~\cite{Byrnes:2014pja}
reviews how Non-Gaussianities are produced in non-vanilla inflationary
models while, here, we restrict ourselves to simple scenarios for
which Non-Gaussianities are very small. The two lectures are therefore
complementary. The lecture notes by D.~Wands explain in details how
CMB anisotropies are generated while, here, we just take it as a known
fact (see also the recent review~\cite{Bucher:2015eia}). Therefore,
the present article and the one on CMB physics are also complementary.

\par

These lecture notes are organized as follows. In the next section,
Sec.~\ref{sec:considerations}, we present general considerations on
inflation. Rather than discussing inflation in detail, which can be
found in many review articles, we just give the basics and choose to
focus on the fundamental principles at the basis of the inflationary
mechanism and its experimental justifications. In
Sec.~\ref{sec:paraminf}, we discuss how inflation can be realized in
practice. In particular, in Sec.~\ref{subsec:hepinf}, we review how
inflation can be embedded in high energy physics. Recently,
alternative parametrizations have been considered and we discuss them
in Sec.~\ref{subsec:otherparam}. In Sec.~\ref{subsec:reheatparam}, we
also review how the reheating phase can be described. Then, in
Sec.~\ref{sec:pert}, we discuss the theory of inflationary
cosmological perturbations of quantum-mechanical origin. This part of
the inflationary scenario is especially important because this is how
one can relate theoretical predictions to astrophysical
observations. In Sec.~\ref{subsec:correl2}, we present the calculation
of the two-point correlation functions, or power spectra, for scalar
and tensor perturbations in the slow-roll approximation. In
Sec.~\ref{subsec:correl3}, we review the calculation of the
three-point correlation function, or bispectrum, and in
Sec.~\ref{subsec:correl4}, the calculation of the four-point
correlation function, or tri-spectrum. All these considerations are
made in the slow-roll approximation and for single-field models with
minimal kinetic terms. In Sec.~\ref{subsec:iso}, we discuss the
isocurvature perturbations and how they can be produced in the
framework of inflation. In Sec.~\ref{sec:observ}, we use the tools
introduced before and compare the inflation predictions to the high
accuracy CMB Planck data. In Sec.~\ref{subsec:curvplanck}, we consider
the measurements of spatial curvature, in Sec.~\ref{subsec:isoplanck}
the measurements of isocurvature perturbations and, in
Sec.~\ref{subsec:ngplanck}, those of Non-Gaussianities. Since these
data indicate that single field models are preferred, we then focus on
this class of scenarios. In Sec.~\ref{subsec:sr}, we give the
constraints on the slow-roll parameters and on the derived power-law
parameters, such as the spectral index, the running or the
tensor-to-scalar ratio. We also discuss the implications of the recent
joint analysis made by the BICEP2/Keck Array team and the Planck
collaboration. In Sec.~\ref{subsec:compar}, we carry out a Bayesian
analysis to do model comparison and determine what are the best models
of inflation. In Sec.~\ref{subsec:reheat}, we present the constraints
on reheating that can be inferred from the Planck data. Finally, in
Sec.~\ref{sec:conclusion}, we recap our main results and discuss which
lesson can be drawn for our understanding of inflation and primordial
cosmology.

\section{General Considerations on Inflation}
\label{sec:considerations}

The motivations for introducing a phase of inflation, \ie a phase of
accelerated expansion, are well-known: postulating $\ddot{a}>0$ ($a$
is the Friedmann-Lema\^{\i}tre-Roberston-Walker -FLRW- scale factor)
allows us to avoid the puzzles of the standard hot Big Bang theory
(for a detailed discussion of these issues, see
Refs.~\cite{Martin:2003bt,Mukhanov:2005sc,PeterUzan2009}). If gravity
is described by General Relativity (GR), then, in a homogeneous and
isotropic Universe, the equations of motion are given by
\begin{eqnarray}
\label{eq:fried}
H^2+\frac{{\cal K}}{a^2}&=&\left(\frac{\dot{a}}{a}\right)^2
+\frac{{\cal K}}{a^2}=
\frac{1}{3\Mp^2}\sum _i\rho_i\equiv \frac{1}{3\Mp^2}\rho,
\\
\label{eq:eqpressure}
-\left(2\frac{\ddot{a}}{a}+\frac{\dot{a}^2}{a^2}
+\frac{{\cal K}}{a^2}\right)
&=& \frac{1}{\Mp^2}\sum _i p_i\equiv \frac{1}{\Mp^2}p,
\end{eqnarray}
where $\rho_i$ and $p_i$ are respectively the energy density and the
pressure of the fluid ``$i$''. In the standard model of Cosmology, we
have indeed a collection of different fluids, pressure-less matter
(made of baryons and cold dark matter), radiation (made of photons and
neutrinos) and dark energy. These different types of matter source the
Einstein equations and control the dynamics of the expansion. Notice
that the expansion rate of the Universe is given by the Hubble
parameter which, according to the above equations, is defined by
$H\equiv \dot{a}/a$ where a dot means a derivative with respect to
cosmic time. The quantity $\Mp$ is the reduced Planck mass and, in the
following, we will also use the quantity $\kappa\equiv
1/\Mp^2=8\pi G_{_{\rm N}}$, $G_{_{\rm N}}$ being the Newton
constant. Finally, the quantity ${\cal K}$, that can always be
normalized to $0$ or $\pm 1$, represents the curvature of the spatial
sections. Notice that one can also define an effective curvature
energy density by $\rho_{\rm curv}\equiv -3{\cal K}/(\kappa a^2)$ such
that the Friedmann equation takes the form $H^2=(\kappa/3)\sum_i\rho_i
+(\kappa/3)\rho_{\rm curv}$. Defining $\Omega_i\equiv \rho/\rho_{\rm
  cri}$ and $\Omega_{\cal K}=\rho_{\rm curv}/\rho_{\rm cri}$, where
the critical energy density is $\rho_{\rm cri}=3H^2/\kappa$, the
Friedmann equation can be rewritten as $\sum_i\Omega_i+\Omega _{\cal
  K}=1$.

\par

Let us now discuss under which physical conditions inflation can be
obtained. The above equations can be combined and lead to the
following formula which relates the acceleration of the expansion to
the matter content of the Universe
\begin{equation}
\label{eq:acceleration}
\frac{\ddot{a}}{a}=-\frac{1}{6\Mp^2}\sum_i\left(\rho_i +3p_i\right)
\end{equation}
This immediately implies that, in order to have inflation, the
pressure must be negative, \ie $p<-\rho/3$ where $\rho$ and $p$ are
defined in Eqs.~(\ref{eq:fried}) and~(\ref{eq:eqpressure}). Having
realized that we need a negative pressure, the next question is of
course which kind of matter can possess this property and this will be
the subject of the two next sections. Of course, as is well-known, we
will see that scalar fields are ideal candidates.

\par

But before starting this discussion, it is interesting to notice that
inflation is a genuine relativistic effect since it involves the term
$3p$ in the above equation~(\ref{eq:acceleration}), which is absent in
Newtonian physics. Indeed, let us consider a sphere of radius $R(t)$
and of uniform density $\rho$. A galaxy of mass $m$, located at the
edge of the sphere, feels a gravitational field ${\bm G}$ that can be
simply evaluated by means of the Gauss's law, $\int {\bm G}\cdot {\rm
  d}{\bm S}=4\pi G_{_{\rm N}} M$, where $M$ is the mass of the
sphere. This gives $G=G_{_{\rm N}}M/R^2$. As a consequence, the
acceleration of the galaxy can be written as
\begin{equation}
m\ddot{R}=-m \frac{G_{_{\rm N}}M}{R^2},
\end{equation}
or
\begin{equation}
\frac{\ddot{R}}{R}=-\frac{4\pi G_{_{\rm
      N}}}{3}\rho=-\frac{\rho}{6\Mp^2},
\end{equation}
where we have used $M=4\pi \rho R^3/3$. This equation is similar to
Eq.~(\ref{eq:acceleration}) except that the term $3p$ is not
present. The physical reason behind the presence of this term is
deeply rooted in the fundamental principles of GR: in GR, every form
of energy weighs, including pressure.

\par

The term $3p$ is so important for inflation that it is interesting to
ask whether it plays a role in other physical situations and if its
appearance has been tested experimentally and/or observationally. This
is a difficult question since, in ordinary cases, the contribution of
pressure is usually negligible, $p\ll \rho$. In fact, four situations
where a gravitating pressure is important can be identified:
inflation, dark energy but in some sense this is the same as
inflation, neutron stars and Big Bang Nucleosynthesis (BBN). In
particular, it is interesting to see what can be said about the $3p$
terms in the last two examples.

\par

\begin{figure}
\begin{center}
\includegraphics[width=10cm]{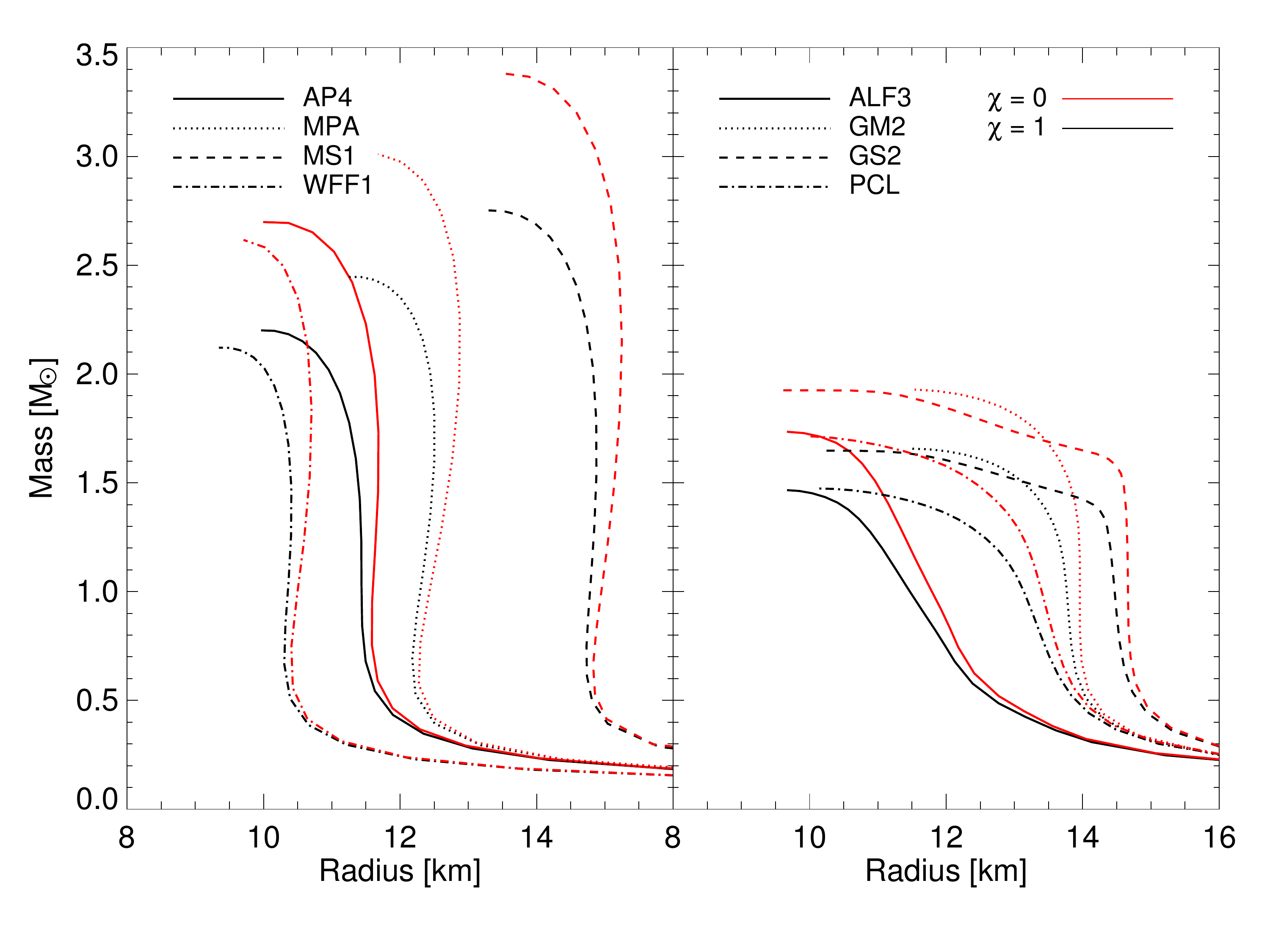}
\end{center}
\caption{Mass-radius relations of neutron stars for different
  equations of state (``standard'' in the left panel, more ``exotic''
  in the right panel). Black curves correspond to the standard GR
  calculation while red curves represent the case where self-gravity
  of pressure is absent. Figure taken from
  Ref.~\cite{Schwab:2008ce}.}
\label{fig:mr}
\end{figure}

Let us start with the internal structure of a neutron
star~\cite{Schwab:2008ce}. As is well-known, it is controlled by the
Tolman-Oppenheimer-Volkoff equations that can be obtained in the
following way. The metric for a static and spherically symmetric
solution can be written as
\begin{equation}
{\rm d}s^2=-{\rm e}^{2\Phi}{\rm d}t^2+{\rm e}^{2\lambda}{\rm d}r^2
+r^2\left({\rm d}\theta^2+\sin ^2 \theta {\rm d}\varphi^2\right),
\end{equation}
where $t$ is time, $r$ a radial coordinate and $\theta$ and $\varphi$
angular coordinates. The quantities $\Phi$ and $\lambda $ are
functions of $r$ only. Matter is assumed to be described by a perfect
fluid, the stress energy tensor of which can be expressed as
\begin{equation}
T_{\mu \nu}=\left(\rho+p\right)u_{\mu}u_{\nu }+p g_{\mu \nu},
\end{equation}
where $g_{\mu \nu}$ is the metric tensor and the normalized
$4$-velocity reads $u_{\mu}=\left(-{\rm e}^{\Phi},0,0,0\right)$. Then,
the time-time and $r-r$ component of the Einstein equations read
\begin{eqnarray}
\label{eq:Grr}
\frac{1}{r^2}{\rm e}^{-2\lambda}
\left(2r\frac{{\rm d}\Phi}{{\rm d}r}+1-{\rm e}^{2\lambda}
\right) &=& \frac{1}{\Mp^2}p, \\
\label{eq:Gtt}
\frac{1}{r^2}{\rm e}^{-2\lambda}\left(-1+2r\frac{{\rm d}\lambda}{{\rm d}r}
+{\rm e}^{2\lambda}\right) &=& \frac{1}{\Mp^2}\rho.
\end{eqnarray}
On the other hand, energy conservation, $\nabla _{\mu}T^{\mu \nu}=0$,
more precisely its radial component, implies that
\begin{equation}
\label{eq:conserv}
\frac{{\rm d}p}{{\rm d}r}=-\left(\rho+p\right)\frac{{\rm d}\Phi}{{\rm
    d}r}.
\end{equation}
The other components lead to the fact that $\rho$ does not depend on
time, $\theta $ or $\varphi$, that is to say $\rho=\rho(r)$. If we now
define the gravitational mass $m(r)$ by
\begin{equation}
\label{eq:defmass}
G_{_{\rm N}}m(r)=\frac{r}{2}\left(1-{\rm e}^{-2\lambda}\right),
\end{equation}
then Eq.~(\ref{eq:Gtt}) implies that
\begin{equation}
\label{eq:dmdr}
\frac{{\rm d}m}{{\rm d}r}=4\pi \rho(r)r^2.
\end{equation}
Introducing the expression of the mass~(\ref{eq:defmass}) in
Eq.~(\ref{eq:Grr}) in order to express ${\rm d}\Phi/{\rm d}r$ and,
then, inserting the corresponding expression in the conservation
equation~(\ref{eq:conserv}) leads to
\begin{equation}
\frac{{\rm d}p}{{\rm d}r}=-(\rho+p)
\frac{G_{_{\rm N}}}{r^2(1-2mG_{_{\rm N}}/r)}
\left[m(r)+3p(r)\left(\frac{4}{3}\pi r^3\right)\right].
\end{equation}
The important point in this formula is that the term $3p$ participates
to this expression. This means that self-gravity of pressure affects
the internal structure of the neutron stars. In practice, in order to
calculate this internal structure, one has first to choose an equation
of state $\rho=\rho(p)$. Once this is done, one can integrate the two
following equations
\begin{equation}
\frac{{\rm d}\rho}{{\rm d}r}=\frac{{\rm d}\rho}{{\rm d}p}
\frac{{\rm d}p}{{\rm d}r}, \quad \frac{{\rm d}m}{{\rm d}r}
=4\pi \rho(r)r^2, 
\end{equation}
the last equation being nothing but Eq.~(\ref{eq:dmdr}). This leads to
the functions $\rho(r)$ and $m(r)$. The radius of the star, $R_{\rm
  star}$, is defined by $\rho\left(R_{\rm star}\right)=0$ and its mass
is given by $M_{\rm star}\equiv m\left(R_{\rm star}\right)$. One can
then plot the mass-radius relation $M_{\rm star}(R_{\rm star})$. Of
course, one obtains different mass-radius relations for different
equations of state. Let us also notice that, at fixed equation of
state, one obtains a curve, and not a unique prediction, because one
needs to specify $\rho\left(r=0\right)$ to be able to integrate the
above equations. One thus has a family of points parametrized by
$\rho\left(r=0\right)$. Several examples are displayed in
Fig.~\ref{fig:mr} (black lines).

The fact that the structure of a neutron star depends on the general
relativistic $3p$ term opens the possibility to experimentally test
it. In order to do so, the idea of Ref.~\cite{Schwab:2008ce} is to
study an ad-hoc modification of the Tolman-Openheimer-Volkoff equation
such that
\begin{equation}
\label{eq:modifiedneutron}
\frac{{\rm d}p}{{\rm d}r}=-(\rho+p)
\frac{G_{_{\rm N}}}{r^2(1-2mG_{_{\rm N}}/r)}
\left[m(r)+3\chi p(r)\left(\frac{4}{3}\pi r^3\right)\right],
\end{equation} 
where $\chi$ is a new, phenomenological, parameter introduced by
hand. The term $3p$ weighs normally when $\chi=1$ and does not weigh
at all when $\chi=0$. Notice that $\chi=0$ is not the Newtonian limit
because there are other relativistic terms in
Eq.~(\ref{eq:modifiedneutron}) (for instance $1-2mG_{_{\rm N}}/r$ at
the denominator). So the idea is now to re-derive the mass-radius
relation for neutron stars and to see the influence of a parameter
$\chi\neq 1$, the hope being to be able to put constraints on $\chi$
from astronomical observations. The results are shown in
Fig.~\ref{fig:mr}. The fact that red curves (namely those obtained
with $\chi=0$) are different from the black ones (those obtained in
the standard GR case) confirms that the $3p$ term has a significant
influence of the mass-radius relation.

\par

\begin{figure}
\begin{center}
\includegraphics[width=10cm]{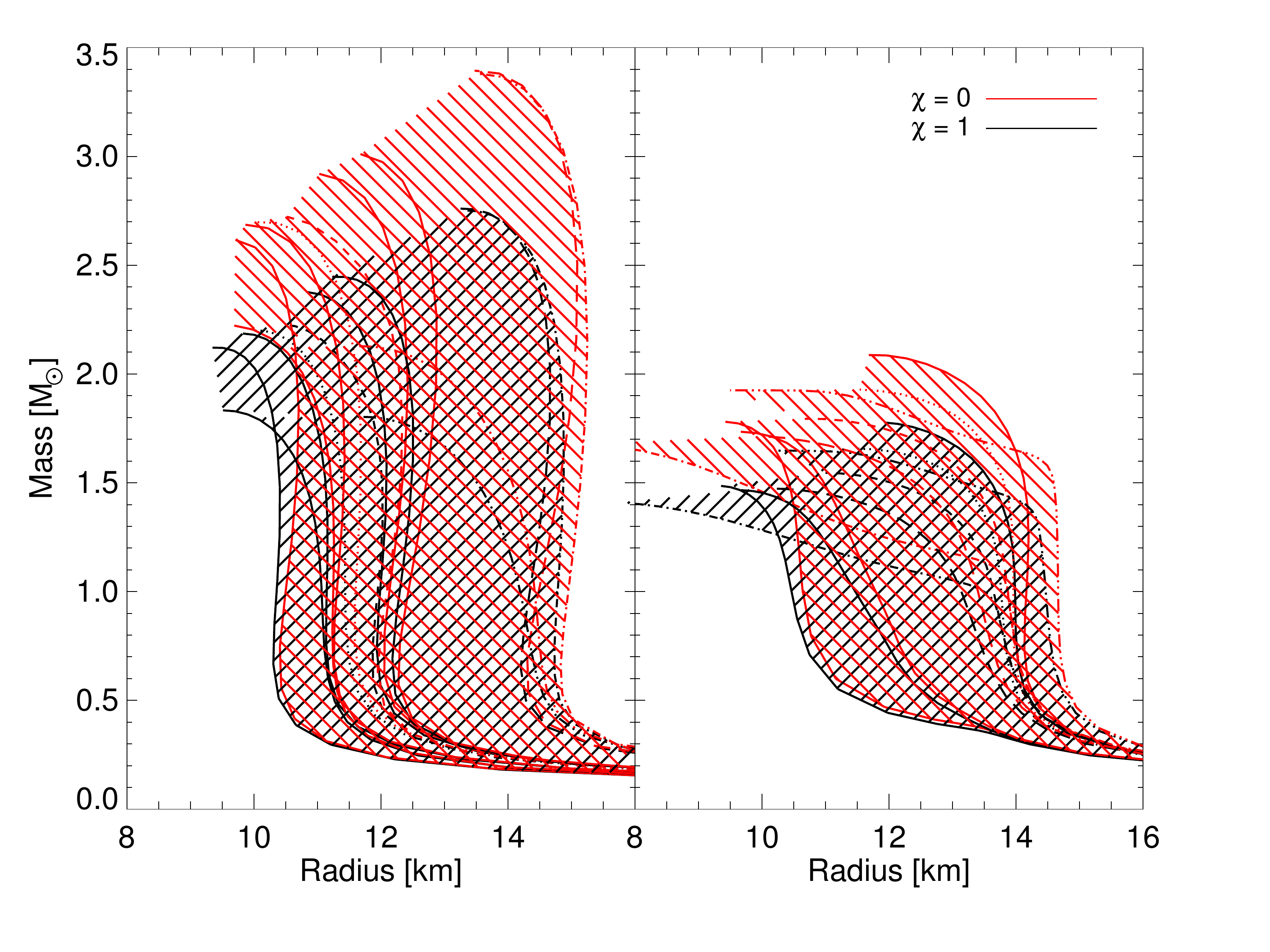}
\end{center}
\caption{Mass radius relations for different equations of state and
  associated theoretical uncertainties. In black are represented the
  mass radius relations obtained when $\chi=1$ (standard GR
  calculation) while, in red, are represented the mass radius
  relations obtained without self-gravity pressure (namely
  $\chi=0$). The hatched regions show the theoretical uncertainty
  associated with the fact that the equation of state is in fact
  unknown. It is clear from the plot that this completely dominates the
  differences between the $\chi=1$ and $\chi=0$ situations. Figure
  taken from Ref.~\cite{Schwab:2008ce}.}
\label{fig:mr_inder}
\end{figure}

However, as shown in Fig.~\ref{fig:mr_inder}, the fact that the
equation of state is not known accurately completely blurs the
effect. Indeed, one sees that the corresponding uncertainty is
typically of the same order of the effect we try to detect. Therefore,
the conclusion is that, although it is true that self-gravity is
crucial in order to predict correctly their mass-radius relation, at
least for the moment, neutron stars cannot be used to experimentally
test the $3p$ term.

\begin{figure}
\begin{center}
\includegraphics[width=10cm]{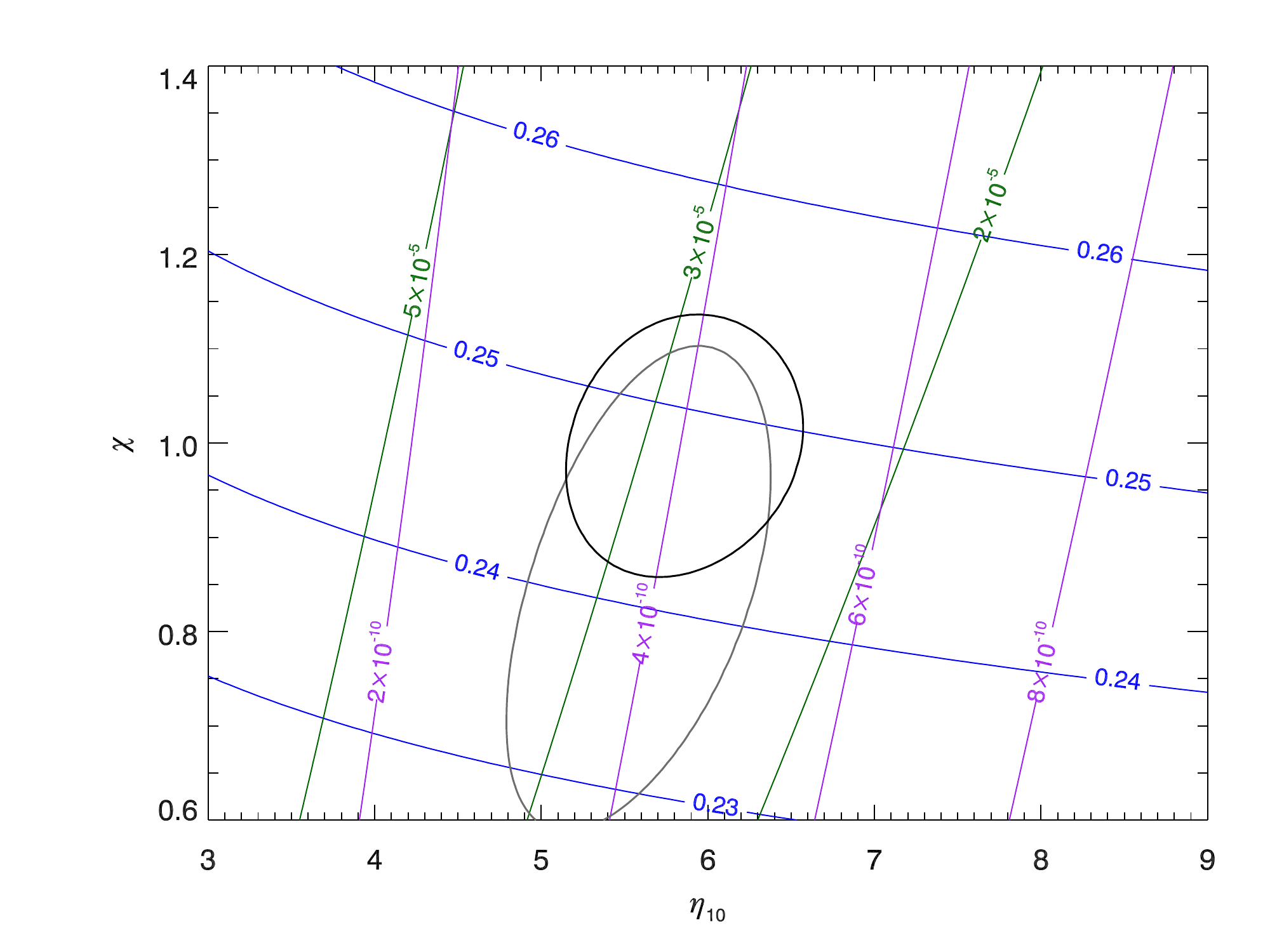}
\end{center}
\caption{Light elements abundances calculated when the Friedmann
  equation is modified according to
  Eq.~(\ref{eq:friedmodified}). Greens contours are for deuterium
  abundance, blue ones for helium-$4$ and purple ones for
  lithium-$7$. The two gray ellipses indicate the region in parameter
  space allowed by observations. Figure taken from
  Ref.~\cite{Rappaport:2007ct}.}
\label{fig:bbn}
\end{figure}

Let us now turn to the other possibility, namely
BBN~\cite{Rappaport:2007ct}. Since BBN takes place during the
radiation dominated era for which $p=\rho/3$, it is clear that the
$3p$ term should have an important impact on BBN. In order test the
influence of the $3p$ term, we follow the same strategy as for neutron
stars and introduce an ad-hoc modification of GR characterized by the
$\chi$ parameter, namely
\begin{equation}
\label{eq:accelerationmodi}
\frac{\ddot{a}}{a}=-\frac{1}{6\Mp^2}(\rho +3\chi p). 
\end{equation}
This equation should be compared to Eq.~(\ref{eq:acceleration}). In
order to derive the Friedmann equation, we need another equation and
we can use the first law of thermodynamics for an adiabatic expansion,
namely ${\rm d}(a^3\rho)=-p{\rm d}(a^3)$, written for a co-moving
volume or, equivalently, $\dot{\rho}+3\dot{a}(\rho+p)/a=0$. Then,
noticing that $\ddot{a}/a=1/(2\dot{a}a){\rm d}(\dot{a}^2)/{\rm d}t$
and using the conservation equation, it is straightforward to derive
the following relation
\begin{equation}
{\rm d}\left(\dot{a}^2\right)=-\frac{1}{3\Mp^2}
\left[\left(1-3\chi\right)\rho a{\rm d}a-\chi a^2{\rm d}\rho\right].
\end{equation}
If $\chi=1$, it is easy to check that
\begin{equation}
{\rm d}\left(\dot{a}^2\right)=\frac{1}{3\Mp^2}{\rm d}\left(\rho a^2\right),
\end{equation}
which gives
\begin{equation}
H^2=\frac{\rho}{3\Mp^2}+\frac{C}{a^2},
\end{equation}
where $C$ is an integration constant leading to a curvature term. Now,
if $\chi \neq 1$ and $p=w\rho$, where $w$ is a constant equation of
state parameter, then one obtains
\begin{equation}
H^2=\frac{1+3\chi w}{1+3w}\frac{\rho}{3\Mp^2}+\frac{C}{a^2}.
\end{equation}
Using this modified Friedmann equation with $w=1/3$ and ignoring the
curvature term (which is sub-dominant in presence of radiation as
shown by the cosmological data), one obtains
\begin{equation}
\label{eq:friedmodified}
H^2=\frac{1+\chi}{2}\frac{\rho_{\rm rad}}{3\Mp^2}.
\end{equation}
Therefore, the effect of the term proportional to $\chi$ is to modify
the expansion rate of the Universe in the radiation dominated era. Or,
if one uses the fact that the energy density of radiation is
$\pi^2g_*T^4/30$, we see that this is also equivalent to changing the
effective number of relativistic degrees of freedom, namely
$g_*'=g_*(1+\chi)/2$.

\par

Ref.~\cite{Rappaport:2007ct} has performed BBN calculations, assuming
Eq.~(\ref{eq:friedmodified}), and computed the abundance of deuterium,
helium-$4$ and lithium-$7$. The isocontours are represented in
Fig.~\ref{fig:bbn} in the plane $(\eta_{10},\chi)$. The parameter
$\eta_{10}$ is defined by $\eta_{10}\equiv 10^{10}\eta$ where
$\eta\equiv n_{_{\rm B}}/n_{\gamma}=\Omega_{_{\rm B}}h^2\pi^2\rho_{\rm
  cri}/[m_{_{\rm B}}h^22T^3\zeta(3)] \simeq 2.73\times
10^{-8}\Omega_{_{\rm B}}h^2$ [in the last expression, $m_{_{\rm
      B}}\simeq 939.6\mbox{MeV}$ is the baryon (neutron) mass,
  $\rho_{\rm cri}\simeq 8.099\times 10^{-47}\mbox{GeV}^4$ is the
  critical energy density today, $T\simeq 2.7255\mbox{K}$ is the CMB
  temperature and $\zeta(3)\simeq 1.20206$]. Green contours represent
the deuterium abundance $(D/H)_{_{\rm P}}$, blue contours are
helium-$4$ abundance $Y_{_{\rm P}}$ and purple contours are
lithium-$7$ abundance. We see that deuterium abundance mainly
determines $\eta_{10}$ while helium-$4$ abundance gives good
constraints on the new parameter $\chi$. Observations indicate that
$\log\left(D/H\right)_{_{\rm P}}=-4.55\pm 0.04$~\cite{O'Meara:2006mj}
and $Y_{_{\rm P}}=0.24\pm 0.006$~\cite{Steigman:2007xt}. Then, one can
identify the region in the space $(\eta_{10},\chi)$ which is
consistent with those observations. This is indicated in
Fig.~\ref{fig:bbn} by the two gray ellipses (corresponding to two
slightly different assumptions about the abundances inferred from the
observations). Without entering a detailed discussion, the conclusion
is that $\chi\simeq 1$ is compatible with observations and that the
value $\chi=0$ is strongly ruled out. Therefore, self-gravity of
pressure is, in some sense, confirmed by cosmological observations.

\par

The previous considerations ``validate'' the mechanism on which
inflation is based. Inflation thus appears as a well-justified
theory. In the next section, we therefore describe this theory in more
detail and discuss the micro-physics of inflation.

\section{The Micro-Physics of Inflation or How to Parametrize Inflation}
\label{sec:paraminf}

\subsection{Inflation and High Energy Physics}
\label{subsec:hepinf}

We have seen in the last section that, in order to have a phase of
inflation, we need a situation where the fluid dominating the matter
content of the Universe has a negative pressure. The next question is
of course which type of matter can have this property. In order to
answer this question, let us first remark that inflation is a high
energy phenomenon by particle physics standards since it is supposed
to occur in the early Universe. In this situation, the relevant
framework to describe matter is not fluid mechanics but field
theory. And the simplest field, compatible with isotropy and
homogeneity, is a time dependent scalar field $\phi(t)$ since it has
no preferred direction and is space-independent. Moreover, in a FLRW
Universe, the energy density and pressure of a scalar field are given
by
\begin{equation}
\rho =\frac{\dot{\phi}^2}{2}+V(\phi), \quad 
 p=\frac{\dot{\phi}^2}{2}-V(\phi). 
\end{equation}
As a consequence, in a situation where the potential energy dominates
over the kinetic energy, namely when the field moves slowly or,
equivalently, when the potential is flat, one obtains a negative
pressure and, hence, inflation. The field which drives inflation is
called the ``inflaton''.

\par

Let also notice that, when $V(\phi)\gg \dot{\phi}^2$, the equation of
state is $p\simeq -\rho$ which, using the conservation equation,
immediately implies that the energy density, and therefore the Hubble
parameter $H$, is almost a constant. The Friedmann equation then leads
to a scale factor $a(t)\propto {\rm e}^{Ht}$. In other words inflation
is also a phase of quasi-exponential expansion. Moreover, using the
expressions established above, one also has
\begin{equation}
\label{eq:omegak}
\vert \Omega _{\cal K}\vert \equiv \left\vert \frac{\rho_{\rm
    curv}}{\rho_{\rm cri}} \right \vert =\frac{\vert {\cal
    K}\vert}{a^2 H^2},
\end{equation}
and we see that $\Omega _{\cal K}$ goes exponentially to zero during
inflation. We therefore expect to measure a vanishing spatial
curvature: this is a first generic prediction of inflation and we will
see in Sec.~\ref{sec:observ} that it is good agreement with the most
recent cosmological observations.

\par

As mentioned before, inflation is a high energy phenomenon and,
therefore, a concrete implementation necessarily rests on high energy
physics. In the modern view, the micro-physics of inflation should
therefore be described by an effective field theory characterized by a
cutoff $\Lambda$. If the gravitational sector is described by GR,
which itself is viewed as an effective theory with a cutoff at the
Planck scale, then $\Lambda <\Mp$. On the other hand, we know that the
Hubble parameter during the part of the inflationary phase we have
observationally access to can be as large as $10^{15} \GeV$ and this
suggests $\Lambda >10^{15}\GeV$. Clearly, at those energy scales,
particle physics remains speculative and this is the reason why there
is currently a plethora of different inflationary scenarios. A priori,
without any further theoretical guidance, the effective action can
therefore be written as
\begin{eqnarray}
\label{eq:generalS}
S &=& \int {\rm d}^4x\sqrt{-g}
\biggl[\Mp^2\Lambda _{_{\rm B}}+\frac{\Mp^2}{2}R+
aR^2+b R_{\mu \nu}R^{\mu \nu}+\frac{c}{\Mp^2}R^3+\cdots
\nonumber \\ & & 
-\frac{1}{2}\sum _i g^{\mu
  \nu}\partial_\mu\phi_i\partial_{\nu }\phi_i
-V(\phi_1, \cdots ,\phi_n)
+\sum_i d_i 
\frac{{\cal O}_i}{\Lambda^{n_i-4}}\biggr]
\nonumber \\ & &
+S_{\rm int}(\phi_1, \cdots, \phi_n,A_{\mu},\Psi)
+\cdots
\end{eqnarray}
In the above equation, the first line represents the effective
Lagrangian for gravity ($\Lambda_{_{\rm B}}$ is the cosmological
constant). In practice, we will mainly work with the Einstein-Hilbert
term only. The second line represents the contribution of matter. We
assumed that several scalar field are present (a priori, there is no
reason to assume that only one field plays a role). The two first
terms are the canonical Lagrangian while ${\cal O}_i$ represents a
higher order operator of dimension $n_i>4$, the amplitude of which is
determined by the coefficient $d_i$. Those corrections can modify the
potential but also the (standard) kinetic
term~\cite{Chialva:2014rla}. The last term encodes the interaction
between the inflaton fields and the rest of the world, \ie the gauge
fields $A_{\mu}$ and the fermions $\Psi$. The dots stand for the rest
of the terms such as the Lagrangians of $A_{\mu}$, of $\Psi$, the
corresponding higher order operators etc ... . Notice that the above
description is not completely general. For instance, suppose that the
action of the inflation field is of the Dirac-Born-Infeld (DBI)
type~\cite{Alishahiha:2004eh}, namely
\begin{equation}
S = \int {\rm d}^4x\sqrt{-g}
\Biggl[\frac{\Mp^2}{2}R-T(\phi)
\sqrt{1-2\frac{X}{T(\phi)}}+T(\phi)-V(\phi)\Biggr],
\end{equation}
where $X\equiv -1/2g^{\mu
  \nu}\partial_{\mu}\phi\partial_{\nu}\phi$. An expansion in $X$ gives
\begin{equation}
S = \int {\rm d}^4x\sqrt{-g}
\biggl(\frac{\Mp^2}{2}R-X+V+\frac{X^2}{2T(\phi)} +\cdots \biggr),
\end{equation}
and we see that the higher order terms are not suppressed by a fixed
cutoff $\Lambda$ but by $T(\phi)$. In this case, in some sense, the
cutoff has become field dependent. As a consequence, the canonical
Lagrangian $X-V$ is not necessarily always the first term of the
series and it makes sense to also consider more complicated cases,
even at ``leading order''.

\par

Another, but related, question is whether the higher order operators
can be neglected during inflation. Firstly, it is necessary that the
field excursion $\Delta \phi$ be small in comparison with the cutoff
scale, \ie $\Delta \phi<\Lambda$. Whether this is the case or not
depends on the model. Second, the tree level potential $V$ can
receive corrections that can be difficult to control. For instance,
if there is a mass term, then typically the mass $m$ becomes
\begin{equation}
m^2\rightarrow m^2+gM^2\ln \left(\frac{\Lambda}{\mu}\right),
\end{equation}
where $\mu $ is a renormalization scale, $M$ the mass of a heavy field
and $g$ the coupling between $\phi$ and the heavy field. If
$M>\Lambda$ then one has $m>H$ since we have $\Lambda >H$. This means
that the potential is no longer flat enough to support inflation, an
embarrassing problem indeed! Ways out consist in assuming that the
coupling $g$ is small or, more convincingly, that symmetries forbid
this type of corrections.

\par

Finally, let us say a few words about the interaction term. Usually,
it is considered to be negligible during the slow-roll phase. If this
is not the case, it leads to warm
inflation~\cite{Berera:1995ie,Yokoyama:1998ju,Berera:2008ar}. Even if
it does not play a role during the accelerated phase, the interaction
term is of fundamental importance for inflation since it is
responsible for the reheating stage, that is to say it explains how
inflation is smoothly connected to the standard hot Big Bang epoch.

\par

We see that, using theoretical considerations only, it is difficult to
restrict the Lagrangian of inflation to a simple form. But, in fact,
the point is that the CMB Planck data can do the job and can constrain
the Lagrangian~(\ref{eq:generalS}). For instance, we will see in the
following that the perturbations are adiabatic (at least for the
moment; this could of course very well change when more accurate data
are collected) and this supports the idea that only one scalar field
is at play during inflation. Moreover, we will also show that
Non-Gaussianities have been measured to be compatible with zero and
this supports the fact that the kinetic term must be standard. We are
therefore led to consider that inflation is described by the simplest
scenario, namely single-field slow-roll with a standard kinetic
term. It is important to emphasize that we are pushed to this class of
models, which is clearly easier to analyze than
Eq.~(\ref{eq:generalS}), not because we want to simplify the scenario
but because this is what the CMB data suggest. In this framework, the
inflationary Lagrangian can be written as
\begin{equation}
\label{eq:normallagrange}
  \calL=-\frac{1}{2}g^{\mu \nu}\partial_{\mu}\phi\partial_{\nu }\phi
-V(\phi)+\calL_{\rm int}(\phi,A_{\mu},\Psi).
\end{equation}
In the following, we will ignore the interaction term during the
accelerated phase and will consider its effect only at the end of
inflation (the ``reheating'' phase). We see that we are left with a
model that contains only one arbitrary function, the potential
$V(\phi)$. Therefore, what remains to be done in order to completely
characterize inflation is to constrain this a priori arbitrary
function with cosmological data. This line of research has played a
dominant role in the recent years.

\par

Let us now describe the slow-roll formalism which is used in practice
to derive the inflationary predictions of the models mentioned
above. As already remarked previously, one can distinguish two
different phases of evolution: the slow-roll phase and the reheating
phase. In principle, once $V(\phi)$ and $\calL_{\rm
  int}(\phi,A_{\mu},\Psi)$ are known, the model is completely
specified. In practice, however, one proceeds in a slightly different
way. The function $V(\phi)$ is considered to be relevant for a limited
range of field values only, corresponding to our observable
window. Then, the evolution of the system is controlled by the
Friedmann and Klein-Gordon equations, namely
\begin{align}
\label{eq:friedman}
H^2 &=\frac{1}{3\Mp^2}\left[\frac{\dot{\phi}^2}{2}+V(\phi)\right], \\
\label{eq:kg}
\ddot{\phi} & +3H\dot{\phi}+V_{\phi} = 0,
\end{align}
where we remind that $H\equiv \dot{a}/a$ denotes the Hubble parameter
and where a subscript $\phi$ means a derivative with respect to the
inflaton field. It is also interesting to introduce the Hubble flow
functions $\epsilon_n$ defined by~\cite{Schwarz:2001vv,Leach:2002ar}
\begin{equation}
\label{eq:defhf}
\epsilon_{n+1} \equiv \frac{\dd \ln \left \vert 
\epsilon_n \right \vert}{\dd N}, 
\quad n\ge 0,
\end{equation}
where $\epsilon_0\equiv H_\uini/H$ starts the hierarchy and $N\equiv
\ln(a/a_\uini)$ is the number of e-folds. These parameters provide
useful information about the inflationary dynamics. For instance, the
first slow-roll parameter can be expressed as
\begin{equation}
\label{eq:eps1}
\epsilon_1=-\frac{\dot{H}}{H^2}=1-\frac{\ddot{a}}{aH^2},
\end{equation}
and, therefore, inflation ($\ddot{a}>0$) occurs if $\epsilon_1<1$. In
fact, since the parameters $\epsilon_n$ are defined in terms of $H$
and since $H$ is determined once $V(\phi)$ is known, see
Eqs.~(\ref{eq:friedman}) and~(\ref{eq:kg}), it follows that one can
also express them in terms of the potential. For instance,
$\epsilon_1$ is given by
\begin{equation}
\epsilon_1=\frac{3\dot{\phi}^2}{2}\frac{1}{\dot{\phi}^2/2+V(\phi)}.
\end{equation}
In fact, it is not sufficient to have $\epsilon_1<1$ but one also
needs $\epsilon_1\ll 1$. Indeed, from the above expression, we see
that this corresponds to a situation where $\dot{\phi}^2/2\ll V(\phi)$
or, in other words, to a situation where the potential is very flat
since the field must roll very slowly. We just recover the case
considered in the previous section. In this situation, referred to as
the slow-roll approximation, one has in fact $\epsilon_n\ll 1$ for any
$n$. If this is the case, then the Hubble flow functions can be
expressed as~\cite{Liddle:1994dx}
\begin{align} 
\label{eq:epsfirst}
\epsilon_1 & \simeq
\frac{\Mp^2}{2}\left(\frac{V_\phi}{V}\right)^2, \\ 
\label{eq:eps2}
\epsilon_2 & \simeq
2\Mp^2\left[\left(\frac{V_\phi}{V}\right)^2
-\frac{V_{\phi \phi}}{V}\right], \\
\label{eq:eps3}
\epsilon_2\epsilon_3 & \simeq 2\Mp^4\left[
\frac{V_{\phi \phi \phi}V_\phi}{V^2}-3\frac{V_{\phi \phi}}{V}
\left(\frac{V_\phi}{V}\right)^2
+2\left(\frac{V_\phi}{V}\right)^4\right].
\end{align} 
The slow-roll approximation allows us to simplify the equations of
motion and to analytically integrate the inflaton trajectory. Indeed,
Eqs.~(\ref{eq:friedman}) and~(\ref{eq:kg}), which control the
evolution of $\phi$, can be rewritten as
\begin{eqnarray}
\label{eq:exactfried}
H^2 &=& \frac{V}{\Mp^2(3-\epsilon_1)}, \\
\left(1 + \dfrac{\epsilon_2}{6 - 2\epsilon_1}\right)
\dfrac{\ud \phi}{\ud N} &=& - \Mp^2 \dfrac{\ud \ln V}{\ud \phi}\,.
\end{eqnarray}
As a consequence, in the slow-roll approximation, one has $H^2\simeq
V/(3\Mp^2)$ and ${\rm d}\phi/{\rm d}N\simeq -\Mp^2{\rm d}\ln V/{\rm
  d}\phi$, from which one obtains
\begin{equation}
\label{eq:srtrajectory}
N-\Nini=-\frac{1}{\Mp^2}\int_{\phiini}^{\phi}\frac{V(\chi)}
{V_\chi(\chi)}\, \ud \chi \, ,
\end{equation} 
$\phiini$ being the initial vacuum expectation value of the field. It
is clear from the above considerations that the inflaton dynamics is
entirely determined once the potential $V(\phi)$ has been
specified. Since, in addition, the function $V(\phi)$ allows us to
make the connection with high energy physics, it appears as a natural
tool to parametrize inflation.

\subsection{Other parametrizations?}
\label{subsec:otherparam}

Recently, other parametrizations of inflation have been
considered. The motivation of these works was to establish a general
framework in order to characterize what the generic or typical
predictions of cosmic inflation are. In this section, we discuss them
and show that, in fact, they all boil down to choosing a specific
potential.

\par

The first alternative parametrization that we discuss is the so-called
``horizon-flow
approach''~\cite{Hoffman:2000ue,Kinney:2002qn,Liddle:2003py,Ramirez:2005cy,Chongchitnan:2005pf}. It
has been recently discussed in detail in
Ref.~\cite{Vennin:2014xta}. Let us define a new set of parameters
${}^{\ell }\lambda$ given by
\begin{equation}
\label{eq:deflambda}
{}^{\ell}\lambda=\left(2\Mp^2\right)^{\ell}
\frac{\left(H'\right)^{\ell-1}}{H^{\ell}}
\frac{{\rm d}^{\ell+1}H}{{\rm d}\phi^{\ell+1}}.
\end{equation}
Of course, this new definition does not bring any new information. The
new parameters can be expressed in terms of the previous ones, for
instance ${}^1\lambda=\epsilon_1-\epsilon_2/2$,
${}^2\lambda=\epsilon_1^2-3\epsilon_1\epsilon_2/2+\epsilon_2\epsilon_3/2$,
etc \dots . It only shows that, if the $\epsilon_n$'s are all of the
same order in slow-roll, the ${}^{\ell }\lambda $ are of increasing
order. Then, the simple equation, see Eq.~(\ref{eq:defhf})
\begin{equation}
\frac{{\rm d}\epsilon_{n}}{{\rm d}N}=\epsilon_n\epsilon_{n+1}
\end{equation}
is replaced with 
\begin{eqnarray}
\frac{{\rm d}\epsilon_1}{{\rm d}N}&=& \epsilon_1\epsilon_2, \\
\frac{{\rm d}\epsilon_2}{{\rm d}N}&=& 
2{}^2\lambda-2\epsilon_1^2-3\epsilon_1\epsilon_2, \\
\frac{{\rm d}{}^{\ell}\lambda }{{\rm d}N}&=& 
-{}^{\ell+1}\lambda-{}^{\ell}\lambda
\left(\frac{\ell-1}{2}\epsilon_2-\epsilon_1\right).
\end{eqnarray}
The idea is now to truncate this hierarchy at some order $M$, \ie to
assume that ${}^{\ell}\lambda =0$ for $\ell >M$, maybe motivated by
the fact that higher order equations deal with higher order slow-roll
parameters and are thus, in some sense, negligible. Then, this finite
set of equations (in practice, the case $M=5$ has been considered) is
numerically integrated many times with different initial
conditions~\cite{Hoffman:2000ue}. In this way, one obtains different
values of the slow-roll parameters at Hubble radius crossing and,
since the observables such as the spectral index $\nS$ or the
tensor-to-scalar ratio $r$ can be expressed in terms of these
parameter (see below), different inflationary predictions. The next
step consists in searching systematic patterns in these predictions
which, as a consequence, would be considered as ``typical'' of
inflation. In particular, it has been claimed that the different
predictions for $\nS$ and $r$ obtained in this way cluster around the
relation~\cite{Kinney:2002qn,Ramirez:2005cy}
\begin{equation}
\label{eq:1/3}
r_{16}\simeq \frac{1}{3}(1-\nS),
\end{equation}
where $r_{16}\equiv r/16$. The above equation is then viewed as a
generic prediction of inflation, obtained without the need to specify
a particular potential $V(\phi)$. 

\par

However, the above claim is not correct~\cite{Vennin:2014xta}. Indeed,
truncating the hierarchy at order $M$ clearly means that one assumes,
see Eq.~(\ref{eq:deflambda}),
\begin{equation}
\frac{{\rm d}^{M+2}H}{{\rm d}\phi^{M+2}}=0,
\end{equation}
an equation which can be easily integrated (!) and leads
to~\cite{Liddle:2003py}
\begin{equation}
H(\phi)=H_0\left[1+\sum_{i=1}^{M+1}A_i
\left(\frac{\phi}{\Mp}\right)^i\right].
\end{equation}
Then, from this expression of the Hubble parameter, one can easily
calculate the corresponding inflationary potential and one obtains
\begin{eqnarray}
\label{eq:potHF}
V(\phi)&=&3\Mp^2H^2(\phi)-2\Mp^4H'(\phi)
\\
&=&3\Mp^2H_0^2\left[1+\sum_{i=1}^{M+1}A_i
\left(\frac{\phi}{\Mp}\right)^i\right]^2-2\Mp^3H_0
\sum_{i=1}^{M+1}i A_i  \left(\frac{\phi}{\Mp}\right)^{i-1}.
\end{eqnarray}
The whole procedure is therefore nothing but a particular choice of a
potential $V(\phi)$ depending on $M+1$ parameters, $A_i$. Moreover,
Ref.~\cite{Vennin:2014xta} has shown that the ``mysterious''
coefficient $1/3$ in Eq.~(\ref{eq:1/3}) can be easily recovered if one
carries out a standard slow-roll analysis of the
potential~(\ref{eq:potHF}). We conclude that this approach is not
generic at all and only consists in studying a very particular
potential.

\par

More recently, it has also been argued that, rather than choosing a
potential $V(\phi)$, it is more generic to choose the equation of
state during inflation, see Refs.~\cite{Mukhanov:2013tua,
  Binetruy:2014zya,Mukhanov:2014uwa}. So, in practice, what is done is
an educated guess for $w(N)=p/\rho$. Notice that, since
\begin{equation}
1+w(N)=\frac{2}{3}\epsilon_1(N),
\end{equation}
this is also equivalent to choosing a particular function
$\epsilon_1(N)$, which is the strategy followed in
Refs.~\cite{Roest:2013fha,Garcia-Bellido:2014gna,Binetruy:2014zya}. Concretely,
one takes
\begin{equation}
\label{eq:w}
1+w(N)=\frac{\beta}{\left(\Nend-N\right)^{\alpha}},
\end{equation}
where $\alpha$ and $\beta $ are two free and positive parameters and
$\Nend$ is the number of e-folds at the end of inflation. However,
again, this choice is in fact a choice of $V(\phi)$. Indeed, the
slow-roll trajectory~(\ref{eq:srtrajectory}), ${\rm d}N=-V{\rm
  d}\phi/(\Mp^2 V')$ can be re-written as
\begin{equation}
\Mp^2\frac{{\rm d}}{{\rm d}N}\left(\ln V\right)
\simeq -\left(\frac{{\rm d}\phi}{{\rm d}N}\right)^2,
\end{equation}
and, from the exact formula
\begin{equation}
\epsilon_1=\frac{1}{2\Mp^2}\left(\frac{{\rm d}\phi}{{\rm d}N}\right)^2,
\end{equation}
one obtains the following system of equations
\begin{eqnarray}
\label{eq:phieqstate}
\left(\frac{{\rm d}\phi}{{\rm d}N}\right)^2 
&=& 3\Mp^2\left[1+w(N)\right], \\
\label{eq:Veqstate}
\frac{{\rm d}}{{\rm d}N}\left(\ln V\right)
&=& -3\left[1+w(N)\right].
\end{eqnarray}
When the above set of equations is solved one obtains $\phi(N)$ and
$V(N)$ and, eventually eliminating $N$, the function $V(\phi)$. We
conclude that giving $w(N)$ and/or $\epsilon_1(N)$ is not a new
generic parametrization but just a particular choice of a
potential. In order to illustrate this point, let us see how it works
in practice for the case of Eq.~(\ref{eq:w}). The trajectory, given by
Eq.~(\ref{eq:phieqstate}), can be written as
\begin{equation}
\label{eq:phiw}
\frac{\phi}{\Mp}=C_1\pm \sqrt{3\beta}\frac{2}{\alpha-2}
\left(\Nend-N\right)^{(2-\alpha)/2},
\end{equation}
where $C_1$ is an integration constant. For the potential, the
integration of Eq.~(\ref{eq:Veqstate}) is also straightforward and one
finds
\begin{equation}
\label{eq:Vw}
\ln V=C_2+\frac{3\beta}{1-\alpha}\left(\Nend-N\right)^{1-\alpha},
\end{equation}
where $C_2$ is another integration constant. Then, from
Eq.~(\ref{eq:phiw}), one arrives at
\begin{equation}
\Nend-N=\left[\pm \frac{\alpha -2}{2\sqrt{3\beta}}
\left(\frac{\phi}{\Mp}-C_1\right)\right]^{2/(2-\alpha)},
\end{equation}
and, inserting this result in Eq.~(\ref{eq:Vw}), one obtains
\begin{equation}
\ln V=C_2+\frac{3\beta}{1-\alpha}
\left[\pm \frac{\alpha -2}{2\sqrt{3\beta}}
\left(\frac{\phi}{\Mp}-C_1\right)\right]^{2(1-\alpha)/(2-\alpha)}.
\end{equation}
This shows that Eq.~(\ref{eq:w}) is, in the slow-roll approximation, 
completely equivalent to the choice 
\begin{equation}
V(\phi)=M^4{\rm e}^{\delta \phi^{\gamma}},
\end{equation}
where $\delta $ and $\gamma$ are
constants~\cite{Mukhanov:2014uwa}. This potential is almost identical
to Logamediate inflation, LMI in the terminology of
Refs.~\cite{Martin:2014vha,Martin:2013nzq}, $V(\phi)=M^4x^{\alpha}{\rm
  e}^{\delta x^{\gamma}}$, with $x=(\phi-\phi_0)/\Mp$, in the case
where $\alpha=0$. This model was studied in detail in
Refs.~\cite{Martin:2014vha,Martin:2013nzq}. The only difference is
that, for LMI, one has $\alpha=4(1-\gamma)$ implying $\gamma=1$ when
$\alpha=0$, which is not the case here (\ie $\alpha =0$ but $\gamma$
is still free).
 
\par

We conclude that all the so-called ``alternative'' parametrizations of
inflation considered so far are in fact strictly equivalent to
specifying a potential. Claiming that it is either new or different or
better seems definitively far-fetched. In addition, discussing
inflation in terms of $V(\phi)$ has the advantage to make the link
with high energy physics explicit. For these reasons, we conclude that
working in terms of $V(\phi)$ and scanning the inflationary landscape
by considering all possible models seems to be the most efficient
method to learn about inflation.

\subsection{Parametrization of Reheating}
\label{subsec:reheatparam}

Let us now consider the end of inflation, namely the reheating phase,
and how one can describe it. When $\epsilon_1=1$, the potential is no
longer flat enough to support an accelerated phase and inflation
stops. Usually, this happens in the vicinity of the ground state
(concretely, the minimum of the potential). At this time, the inflaton
field starts oscillating and decaying. Then, these decay products
thermalize~\cite{Podolsky:2005bw} and the radiation dominated epoch of
the hot Big Bang phase commences. The micro-physics of reheating is
described by the term $\calL_{\rm int}(\phi,A_{\mu},\Psi)$ in
Eq.~(\ref{eq:normallagrange}). But, in fact, in order to parametrize
reheating, we do not need to have such a detailed description. Indeed,
as we will see in the following, the inflationary observational
predictions are expressed in terms of $\epsilon_{n*}\equiv
\epsilon_n(\phi_*)$, where $\phi_*$ is the value of $\phi$ when the
pivot scale $k_{_{\rm P}}$ leaves the Hubble radius during inflation
(the pivot scale is conveniently chosen in the middle of the
observable window). Since, in the slow-roll approximation, we know the
trajectory $\phi=\phi(N)$, we just need to determine $\Nstar$ such
that $\phi_*=\phi(\Nstar)$. This can be done as follows. The physical
pivot scale during inflation is given by
\begin{equation}
\label{eq:pivot}
\frac{k_{_{\rm P}}}{a(N)}=\frac{k_{_{\rm P}}}{a_{\rm now}}
\frac{a_{\rm now}}{a_{\rm reh}}\frac{a_{\rm reh}}{a_{\rm end}}
\frac{a_{\rm end}}{a(N)}=\frac{k_{_{\rm P}}}{a_{\rm now}}
\frac{a_{\rm now}}{a_{\rm reh}}\frac{a_{\rm reh}}{a_{\rm end}}
{\rm e}^{\Nend-N},
\end{equation}
where $a_{\rm end}$ denotes the scale factor at the end of inflation
and $a_{\rm reh}$ the scale factor at the end of reheating. In the
above expression, $k_{_{\rm P}}/a_{\rm now}$ is known and, concretely,
we take $k_{_{\rm P}}/a_{\rm now}=0.05 \Mpc^{-1}$. The quantity
$a_{\rm now}/a_{\rm reh}$ is also known since it only involves the
standard thermal history of the Universe. On the other hand, the ratio
$a_{\rm reh}/a_{\rm end}$ depends on what happens during reheating and
this is precisely the reason why the inflationary predictions are
sensitive to this phase of evolution. To go further, we write the
above equation at the time $N=\Nstar$. Since, by definition, $k_{_{\rm
    P}}/a(\Nstar)=H(\Nstar)$, Eq.~(\ref{eq:pivot}) becomes
\begin{equation}
\label{eq:Hreheat}
H(\Nstar)=\frac{1}{\Mp}\sqrt{\frac{V(\Nstar)}{3-\epsilon_1(\Nstar)}}
=\frac{k_{_{\rm P}}}{a_{\rm now}}
\frac{a_{\rm now}}{a_{\rm reh}}\frac{a_{\rm reh}}{a_{\rm end}}
{\rm e}^{\Nend-\Nstar},
\end{equation}
the first expression being just the Friedmann equation, see
Eq.~(\ref{eq:exactfried}). We see that this is a transcendental
equation for $N_*$ which, therefore, needs to be solved
numerically. We also see that it depends on the potential $V(\phi)$
and, hence, on the model under consideration. Finally, in order to
solve this equation, one needs to estimate the quantity $a_{\rm
  reh}/a_{\rm end}$. Let $\rho $ and $p$ be the total energy density
and pressure during reheating. Notice that one can have several
fluids, possibly interacting which each others. The treatment
presented here is therefore completely general. Conservation of total
energy density (we emphasize again that it is not necessary to assume
that the energy density of each fluid is separately conserved) implies
that
\begin{equation}
\label{eq:rhorehend} 
\rho\left(N\right)=\rhoend
\exp\left\{-3\int_{\Nend}^{N}\left[1+ \wreh \left(n\right)\right]{\ud}
n\right\},
\end{equation} 
where $\wreh\equiv p/\rho$ is the ``instantaneous'' equation of state
during reheating. Then, let us define the mean equation of state
parameter, $\wrehbar$, by
\begin{equation} 
\label{eq:wrehbar}
\wrehbar \equiv \frac{1}{\Delta N}\int_{\Nend}^{\Nreh} \wreh(n)\dd n,
\end{equation}
where $\Delta N \equiv \Nreh - \Nend$
is the total number of e-folds during reheating. It follows that
\begin{equation}
\rho_{\rm reh}=\rho_{\rm end}{\rm e}^{-3(1+\wrehbar)\Delta N},
\end{equation}
and, therefore, 
\begin{equation}
\label{eq:ratioreheat}
{\rm e}^{\Delta N}=\frac{a_{\rm reh}}{a_{\rm end}}
=\left(\frac{\rho_{\rm reh}}{\rho_{\rm
    end}}\right)^{-1/(3+3\wrehbar)}.
\end{equation}
As a consequence, the ratio $a_{\rm reh}/a_{\rm end}$ depends on two
quantities only: the energy density at the end of reheating,
$\rho_{\rm reh}$, and the mean equation of state during reheating,
$\wrehbar$. Once a model of inflation is known, $\rho_{\rm end}$ can
be calculated so this is not a new quantity (but, again, it introduces
an additional dependence on the inflationary potential). Inserting
Eq.~(\ref{eq:ratioreheat}) into the above
expression~(\ref{eq:Hreheat}) leads to
\begin{equation}
H(\Nstar)=\frac{1}{\Mp}\sqrt{\frac{V(\Nstar)}{3-\epsilon_1(\Nstar)}}
=\frac{k_{_{\rm P}}}{a_{\rm now}}
\frac{a_{\rm now}}{a_{\rm reh}}
\left(\frac{\rho_{\rm reh}}{\rho_{\rm end}}\right)^{-1/(3+3\wrehbar)}
{\rm e}^{\Nend-\Nstar}.
\end{equation}
The above formula still contains $a_{\rm reh}$, a quantity that we
would like to eliminate from the final expression. For this purpose,
we write $a_{\rm now}/a_{\rm reh}$ as $a_{\rm now}/a_{\rm eq}\times
a_{\rm eq}/a_{\rm reh}$, where $a_{\rm eq}$ is the scale factor at
matter-radiation equality. Then, we use the fact that, during the
radiation dominated era, $a\propto \rho^{-1/4}$, to write
\begin{eqnarray}
H(\Nstar)&=&\frac{1}{\Mp}\sqrt{\frac{V(\Nstar)}{3-\epsilon_1(\Nstar)}}
=\frac{k_{_{\rm P}}}{a_{\rm now}}
\frac{a_{\rm now}}{a_{\rm eq}}
\left(\frac{\rho_{\rm reh}}{\rho_{\rm eq}}\right)^{1/4}
\left(\frac{\rho_{\rm reh}}{\rho_{\rm end}}\right)^{-1/(3+3\wrehbar)}
{\rm e}^{\Nend-\Nstar} \nonumber \\
\\
\label{eq:interHreheat}
&=& 
\frac{k_{_{\rm P}}}{a_{\rm now}}
\frac{a_{\rm now}}{a_{\rm eq}}
\frac{\Mp}{\rho_{\rm eq}^{1/4}}
\frac{\rho_{\rm end}^{1/2}}{\Mp^2}
\frac{\Mp}{\rho_{\rm end}^{1/4}}
\left(\frac{\rho_{\rm reh}}{\rho_{\rm end}}\right)^{1/4-1/(3+3\wrehbar)}
{\rm e}^{\Nend-\Nstar}.
\end{eqnarray}
Except the quantities that are known from standard cosmology (since
they only depend on post-inflationary physics), such as $a_{\rm
  now}/a_{\rm eq}\times \Mp/\rho_{\rm eq}^{1/4}$, we see that this
equation singles out the following combination (by definition, the
``reheating''
parameter)~\cite{Martin:2006rs,Lorenz:2007ze,Martin:2010kz,Martin:2014nya}
\begin{equation}
\label{eq:defR}
R\equiv \frac{\rho_{\rm end}^{1/4}}{\Mp}\Rrad,
\end{equation}
with 
\begin{equation}
\label{eq:defRrad}
\Rrad\equiv \left(\frac{\rho_{\rm reh}}
{\rho_{\rm end}}\right)^{-1/4+1/(3+3\wrehbar)}
=\left(\frac{\rho_{\rm reh}}
{\rho_{\rm end}}\right)^{(1-3\wrehbar)/(12+12\wrehbar)}.
\end{equation}
Notice that we have a term $\rho_{\rm end}^{1/2}/\Mp^2$ left in
Eq.~(\ref{eq:interHreheat}). It is introduced because it produces a
term proportional to the square root of the potential at the end of
inflation and combines nicely with the $\sqrt{V_*}$ on the left hand
side of Eq.~(\ref{eq:interHreheat}). The arguments presented above can
be easily generalized to take into account a change of relativistic
degrees of freedom between the reheating epoch and today, see
Ref.~\cite{Martin:2014vha}.

\par

The reheating parameter encodes what can be learned about reheating
from the CMB. In sec.~\ref{sec:observ}, we will see that the Planck
data already put constraints on its value.

\section{Inflationary perturbations}
\label{sec:pert}

In this section, we review the theory of inflationary
perturbations~\cite{Mukhanov:1990me,Bardeen:1980kt,Peter:2013woa}. This
part of the inflationary scenario is very important because it allows
us to use astrophysical data to put constraints on cosmic inflation. In
the following, we pay special attention to the question of how one can
calculate the correlation functions of the perturbations and to the
concept of adiabatic and isocurvature perturbations. As will be seen
in Sec.~\ref{sec:observ}, these quantities carry useful information
about the type of inflationary model that is realized in Nature. This
section can therefore be viewed as a preparation to
Sec.~\ref{sec:observ} in the sense that we discuss in some detail the
meaning of the quantities that have been measured recently by the
Planck experiment.

\par

To describe CMB anisotropies and large scale structures, one must go
beyond the cosmological principle. This is a priori a technically
difficult task but since the inhomogeneities are small in the early
Universe, one can use a perturbative approximation which, obviously,
greatly simplifies the problem. Then, the idea is to write the metric
tensor as $g_{\mu \nu}(\eta, {\bm x})=g_{\mu \nu}^{_{\rm FLRW}}(\eta)
+\delta g_{\mu \nu}(\eta, {\bm x})+\cdots $, where $g_{\mu \nu}^{_{\rm
    FLRW}}(\eta)$ represents the metric tensor of the FLRW Universe
and where $\delta g_{\mu \nu}(\eta, {\bm x})\ll g_{\mu \nu}^{_{\rm
    FLRW}}(\eta)$. In fact, $\delta g_{\mu \nu}(\eta, {\bm x})$ can be
expressed in terms of three types of perturbations, scalar, vector and
tensor. In the context of inflation, only scalar and tensor are
important. Scalar perturbations are directly coupled to the perturbed
stress-energy tensor while tensor fluctuations are independent of
$\delta T_{\mu \nu}$ and, in fact, are nothing but gravity waves. The
equations of motion of each type of fluctuations are given by the
perturbed Einstein equations, namely $\delta G_{\mu \nu}=\kappa \delta
T_{\mu \nu}$.

\par

In order to calculate the behavior of the fluctuations, we also need
to specify the initial conditions. This is done by postulating that
the perturbations are of quantum-mechanical origin and that,
initially, their quantum state is the vacuum. This is possible
because, at the beginning of inflation, the physical wavelengths of
the Fourier modes of the perturbations are smaller than the Hubble
radius. This means that, initially, space-time curvature is not felt
and that, as a consequence, a well-motivated vacuum state can be
defined.

\subsection{Inflationary two-point Correlation Functions}
\label{subsec:correl2}

Once the equations of motion have been derived and the initial
conditions specified, one can determine all the statistical properties
of the fluctuations, in particular their two-point correlation
functions or, in Fourier space, power spectra. The scalar
perturbations are curvature perturbations defined by $\zeta(\eta, {\bm
  x}) \equiv \Phi+2({\cal H}^{-1}\Phi'+\Phi)/(3+3w)$, with $w=p/\rho$
the equation of state during inflation and $\Phi$ being the Bardeen
potential~\cite{Bardeen:1980kt} (not to be confused with the scalar
field $\phi$). As usual in a linear theory, it is convenient to work
in Fourier space and, therefore, we write
\begin{equation}
\zeta(\eta, {\bm x})=\frac{1}{(2\pi)^{3/2}}\int {\rm d}{\bm k}\, 
\zeta_{\bm k}(\eta)\, {\rm e}^{-i{\bm k}\cdot {\bm x}}.
\end{equation}
As explained before, in the framework of the theory of cosmological
perturbations of quantum-mechanical origin, the source of the
perturbations is the unavoidable zero-point vacuum fluctuations. As a
consequence, $\zeta(\eta ,{\bm x})$ must in fact be viewed as a
quantum operator and can be expressed as
\begin{equation}
\label{eq:qfieldzeta}
\hat{\zeta}(\eta, {\bm x})=\int \frac{{\rm d}^3{\bm k}}{(2\pi)^{3/2}}
\left[a_{\bm k}g_{\bm k}(\eta){\rm e}^{i{\bm k}\cdot {\bm x}}
+a_{\bm k}^{\dagger}g_{\bm k}^*(\eta){\rm e}^{-i{\bm k}\cdot {\bm x}}\right],
\end{equation}
where $a_{\bm k}$ and $a_{\bm k}^{\dagger}$ are respectively the
annihilation and creation operators satisfying $[a_{\bm k},a_{\bm
    p}^{\dagger}]=\delta ^{(3)}({\bm k}-{\bm p})$. The quantum state
of the perturbations is the vacuum $\vert 0\rangle$ which is, by
definition, annihilated by the operator $a_{\bm k}$, namely $a_{\bm
  k}\vert 0\rangle =0$. The function $g_{\bm k}(\eta)$ is the mode
function and the Fourier transform of $\zeta(\eta, {\bm x})$ is given
by $\zeta_{\bm k}(\eta)=a_{\bm k}g_{\bm k}(\eta)+a_{-{\bm
    k}}^{\dagger}g_{\bm k}^*(\eta)$. This last equation leads to
$\langle 0\vert \zeta_{\vka} \zeta_{\vkb}\vert 0\rangle =\vert g_{{\bm
    k}_1}\vert^2\delta^{(3)}\l(\vka+\vkb\r)$. From the previous
considerations, it follows that the two-point correlation function is
given by
\begin{equation}
\label{eq:meanzetasquare}
  \langle \zeta^2(\eta, {\bm x})\rangle =\int \frac{{\rm d}k}{k}
  \calP_{\zeta}(k)=\int \frac{{\rm d}k}{k}
  \frac{k^3}{2\pi^2}\vert g_{\bm k}\vert^2,
\end{equation}
where $\calP_{\zeta}(k)$ is, by definition, the power spectrum of
scalar perturbations. An exact calculation of this power spectrum is
rarely available but a perturbative expansion into the slow-roll
parameters (since they are small parameters) can be done and results
in
\begin{equation} 
\label{spectrumsr}
\frac{\calP_\zeta(k)}{\calP_{\zeta 0}(k_{_{\rm P}})} = a_0^{_{({\rm S})}} + 
a_1^{_{({\rm S})}} \ln \left(\dfrac{k}{k_{_{\rm P}}}\right) 
+ \frac{a_2^{_{({\rm S})}}}{2} \ln^2\left(\dfrac{k}{k_{_{\rm P}}}\right)
+ \dots \, ,
\end{equation}
where, as already mentioned, $k_{_{\rm P}}$ is the pivot scale and the
overall amplitude can be written as
\begin{equation}
\label{eq:scalaramp}
\calP_{\zeta \zero} =\frac{H_*^2}{8 \pi^2 \epsilon_{1*} \Mp^2}\,,
\end{equation}
a star meaning that a quantity is evaluated at the time at which the
pivot scale crossed out the Hubble radius during inflation. We see
that the amplitude of the power spectrum depends on $H_*$ but also on
the first slow-roll parameter, $\epsilon_{1*}$. The coefficients
$a_i^{_{({\rm S})}}$ can be expressed in terms of the Hubble flow
functions. For scalar perturbations, at second order in the slow-roll
approximation, one
gets~\cite{Schwarz:2001vv,Martin:2002vn,Casadio:2004ru,Casadio:2005xv,
  Casadio:2005em,Gong:2001he,Choe:2004zg,Leach:2002ar,Lorenz:2008et,
Martin:2013uma,Jimenez:2013xwa}
\begin{eqnarray}
\label{eqn:as0}
a_{0}^{\usssPSP} &=& 1 - 2\left(C + 1\right)\epsilon_{1*} - C \epsilon_{2*}
+ \left(2C^2 + 2C + \frac{\pi^2}{2} - 5\right) \epsilon_{1*}^2 \nonumber
\\ & + & \left(C^2 - C + \frac{7\pi^2}{12} - 7\right)
\epsilon_{1*}\epsilon_{2*} + \left(\frac12 C^2 + \frac{\pi^2}{8} -
1\right)\epsilon_{2*}^2 \nonumber \\ & + & \left(-\frac12 C^2  +
\frac{\pi^2}{24}\right)  \epsilon_{2*}\epsilon_{3*} +\cdots \, , \\
\label{eqn:as1}
a_{1}^{\usssPSP} & = & - 2\epsilon_{1*} - \epsilon_{2*} 
+ 2(2C+1)\epsilon_{1*}^2
+ (2C - 1)\epsilon_{1*}\epsilon_{2*} + C\epsilon_{2*}^2 
- C\epsilon_{2*}\epsilon_{3*}
+\cdots \, ,\\ 
a_{2}^{\usssPSP} &=& 4\epsilon_{1*}^2 + 2\epsilon_{1*}\epsilon_{2*} +
\epsilon_{2*}^2 - \epsilon_{2*}\epsilon_{3*} +\cdots \, ,\\
a_{3}^{\usssPSP} &=& \calO(\epsilon_{n*}^3)\, ,
\label{eqn:as2}
\end{eqnarray}
where $C \equiv \gamma_{\usssE} + \ln 2 - 2 \approx -0.7296$, $\gamma
_\usssE$ being the Euler constant.

\par

For tensor fluctuations, the approach is exactly similar to what we
have just described. In particular, the tensor power spectrum
$\calP_h$ can be written in the same way as Eq.~(\ref{spectrumsr}) but
with a global amplitude now given by
\begin{equation}
\label{eq:tensoramp}
\calP_{h \zero} =\frac{2 H_*^2}{\pi^2 \Mp^2}\,.
\end{equation}
This time, the amplitude only depends on the Hubble parameter during
inflation. Moreover, the coefficients $a_i^{\usssPTP}$ have a similar
structure and can be written as
\begin{eqnarray}
a_{0}^{\usssPTP} &=&
 1 - 2\left(C + 1\right)\epsilon_{1*} 
 + \left(2C^2 + 2C + \frac{\pi^2}{2} - 5\right) 
 \epsilon_{1*}^2 \nonumber \\ 
& + & \left(-C^2 - 2C + \frac{\pi^2}{12} - 2\right) 
 \epsilon_{1*}\epsilon_{2*} +\cdots \, ,\\
a_{1}^{\usssPTP} &=& 
 - 2\epsilon_{1*} + 2(2C + 1)\epsilon_{1*}^2 
 - 2(C + 1)\epsilon_{1*}\epsilon_{2*}+\cdots  \, ,
\label{eqn:at1} \\
a_{2}^{\usssPTP} &=& 4\epsilon_{1*}^2 - 2\epsilon_{1*}\epsilon_{2*}
+\cdots  \, , \\
a_{3}^{\usssPTP} &=& \calO(\epsilon_{n*}^3)\, .
\label{eqn:at2}
\end{eqnarray}
The coefficients in front of the $\ln k$ term are related to the
spectral indices and, at first order in the slow-roll parameters (we
will discuss them in more detail in Sec.~\ref{sec:observ}), they can
be expressed as
\begin{equation}
\label{eq:specindices}
\nS=1-2\epsilon_1-\epsilon_2, \quad \nT=-2\epsilon_1,
\end{equation}
where the first expression refers to scalar perturbations while the
second is for tensor perturbations. Notice that, sometimes, the power
spectrum is written as $k^{\nS-1}$. In the context of slow-roll
inflation, this is clearly not justified as it would amount to keep an
infinite number of higher order terms while $\nS$ has been evaluated
at first order only. It is worth stressing that power-law power
spectra are predictions of power-law inflation only, that is to say
the inflationary model for which $V(\phi)\propto \exp(-C
\phi)$~\cite{Lucchin:1984yf}. From Eqs.~(\ref{eq:scalaramp})
and~(\ref{eq:tensoramp}), one can also estimate the relative
contribution of tensor and scalar amplitudes
\begin{equation}
\label{eq:defr}
r \equiv \frac{\calP_h}{\calP_\zeta}=16\epsilon_{1*},
\end{equation}
which means that, since $\epsilon_{1*}\ll 1$, tensor are
sub-dominant. This is of course rather unfortunate since a direct
measurement of gravity wave would directly lead to the energy scale
during inflation, $H_*$.

\subsection{Inflationary three-point Correlation Functions}
\label{subsec:correl3}

We have just derived the slow-roll inflationary two-point correlation
functions but, of course, higher order correlation functions are also
interesting and the field of Non-Gaussianity has played an important
role in the recent years, see
Refs.~\cite{Gangui:1993tt,Gangui:1994yr,Wang:1999vf,Gangui:1999vg,Gangui:2000gf,Gangui:2002qc}
for original works on this question and
Refs.~\cite{Maldacena:2002vr,Seery:2005wm,Chen:2006nt,Hotchkiss:2009pj,Chen:2010xka,Martin:2011sn,Hazra:2012kq,Hazra:2012yn,Sreenath:2013xra,Martin:2014kja,Sreenath:2014nka,Sreenath:2014nca}
for later works. For a complete overview of the subject, we refer to
the lecture notes by C.~Byrnes~\cite{Byrnes:2014pja}. Here, in order
to be able to fully appreciate the relevance of the Planck data on
Non-Gaussianities, we discuss how the three-point inflationary
correlation functions can be calculated in the case of single-field
slow-roll inflation with a minimal kinetic term.

\par

For the two-point correlation, we have seen that it is convenient to
work in Fourier space and to define the power spectrum. In the same
way, for the three-point correlation function, we can define the
bispectrum as a correlator in Fourier space, namely $\langle
\zeta_{\vka} (\eta)\, \zeta_{\vkb} (\eta)\, \zeta_{\vkc} (\eta)\rangle
$. In fact, we will rather calculate the quantity $\langle \cR_{\vka}
(\eta)\, \cR_{\vkb} (\eta)\, \cR_{\vkc} (\eta)\rangle $ where $\cR
\equiv -\Psi-{\cal H}\delta \phi^{\rm (gi)}/\phi'$, $\Psi=\Phi$ (valid
if a scalar field dominates the matter content of the Universe) being
another Bardeen potential and $\delta \phi^{\rm (gi)}$ being the gauge
invariant scalar field fluctuation~\cite{Mukhanov:1990me}. This
amounts to a simple change of sign of the three-point function (and no
change at the power spectrum level, namely
$\calP_{\zeta}=\calP_{\cR}$, because the power spectrum is quadratic
in the Fourier amplitudes) since $\cR=-\zeta$\footnote{Indeed, the
  space time component of the perturbed Einstein equation reads
\begin{equation}
-\frac{2}{a^2}\partial_i\left({\cal H}\Phi+\Phi'\right)
=\kappa (\rho+p)\partial _i v^{\rm (gi)}, 
\end{equation}
where, for a scalar field, $v^{\rm (gi)}=-\delta \phi^{\rm (gi)}/\phi'$. As a 
consequence
\begin{equation}
\Phi+{\cal H}^{-1}\Phi'=\frac{\kappa a^2}{2{\cal H}}(\rho+p)
\frac{\delta \phi^{\rm (gi)}}{\phi'}.
\end{equation}
Using this last expression in the definition of $\zeta$ and the Friedmann 
equation ${\cal H}^2=\kappa a^2\rho/3$, one obtains
\begin{equation}
\zeta=\Phi+\frac23\frac{{\cal H}^{-1}\Phi'+\Phi}{1+w}
=\Phi+{\cal H}\frac{\delta \phi^{\rm (gi)}}{\phi'}=-\cR,
\end{equation}
namely the equation mentioned in the text.}. Concretely one has
\begin{align}
\langle {\cR}(\eta,\vx)\, {\cR}(\eta,\vx)\, 
{\cR}(\eta,\vx)\rangle
=& \int \f{\d^3 \vka}{(2\,\pi)^{3/2}}\; 
\int\! \f{\d^3 \vkb}{(2\,\pi)^{3/2}}\;
\int \f{\d^3 \vkc}{(2\,\pi)^{3/2}}\; 
\langle \cR_{\vka} (\eta)\, \cR_{\vkb} (\eta)\, 
\cR_{\vkc} (\eta)\rangle\; 
\nonumber \\ & \times 
{\rm e}^{i\,\l(\vka+\vkb+\vkc\r)\cdot \vx}.
\end{align}
In the above expression $\cR_{\vk}(\eta)$ obviously represents the
Fourier transform of the curvature (scalar) perturbation $\cR (\eta,
{\bm x})$, namely
\begin{equation}
\label{eq:fourierR}
  \cR(\eta, {\bm x})=\frac{1}{(2\pi)^{3/2}}\int {\rm d}{\bm k}\, 
\cR_{\bm k}(\eta)
  \, {\rm e}^{-i{\bm k}\cdot {\bm x}}.
\end{equation}
As explained before, in the framework of the theory of cosmological
perturbations of quantum-mechanical origin, it is an operator and it
can be expressed as
\begin{equation}
\label{eq:qfieldR}
\hat{\cR} (\eta, {\bm x})=\int \frac{{\rm d}^3{\bm k}}{(2\pi)^{3/2}}
\left[a_{\bm k}f_{\bm k}(\eta){\rm e}^{i{\bm k}\cdot {\bm x}}
+a_{\bm k}^{\dagger}f_{\bm k}^*(\eta){\rm e}^{-i{\bm k}\cdot {\bm x}}\right],
\end{equation}
leading to $\langle \cR_{\vka} \cR_{\vkb}\rangle =\vert f_{{\bm
    k}_1}\vert^2\delta^{(3)}\l(\vka+\vkb\r)$ since $\cR_{\bm k}=a_{\bm
  k}f_{\bm k}+a_{-{\bm k}}^{\dagger}f_{\bm k}^*$. Here, the creation
and annihilation operators are the same as those appearing in
Eq.~(\ref{eq:qfieldzeta}). Of course working in terms of $\cR_{\bm
  k}(\eta)$ instead of $\zeta_{\bm k}(\eta)$ is both harmless and
trivial since $\cR_{\bm k}=-\zeta_{\bm k}$ and $f_{\bm k}(\eta)=-g_{\bm
  k}(\eta)$! We do it since many papers on Non-Gaussianities use this
variable.

\par

At this stage, it may be useful to say a few words about
conventions. In this article, we are using Fourier transforms as
defined in Eq.~(\ref{eq:fourierR}). Another convention, often used in
the literature on Non-Gaussianities, is
\begin{equation}
\label{eq:newconvention}
\cR(\eta, {\bm x})=\frac{1}{(2\pi)^{3}}\int {\rm d}{\bm k}\, 
\bar{\cR}_{\bm k}\, 
{\rm e}^{-i{\bm k}\cdot {\bm x}},
\end{equation}
so that $\bar{\cR}_{\bm k}=(2\pi)^{3/2}\cR_{\bm k}$. This implies that
$\langle \bar{\cR}_{\vka} \bar{\cR}_{\vkb}\rangle =(2\pi)^3\vert
f_{{\bm k}_1}\vert^2\delta^{(3)}\l(\vka+\vkb\r)$. Notice that the
two-point correlation function is sometimes defined as $\langle
\bar{\cR}_{\vka} \bar{\cR}_{\vkb}\rangle \equiv
(2\pi)^3P_{\cR}(k_1)\delta^{(3)}\l(\vka+\vkb\r)$ which leads to the
identification $P_{\cR}(k_1)=\vert f_{{\bm k}_1}\vert^2$ [the quantity
  $P_{\cR}(k_1)$ should not be confused with $\calP_{\cR}(k_1)=k_1^3
  \vert f_{{\bm k}_1}\vert^2/(2\pi^2)]$. These definitions imply that
$\bar{\cR}_{\bm k}=(2\pi)^{3/2}\left(a_{\bm k}f_{\bm k}+a_{-{\bm
    k}}^{\dagger}f_{\bm k}^*\right)$ which can be rewritten as
$\bar{\cR}_{\bm k}=\bar{a}_{\bm k}f_{\bm k}+\bar{a}_{-{\bm
    k}}^{\dagger}f_{\bm k}^*$ with $\bar{a}_{\bm k}=(2\pi)^{3/2}a_{\bm
  k}$. In particular, since $[a_{\bm k},a_{\bm
    p}^{\dagger}]=\delta^{(3)}\left({\bm k}-{\bm p}\right)$, we now
have $[\bar{a}_{\bm k},\bar{a}_{\bm
    p}^{\dagger}]=(2\pi)^3\delta^{(3)}\left({\bm k}-{\bm
  p}\right)$. Different conventions basically correspond to different
choices for where the factors $2\pi$ appear in the equations. In
principle straightforward, it can sometimes be confusing when one
tries to check a result in the existing literature.

\par

The bispectrum can be evaluated using the standard rules of quantum
field theory. It is given by~\cite{Maldacena:2002vr,Seery:2005wm}
\begin{eqnarray}
\label{eq:inin}
\langle \cR_{\vka}(\eta)\, \cR_{\vkb}(\eta)\, \cR_{\vkc}(\eta)\rangle
= -i\int_{\eta_{\rm ini}}^{\eta_{\rm e}} \d\tau \, a(\tau)
\l\langle\l[\cR_{\vka}(\eta)\, \cR_{\vkb}(\eta)\, 
\cR_{\vkc}(\eta), H_{\rm int}(\tau)\r]\r\rangle,\nonumber \\ 
\end{eqnarray}
where $\eta_{\rm ini}$ represents an initial time at the beginning of
inflation (in practice we take $\eta_{\rm ini}\rightarrow -\infty$)
and $\eta_{\rm e}$ a final time at the end of inflation when all the
scales relevant to the problem are outside the Hubble radius (in
practice we take $\eta_{\rm e}\rightarrow 0$). The quantity $H_{\rm
  int}$ is the interaction Hamiltonian. It can be obtained from the
action of the system expanded up to third order in $\cR$, the action
of the system being the Einstein-Hilbert action plus that of a scalar
field (the inflaton). A now standard calculation
gives~\cite{Maldacena:2002vr,Seery:2005wm,Chen:2006nt,Chen:2010xka}
\begin{eqnarray}
& &{\cal S}_{3}[\cR] 
=\Mp^2\int \d t\; \d^3{\bm x}\; 
\Biggl[a^3\, \epsilon_{1}^2\, {\cR}\, {\dot {\cR}}^2
+a\,\epsilon_{1}^2\, {\cR}\, (\pa {\cR})^2
-2\,a\, \epsilon_{1}\, {\dot {\cR}}\, 
(\pa^{i}{\cR})\, (\pa_{i}\chi)
\nonumber \\
&+ &\frac{a^3}{2}\,\epsilon_{1}\, {\dot \epsilon_{2}}\, 
{\cR}^2\, \dot{\cR}
+\frac{\epsilon_{1}}{2a}\, (\partial^{i}\cR)\, 
(\pa_{i}\chi)\, (\pa^2 \chi)
+\f{\epsilon_{1}}{4\,a}\, (\pa^2{\cR})\, (\pa \chi)^2
+ {\cal F}\l(\f{\delta {\cal L}_{2}}{\delta \cR}\r)\Biggr],
\label{eq:cubicaction}
\end{eqnarray}
where $\delta {\cal L}_{2}/\delta\cR$ denotes the variation of the 
second order action with respect to $\cR$, and is 
given by
\begin{equation}
\f{\delta {\cal L}_{2}}{\delta\cR}
={\dot \Lambda}+H\, \Lambda 
-\epsilon_{1}\, \partial^2\cR, 
\end{equation}
and the quantities $\Lambda$ and $\chi$ are defined by
\begin{equation}
\Lambda \equiv \frac{a^2\dot\phi^2}{2\Mp^2H^2}\dot 
\cR=a^2\epsilon_1\dot \cR,
\quad \chi \equiv \partial ^{-2}\Lambda .
\end{equation}
The term ${\cal F}(\delta {\cal L}_{2}/\delta \cR)$ introduced in
Eq.~(\ref{eq:cubicaction}) stands for the following complicated
expression
\begin{eqnarray}
\label{eq:termeom}
{\cal F}\l(\f{\delta {\cal L}_{2}}{\delta \cR}\r)
&=& \frac{a}{2}\epsilon_2
\left(\f{\delta {\cal L}_{2}}{\delta\cR}\right)\cR^2
+\frac{2a}{H}\left(\f{\delta {\cal L}_{2}}{\delta\cR}\right)
\dot \cR\cR
\nonumber \\ 
&+&\f{1}{2aH}
\Biggl\{(\partial^{i}\cR)\; (\pa_{i}\chi)\, 
\l(\f{\delta {\cal L}_{2}}{\delta\cR}\r)
+\,\delta^{ij}\, \l[\Lambda\, (\pa_{i}\cR) 
+(\pa^{2}\cR)\, (\pa_{i}\chi)\r]\,
\nonumber \\ 
&\times & \pa_{j}\l[\pa^{-2}\l(\f{\delta {\cal L}_{2}}{\delta\cR}\r)\r]
+
\frac{\delta^{im}\delta^{jn}}{H}(\pa_{i}\cR)\, (\pa_{j}\cR)\;
\pa_{m}\pa_{n}\l[\pa^{-2}\l(\f{\delta {\cal L}_{2}}{\delta\cR}\r)\r]\Biggr\}.
\nonumber \\
\end{eqnarray}
The terms which involves $\delta {\cal L}_{2}/\delta \cR$ can be
removed by a suitable field redefinition of~$\cR$ of the following
form~\cite{Maldacena:2002vr,Seery:2005wm,Chen:2006nt,Chen:2010xka}:
\begin{equation}
\cR \rightarrow \cR_n+\epsilon_2\frac{\cR_n^2}{4}.\label{eq:frd}
\end{equation}
After this redefinition, the perturbed action~(\ref{eq:cubicaction})
becomes a functional of $\cR_n$. In the following, in order to avoid
too complicated notations, we will still use $\cR$ in place of
$\cR_n$. Then, with the redefinition~(\ref{eq:frd}), the interaction
Hamiltonian can be expressed as
\begin{eqnarray}
H_{\rm int}(\eta) 
&=& -\Mp^2\int \d^{3}{\bm x}\;
\biggl[a\, \epsilon_{1}^2\, \cR\, \cR'^2  
+ a\, \epsilon_{1}^2\, \cR\, (\pa \cR)^2 
- 2\, \epsilon_{1}\, \cR'\, (\pa^{i}\cR)\, (\pa_{i} \chi)
\nonumber \\
&+& \frac{a}{2}\, \epsilon_{1}\, \epsilon_{2}'\, 
\cR^2\, \cR'
+ \f{\epsilon_{1}}{2\,a}\, (\pa^{i} \cR)\, 
(\pa_{i}\chi)\, \l(\pa^2 \chi\r)
+ \f{\epsilon_{1}}{4\,a}\, \l(\pa^2 \cR\r)\, 
(\pa \chi)^2\biggr].\label{eq:Hint}
\end{eqnarray}
where we remind that a prime means a derivative with respect to
conformal time. The first three terms are second order in the
slow-roll parameters while the three last ones are third order. As a
consequence, already at this stage, we see that the bispectrum will be
a small quantity. Since we now know the interaction Hamiltonian we can
insert its expression in Eq.~(\ref{eq:inin}) in order to derive the
bispectrum explicitly. One finds that
\begin{align}
\label{eq:threezeta}
\langle \cR_{\vka}(\eta _{\rm e})\, & \cR_{\vkb}(\eta _{\rm e})\,
\cR_{\vkc}(\eta _{\rm e})\rangle 
=\f{\l(2\,\pi\r)^3}{\l(2\pi\r)^{9/2}}\Mp^2
\sum_{C=1}^{6}\; 
\Biggl[f_{{\bm k}_1}(\eta_{\rm e})\,f_{{\bm k}_2}(\eta_{\rm e})\,
f_{{\bm k}_3}(\eta_{\rm e})
\cG_{_{C}}(\vka,\vkb,\vkc)
\nonumber \\ & 
+f_{{\bm k}_1}^{\ast}(\eta_{\rm e})\, f_{{\bm k}_2}^{\ast}(\eta_{\rm e})
\,f_{{\bm k}_3}^{\ast}(\eta_{\rm e})
\cG_{_{C}}^{\ast}(\vka,\vkb,\vkc)\Biggr]
\delta^{(3)}\l(\vka+\vkb+\vkc\r),
\end{align}
where the delta function ensures momentum conservation. Written in
this way, the correlator is obviously real. In the above expression,
the term $\cG_{_{C}}(\vka,\vkb,\vkc)$ with ${\small C}=(1,6)$
correspond to the six terms in the interaction
Hamiltonian~(\ref{eq:Hint}) (the six ``vertices''), and are explicitly
given by~\cite{Maldacena:2002vr}
\begin{eqnarray}
\label{eq:g1}
\cG_{1}(\vka,\vkb,\vkc)
&=&2i\int_{\eta_{\rm ini}}^{\eta_{\rm e}} \d\tau \,a^2\, 
\epsilon_{1}^2\, \l(f_{\vka}^{\ast}\,f_{\vkb}'^{\ast}\,
f_{\vkc}'^{\ast}+{\rm two~permutations}\r), \\
\label{eq:g2}
\cG_{2}(\vka,\vkb,\vkc) 
&=&-2i\int_{\eta_{\rm ini}}^{\eta_{\rm e}} \d\tau a^2\, 
\epsilon_{1}^2\, f_{\vka}^{\ast}\,f_{\vkb}^{\ast}\,
f_{\vkc}^{\ast}\,
\l(\vka\cdot \vkb + {\rm two~permutations}\r), \\
\label{eq:g3} 
\cG_{3}(\vka,\vkb,\vkc)
&=&-2i\int_{\eta_{\rm ini}}^{\eta_{\rm e}} \d\tau \; a^2\,
\epsilon_{1}^2\, \biggl[f_{\vka}^{\ast}\,f_{\vkb}'^{\ast}\,
f_{\vkc}'^{\ast} \l(\f{\vka\cdot\vkb}{\kb^{2}}\r) \\ & & 
+ {\rm five~permutations}\biggr],\\
\label{eq:g4}
\cG_{4}(\vka,\vkb,\vkc)
&=&i\int_{\eta_{\rm ini}}^{\eta_{\rm e}} \d\tau\; a^2\,\epsilon_{1}\,
\epsilon_{2}'\, \l(f_{\vka}^{\ast}\,f_{\vkb}^{\ast}\,
f_{\vkc}'^{\ast}+{\rm two~permutations}\r), \\
\label{eq:g5}
\cG_{5}(\vka,\vkb,\vkc)
&=&\frac{i}{2}\int_{\eta_{\rm ini}}^{\eta_{\rm e}} \d\tau\; 
a^2\, \epsilon_{1}^{3}\, \biggl[f_{\vka}^{\ast}\,f_{\vkb}'^{\ast}\,
f_{\vkc}'^{\ast} \l(\f{\vka\cdot\vkb}{\kb^{2}}\r) \\ & &
+ {\rm five~permutations}\biggr], \\
\label{eq:g6}
\cG_{6}(\vka,\vkb,\vkc)
&=&\frac{i}{2}\int_{\eta_{\rm ini}}^{\eta_{\rm e}} \d\tau\; a^2\, 
\epsilon_{1}^{3}\,
\biggl[f_{\vka}^{\ast}\,f_{\vkb}'^{\ast}\, f_{\vkc}'^{\ast}\, 
\l(\f{\ka^{2}}{\kb^{2}\,\kc^{2}}\r)\, \l(\vkb\cdot\vkc\r)
\nonumber \\ & & 
+\, {\rm two~permutations}\biggr].
\end{eqnarray}
Actually, an additional seventh term arises due to the field
redefinition~(\ref{eq:frd}), and its contribution to the three point
correlation function is found to be
\begin{eqnarray}
\label{eq:seven}
\langle
\cR_{\vka}(\eta _{\rm e})\, \cR_{\vkb}(\eta _{\rm e})\,
\cR_{\vkc}(\eta _{\rm e})\rangle ^{(7)}
&=& \f{\l(2\,\pi\r)^3}{\l(2\pi\r)^{9/2}}\frac{\epsilon_{2}}{2}
\l(\vert f_{\vkb}\vert^{2}\,\vert f_{\vkc}\vert^{2} 
+ {\rm two~permutations}\r)
\nonumber \\ & & \times
\delta^{(3)}\l(\vka+\vkb+\vkc\r).
\end{eqnarray}
The other terms in Eq.~(\ref{eq:termeom}) do not contribute because
they all contain a derivative (time derivative and/or space
derivative) and, at the end of inflation, on super Hubble scales,
$\zeta=-\cR$ is constant.

\par

In order to calculate each of the above terms, one obviously needs to
know the mode function $f_{\bm k}$. Since we evaluate the bispectrum
at leading order in slow roll, it is in fact sufficient to use the de
Sitter mode function, namely $f_{\bm k}=iH(1+ik\eta){\rm
  e}^{-ik\eta}/(2\Mp\sqrt{k^3\epsilon_1})$ (which is properly
normalized). Moreover, we only need to calculate the first three terms
and $\cG_{7}(\vka,\vkb,\vkc)$, the other contributions being of higher
orders in slow-roll. In order to illustrate how the calculation
proceeds, let us explain in detail how $\cG_{2}(\vka,\vkb,\vkc)$ can
be calculated (this term is easier than the others since we do not
have to use the derivative of the mode function). Inserting the de
Sitter mode function into Eq.~(\ref{eq:g2}), one obtains
\begin{align}
\cG_{2}(\vka,\vkb,\vkc)
&=
-2i\frac{(-iH)^3}{8\Mp^3\sqrt{\epsilon_{1*}^3k_1^3k_2^3k_3^3}}
\frac{1}{H^2}\epsilon_{1*}^2
\l(\vka\cdot \vkb + {\rm two~permutations}\r)
\nonumber \\ &  \times 
\int_{\eta_{\rm ini}}^{\eta_{\rm e}}
\frac{{\rm d}\tau}{\tau^2}{\rm e}^{ik_{_{\rm T}}\tau}
(1-ik_1\tau)(1-ik_2\tau)(1-ik_3\tau),
\\ &= 
-2i\frac{(-iH)^3}{8\Mp^3\sqrt{\epsilon_{1*}^3k_1^3k_2^3k_3^3}}
\frac{1}{H^2}\epsilon_{1*}^2
\l(\vka\cdot \vkb + {\rm two~permutations}\r)
\nonumber \\ &  \times 
\int_{\eta_{\rm ini}}^{\eta_{\rm e}}
\frac{{\rm d}\tau}{\tau^2}\biggl[1-ik_{_{\rm T}}\tau
-\left(k_1k_2+k_2k_3+k_1k_3\right)\tau^2
+ik_1k_2k_3\tau^3\biggr]{\rm e}^{ik_{_{\rm T}}\tau},
\end{align}
where $k_{_{\rm T}}\equiv k_1+k_2+k_3$ is the ``total''
wave-number. This expression is made of four integrals that we need to
calculate. The first and the fourth ones can be integrated by parts
and the third one can be directly performed. This leads to
\begin{align}
\cG_{2}(\vka,\vkb,&\vkc)
=
-2i\frac{(-iH)^3}{8\Mp^3\sqrt{\epsilon_{1*}^3k_1^3k_2^3k_3^3}}
\frac{1}{H^2}\epsilon_{1*}^2
\l(\vka\cdot \vkb + {\rm two~permutations}\r)
\nonumber \\ & 
\times 
\Biggl[\frac{-1}{\tau}{\rm e}^{ik_{_{\rm T}}\tau}
\biggr\vert_{\eta_{\rm ini}}^{\eta_{\rm e}}
+ik_{_{\rm T}}\int_{\eta_{\rm ini}}^{\eta_{\rm e}}
\frac{{\rm e}^{ik_{_{\rm T}}\tau}}{\tau }{\rm d}\tau
-ik_{_{\rm T}}\int_{\eta_{\rm ini}}^{\eta_{\rm e}}
\frac{{\rm e}^{ik_{_{\rm T}}\tau}}{\tau }{\rm d}\tau
\nonumber \\ & 
-\left(k_1k_2+k_2k_3+k_1k_3\right)
\frac{{\rm e}^{ik_{_{\rm T}}\tau}}{ik_{_{\rm T}}}
\bigg\vert_{\eta_{\rm ini}}^{\eta_{\rm e}}
+ik_1k_2k_3
\left(\frac{\tau {\rm e}^{ik_{_{\rm T}}\tau}}{ik_{_{\rm T}}}
+\frac{{\rm e}^{ik_{_{\rm T}}\tau}}{k_{_{\rm T}}^2}
\right. \bigg\vert_{\eta_{\rm ini}}^{\eta_{\rm e}}
\Biggr],
\end{align}
and we see that the second integral exactly cancels the term arising
from the integration by parts of the first integral. In principle, at
this stage, it is sufficient to take $\eta_{\rm ini}=-\infty$ in the
above expression in order to get the final result. But, obviously, the
result would be ill-defined. So what is done is to slightly rotate the
integration path in the complex plane and replace $\eta_{\rm ini}$
with $-\infty (1-i\delta)$ where $\delta $ is a small parameter. This
produces a term ${\rm e}^{-ik_{_{\rm T}}\infty-k_{_{\rm T}}\delta
  \infty}$ which, in fact, kills all terms proportional to ${\rm
  e}^{ik_{_{\rm T}}\eta_{\rm ini}}$. It is worth noticing that this
should not be viewed as an arbitrary technical trick but as the
standard method to properly identify the correct vacuum
state~\cite{Peskin:1995ev}. As a result, one obtains the following
  expression
\begin{eqnarray}
\cG_{2}(\vka,\vkb,\vkc)
&=& 
-2i\frac{(-iH)^3}{8\Mp^3\sqrt{\epsilon_{1*}^3k_1^3k_2^3k_3^3}}
\frac{1}{H^2}\epsilon_{1*}^2
\l(\vka\cdot \vkb + {\rm two~permutations}\r)
\nonumber \\ & & 
\times 
\biggl[\frac{-1}{\eta_{\rm e}}{\rm e}^{ik_{_{\rm T}}\eta_{\rm e}}
-\left(k_1k_2+k_2k_3+k_1k_3\right)
\frac{{\rm e}^{ik_{_{\rm T}}\eta_{\rm e}}}{ik_{_{\rm T}}}
\nonumber \\ & &
+ik_1k_2k_3
\left(\frac{\eta_{\rm e} {\rm e}^{ik_{_{\rm T}}\eta_{\rm e}}}{ik_{_{\rm T}}}
+\frac{{\rm e}^{ik_{_{\rm T}}\eta_{\rm e}}}{k_{_{\rm T}}^2}
\right)
\biggr].
\end{eqnarray}
Then, the final step is to take $\eta_{\rm e}\rightarrow 0$. Clearly,
there is a problem with the first term and, therefore, in the
following expressions, we will keep $\eta_{\rm e}$ unspecified. For
the other terms, the above expression simplifies and one is led to
\begin{eqnarray}
\label{eq:curlG2}
\cG_{2}(\vka,\vkb,\vkc)
&=& 
-2i\frac{(-iH)^3}{8\Mp^3\sqrt{\epsilon_{1*}^3k_1^3k_2^3k_3^3}}
\frac{1}{H^2}\epsilon_{1*}^2
\l(\vka\cdot \vkb + {\rm two~permutations}\r)
\nonumber \\ & & 
\times 
\biggl[\frac{-1}{\eta_{\rm e}}{\rm e}^{ik_{_{\rm T}}\eta_{\rm e}}
+\frac{i}{k_{_{\rm T}}}\left(k_1k_2+k_2k_3+k_1k_3\right)
+\frac{i}{k_{_{\rm T}}^2}k_1k_2k_3
\biggr].
\end{eqnarray}
This completes the calculation of $\cG_{2}(\vka,\vkb,\vkc)$. Now, we
insert the above result into Eq.~(\ref{eq:threezeta}) in order to
determine the contribution of $\cG_{2}(\vka,\vkb,\vkc)$ to $ \langle
\cR_{\vka}(\eta _{\rm e})\, \cR_{\vkb}(\eta _{\rm e})\,
\cR_{\vkc}(\eta _{\rm e})\rangle $. To perform this calculation, we
need $f_{\bm k}(\eta_{\rm e})$, which we take to be
$iH/[8\Mp^3\sqrt{\epsilon_1^3(\eta_{\rm e})k_1^3k_2^3k_3^3}]$ since
the limit $\eta_{\rm e}\rightarrow 0$ does not cause any problem in
that case. As a result, one finds that
\begin{align}
  \langle &\cR_{\vka}(\eta _{\rm e})\, \cR_{\vkb}(\eta _{\rm e})\,
  \cR_{\vkc}(\eta _{\rm e})\rangle ^{(2)} =
  \frac{(2\pi)^3}{(2\pi)^{9/2}}\Mp^2 \delta^{(3)}\l(\vka+\vkb+\vkc\r)
\nonumber
\\ & \times
  \biggl[\frac{(iH)^3}{8\Mp^3\sqrt{\epsilon_1^3(\eta_{\rm
        e})k_1^3k_2^3k_3^3}} \cG_{2}(\vka,\vkb,\vkc) 
  +\frac{(-iH)^3}{8\Mp^3\sqrt{\epsilon_1^3(\eta_{\rm
        e})k_1^3k_2^3k_3^3}} \cG_{2}^*(\vka,\vkb,\vkc) \biggr],
\end{align}
which, combined with Eq.~(\ref{eq:curlG2}), leads to
\begin{align}
&\langle
\cR_{\vka}(\eta _{\rm e})\, \cR_{\vkb}(\eta _{\rm e})\,
\cR_{\vkc}(\eta _{\rm e})\rangle ^{(2)}
=\frac{(2\pi)^3}{(2\pi)^{9/2}}\Mp^2 \delta^{(3)}\l(\vka+\vkb+\vkc\r)
\frac{H^3}{8\Mp^3\sqrt{\epsilon_1^3(\eta_{\rm e})k_1^3k_2^3k_3^3}}
\nonumber \\ & \times
\frac{2H^3}{8\Mp^3\sqrt{\epsilon_{1*}^3k_1^3k_2^3k_3^3}}
\frac{1}{H^2}\epsilon_{1*}^2
\l(\vka\cdot \vkb + {\rm two~permutations}\r)
\biggl\{
-i(-i)^3i^3 \biggl[\frac{-1}{\eta_{\rm e}}{\rm e}^{ik_{_{\rm T}}\eta_{\rm e}}
\nonumber \\ &
+\frac{i}{k_{_{\rm T}}}\left(k_1k_2+k_2k_3+k_1k_3\right)
+\frac{i}{k_{_{\rm T}}^2}k_1k_2k_3
\biggr]
+
i(i)^3(-i)^3 \biggl[\frac{-1}{\eta_{\rm e}}{\rm e}^{-ik_{_{\rm T}}\eta_{\rm e}}
\nonumber \\ & -\frac{i}{k_{_{\rm T}}}(k_1k_2+k_2k_3
+k_1k_3)
-\frac{i}{k_{_{\rm T}}^2}k_1k_2k_3
\biggr]\biggr\}.
\end{align}
This expression can be simplified further and one obtains the following 
formula
\begin{align}
\langle
\cR_{\vka}(\eta _{\rm e})\, & \cR_{\vkb}(\eta _{\rm e})\,
\cR_{\vkc}(\eta _{\rm e})\rangle ^{(2)}
= \frac{(2\pi)^3}{(2\pi)^{9/2}}\Mp^2 \delta^{(3)}\l(\vka+\vkb+\vkc\r)
\nonumber \\ & \times
\frac{2 H^6\epsilon_{1*}^2}{64H^2\Mp^6\epsilon_{1*}^{3/2}
\epsilon_{1}^{3/2}(\eta_{\rm e})k_1^3k_2^3k_3^3}
\l(\vka\cdot \vkb + {\rm two~permutations}\r)
\nonumber \\ & \times
\biggl[-2k_{_{\rm T}}\frac{\sin(k_{_{\rm T}}\eta_{\rm e})}{k_{_{\rm T}}\eta_{\rm e}}
+\frac{2}{k_{_{\rm T}}}\left(k_1k_2+k_2k_3+k_1k_3\right)
+\frac{2}{k_{_{\rm T}}^2}k_1k_2k_3\biggr].
\end{align}
We see that the limit $\eta_{\rm e}\rightarrow 0$ is now well defined
and can be taken. The term in $\cG_{2}(\vka,\vkb,\vkc)$ was singular
but, combined with its complex conjugate in the correlator, the limit
has become regular. Therefore, the appearance of a singular limit was
just a temporary technical problem and, in the expression of the
physical quantity, the problematic term has disappeared. The final
expression reads
\begin{align}
\label{eq:term2}
\langle
\cR_{\vka}(\eta _{\rm e})\, & \cR_{\vkb}(\eta _{\rm e})\,
\cR_{\vkc}(\eta _{\rm e})\rangle ^{(2)}
= \frac{(2\pi)^3}{(2\pi)^{9/2}}
\frac{H^4}{16\Mp^4\epsilon_1}
\frac{1}{(k_1k_2k_3)^3}
\l(\vka\cdot \vkb + {\rm two~permutations}\r)
\nonumber \\ & \times
\biggl[-k_{_{\rm T}}
+\frac{1}{k_{_{\rm T}}}\left(k_1k_2+k_2k_3+k_1k_3\right)
+\frac{1}{k_{_{\rm T}}^2}k_1k_2k_3\biggr]\delta^{(3)}\l(\vka+\vkb+\vkc\r).
\end{align}
As expected the amplitude is controlled by the Hubble parameter (to
the power four while the amplitude of the power spectrum was quadratic
in $H$) and the (first) slow-roll parameter. We also see that the
scale dependence is quite complicated.

\par

The calculation proceeds exactly the same way for the first and third
terms. Explicitly, one obtains
\begin{align}
\label{eq:term1}
\langle
\cR_{\vka}(\eta _{\rm e})\, & \cR_{\vkb}(\eta _{\rm e})\,
\cR_{\vkc}(\eta _{\rm e})\rangle ^{(1)}
 = \frac{(2\pi)^3}{(2\pi)^{9/2}}
\frac{H^4}{16\Mp^4\epsilon_1}
\frac{1}{(k_1k_2k_3)^3}
\nonumber \\ & \times
\biggl[\left(1+\frac{k_1}{k_{_{\rm T}}}\right)
\frac{k_2^2k_3^2}{k_{_{\rm T}}}
+\left(1+\frac{k_2}{k_{_{\rm T}}}\right)
\frac{k_1^2k_3^2}{k_{_{\rm T}}}
+\left(1+\frac{k_3}{k_{_{\rm T}}}\right)
\frac{k_1^2k_2^2}{k_{_{\rm T}}}\biggr]
\delta^{(3)}\l(\vka+\vkb+\vkc\r),
\\
\label{eq:term3}
\langle
\cR_{\vka}(\eta _{\rm e})\, & \cR_{\vkb}(\eta _{\rm e})\,
\cR_{\vkc}(\eta _{\rm e})\rangle ^{(3)}
 = -\frac{(2\pi)^3}{(2\pi)^{9/2}}
\frac{H^4}{16\Mp^4\epsilon_1}
\frac{1}{(k_1k_2k_3)^3}
\nonumber \\ & \times
\biggl[(\vka \cdot \vkb)\frac{k_3^2}{k_{_{\rm T}}}
\left(2+\frac{k_1+k_2}{k_{_{\rm T}}}\right)
+
(\vka \cdot \vkc)\frac{k_2^2}{k_{_{\rm T}}}
\left(2+\frac{k_1+k_3}{k_{_{\rm T}}}\right)
\nonumber \\ & 
+(\vkb \cdot \vkc)\frac{k_1^2}{k_{_{\rm T}}}
\left(2+\frac{k_2+k_3}{k_{_{\rm T}}}\right)
\biggr]
\delta^{(3)}\l(\vka+\vkb+\vkc\r).
\end{align}
Finally, the seventh term given by Eq.~(\ref{eq:seven}) can be
re-written in terms of the two-point correlation function
\begin{align}
\label{eq:term7}
\langle &
\cR_{\vka}(\eta _{\rm e})\,\cR_{\vkb}(\eta _{\rm e})\,
\cR_{\vkc}(\eta _{\rm e})\rangle ^{(7)}
 = \frac{(2\pi)^3}{(2\pi)^{9/2}}
2\pi^4\epsilon_2
\frac{1}{(k_1k_2k_3)^3}
\nonumber \\ & \times
\biggl[k_1^3\calP_{\cR}(k_2)\calP_{\cR}(k_3)
+k_2^3\calP_{\cR}(k_1)\calP_{\cR}(k_3)
+k_3^3\calP_{\cR}(k_1)\calP_{\cR}(k_2)
\biggr]
\delta^{(3)}\l(\vka+\vkb+\vkc\r),
\end{align}
where, in order to evaluate the last term, we have made use of the
definition introduced before: $\calP_{\cR}(k)=k^3\vert f_{\bm
  k}\vert^2/(2\pi^2)$, see Eq.(\ref{eq:meanzetasquare}).

\par

We have now completed the calculation of the three-point correlation
function in Fourier space.  We notice that, as already mentioned
above, the dependence in $k_1$, $k_2$, $k_3$ is rather non trivial. In
order to emphasize this point, it is interesting to recalculate the
three-point correlation function in the following simple
setup. Suppose that we write the curvature perturbation as
\begin{equation}
\label{eq:deffnl}
\cR(\eta, {\bm x})=\cR_{_{\mathrm{G}}}(\eta, {\bm x})
-\frac{3\,\fnlloc}{5}\, 
\cR_{_{\mathrm{G}}}^2(\eta,{\bm x})+\cdots ,
\end{equation}
where $\cR_{_{\mathrm{G}}}$ denotes a Gaussian quantity, and the
factor of $3/5$ arises due to the relation between the Bardeen
potential and the curvature perturbation during the matter dominated
epoch. The amplitude of the quadratic term is constant and
conventionally called $\fnlloc$. Let us notice that this assumption is
highly non trivial and that, a priori, the coefficient in front of the
quadratic term is expected to be a function of space. Postulating that
it is a constant enforces a particular scale dependence of the
three-point correlation function as we are going to see. In Fourier
space, the Gaussian part is written $\cR_{_{\rm G}}=(2\pi)^{-3/2} \int
{\rm d}{\bm k}\cR_{\bm k}^{_{\rm G}}{\rm e}^{-i{\bm k}\cdot{\bm x}}$
and it follows that
\begin{equation}
\cR^2(\eta,{\bm x})=\frac{1}{(2\pi)^{3/2}}
\int {\rm d}{\bm k} \, (2\pi)^{-3/2}\int {\rm d}{\bm p}\, 
\cR_{\bm p}^{_{\rm G}}\, \cR_{{\bm k}-{\bm p}}^{_{\rm G}}
\, {\rm e}^{-i{\bm k}\cdot{\bm x}},
\end{equation}
from which we can read the Fourier coefficient of the non-linear
curvature perturbation, namely
\begin{equation}
\label{eq:fouriersquare}
\cR_{\bm k}=\cR_{\bm k}^{_{\rm G}}
-\frac{3\,\fnlloc}{5}(2\pi)^{-3/2}\int {\rm d}{\bm p}\, 
\cR_{\bm p}^{_{\rm G}}\, \cR_{{\bm k}-{\bm p}}^{_{\rm G}}.
\end{equation}
Using this expression, one can now evaluate the bispectrum. One obtains
\begin{align}
\left \langle \cR_{{\bm k}_1}(\eta)\cR_{{\bm k}_2}(\eta)
\cR_{{\bm k}_3}(\eta)\right \rangle & =
\biggl \langle 
\left[\cR_{{\bm k}_1}^{_{\rm G}}
-\frac{3\,\fnlloc}{5}(2\pi)^{-3/2}\int {\rm d}{{\bm p}_1}\, 
\cR_{{\bm p}_1}^{_{\rm G}}\, \cR_{{\bm k}_1-{\bm p}_1}^{_{\rm G}}
\right]
\nonumber \\ & \times
\left[\cR_{{\bm k}_2}^{_{\rm G}}
-\frac{3\,\fnlloc}{5}(2\pi)^{-3/2}\int {\rm d}{\bm p}_2\, 
\cR_{{\bm p}_2}^{_{\rm G}}\, \cR_{{\bm k}_2-{\bm p}_2}^{_{\rm G}}
\right]
\nonumber \\ & \times
\left[\cR_{{\bm k}_3}^{_{\rm G}}
-\frac{3\,\fnlloc}{5}(2\pi)^{-3/2}\int {\rm d}{\bm p}_3\, 
\cR_{{\bm p}_3}^{_{\rm G}}\, \cR_{{\bm k}_3-{\bm p}_3}^{_{\rm G}}
\right]
\biggr \rangle,
\end{align}
and, therefore, 
\begin{align}
&\left \langle \cR_{{\bm k}_1}(\eta)\cR_{{\bm k}_2}(\eta)
\cR_{{\bm k}_3}(\eta)\right \rangle =
\left \langle \cR_{{\bm k}_1}^{_{\rm G}}(\eta)\cR_{{\bm k}_2}^{_{\rm G}}(\eta)
\cR_{{\bm k}_3}^{_{\rm G}}(\eta)\right \rangle
\nonumber \\ &
-\frac{3\,\fnlloc}{5}(2\pi)^{-3/2}\int {\rm d}{{\bm p}_3}\
\left \langle \cR_{{\bm k}_1}^{_{\rm G}}(\eta)\cR_{{\bm k}_2}^{_{\rm G}}(\eta)
\cR_{{\bm p}_3}^{_{\rm G}}(\eta)\cR_{{\bm k}_3-{\bm p}_3}^{_{\rm G}}(\eta)
\right \rangle
+{\rm two~permutations}+\cdots ,
\end{align}
where the dots denote the higher order terms. Since the three point
correlation function vanishes for Gaussian statistics, the previous
expression reduces to
\begin{align}
\left \langle \cR_{{\bm k}_1}(\eta)\cR_{{\bm k}_2}(\eta)
\cR_{{\bm k}_3}(\eta)\right \rangle & =
-\frac{3\,\fnlloc}{5}(2\pi)^{-3/2}\int {\rm d}{{\bm p}_3}\
\left \langle \cR_{{\bm k}_1}^{_{\rm G}}(\eta)\cR_{{\bm k}_2}^{_{\rm G}}(\eta)
\cR_{{\bm p}_3}^{_{\rm G}}(\eta)\cR_{{\bm k}_3-{\bm p}_3}^{_{\rm G}}(\eta)
\right \rangle
\nonumber \\ &
+{\rm two~permutations}+\cdots .
\end{align}
As expected, the three-point correlation function is proportional to
the coefficient $\fnlloc$. To proceed, one can evaluate this
expression by means of the Wick's theorem. Then, one obtains
\begin{align}
\langle & \cR_{{\bm k}_1}(\eta) \cR_{{\bm k}_2}(\eta) 
\cR_{{\bm k}_3}(\eta)\rangle =
\nonumber \\ &
-\frac{3\,\fnlloc}{5}(2\pi)^{-3/2}\int {\rm d}{{\bm p}_3}\biggl[
\left \langle \cR_{{\bm k}_1}^{_{\rm G}}(\eta)\cR_{{\bm k}_2}^{_{\rm G}}(\eta)
\right \rangle \left \langle
\cR_{{\bm p}_3}^{_{\rm G}}(\eta)\cR_{{\bm k}_3-{\bm p}_3}^{_{\rm G}}(\eta)
\right \rangle
\nonumber \\ &
+
\left \langle \cR_{{\bm k}_1}^{_{\rm G}}(\eta)\cR_{{\bm p}_3}^{_{\rm G}}(\eta)
\right \rangle \left \langle
\cR_{{\bm k}_2}^{_{\rm G}}(\eta)\cR_{{\bm k}_3-{\bm p}_3}^{_{\rm G}}(\eta)
\right \rangle
+
\left \langle \cR_{{\bm k}_1}^{_{\rm G}}(\eta)
\cR_{{\bm k}_3-{\bm p}_3}^{_{\rm G}}(\eta)
\right \rangle \left\langle
\cR_{{\bm k}_2}^{_{\rm G}}(\eta)\cR_{{\bm p}_3}^{_{\rm G}}(\eta)
\right \rangle
\nonumber \\ &
+{\rm two~permutations}+\cdots \biggr].
\end{align}
Since the two-point correlation functions are nothing but the 
power spectrum, the above expression takes the following form
\begin{align}
\langle \cR_{{\bm k}_1}(\eta) \cR_{{\bm k}_2}(\eta) &
\cR_{{\bm k}_3}(\eta)\rangle =
-\frac{3\,\fnlloc}{5}(2\pi)^{-3/2}
\nonumber \\ & \times
\int {\rm d}{{\bm p}_3}\biggl[
\frac{(2\pi)^2}{2}\frac{\calP_{\cR}(k_1)}{k_1^3}
\delta^{(3)}\l({\bm k}_1+{\bm k}_2\r)
\frac{(2\pi)^2}{2}\frac{\calP_{\cR}(p_3)}{k_3^3}
\delta^{(3)}\l({\bm p}_3+{\bm k}_3-{\bm p}_3\r)
\nonumber \\ & 
+\frac{(2\pi)^2}{2}\frac{\calP_{\cR}(k_1)}{k_1^3}
\delta^{(3)}\l({\bm k}_1+{\bm p}_3\r)
\frac{(2\pi)^2}{2}\frac{\calP_{\cR}(k_2)}{k_2^3}
\delta^{(3)}\l({\bm k}_2+{\bm k}_3-{\bm p}_3\r)
\nonumber \\ &
+\frac{(2\pi)^2}{2}\frac{\calP_{\cR}(k_1)}{k_1^3}
\delta^{(3)}\l({\bm k}_1+{\bm k}_3-{\bm p}_3\r)
\frac{(2\pi)^2}{2}\frac{\calP_{\cR}(k_2)}{k_2^3}
\delta^{(3)}\l({\bm k}_2+{\bm p}_3\r)
\nonumber \\ &
+{\rm two~permutations}+\cdots
\biggr].
\end{align}
Then, the integral over ${\bm p}_3$ can be easily performed, thanks to
the presence of the Dirac delta functions. We see that the first term
in the above expression is different from the two next ones. Indeed,
it leads to a term $\delta^{(3)}({\bm k}_3)$ which can be ignored
since, in some sense, it is homogeneous and only participates to the
background. The two other terms yield a $\delta^{(3)}({\bm k}_1+{\bm
  k}_2+{\bm k}_3)$ which ensures momentum conservation.  The final
expression reads
\begin{eqnarray}
\label{eq:bispecloc}
\langle \cR_{\vka} \cR_{\vkb} \cR_{\vkc} \rangle
&=& -\frac{3\,\fnlloc}{10}\; (2\,\pi)^{4}\; (2\,\pi)^{-3/2}\;
\f{1}{k_{1}^3\, k_{2}^3\,k_{3}^3\,}\,
\delta^{(3)}(\vka+\vkb+\vkc)\nn\\
& &\times\l[k_1^{3}\; {\cal P}_{\cR}(k_2)\; {\cal P}_{\cR}(k_3)
+{\rm two~permutations}\r].
\end{eqnarray}
We see that the scale dependence of the bispectrum for this simple
model does not reproduce what we obtained in the case of inflation,
see Eqs.~(\ref{eq:term2}), (\ref{eq:term1}), (\ref{eq:term3})
and~(\ref{eq:term7}). The inflationary case is clearly much more
complicated. In fact, Eq.~(\ref{eq:bispecloc}) has a similar structure
as $\langle \cR_{\vka}(\eta _{\rm e})\, \cR_{\vkb}(\eta _{\rm e})\,
\cR_{\vkc}(\eta _{\rm e})\rangle ^{(7)}$, see
Eq.~(\ref{eq:term7}). But the three extra terms $\langle
\cR_{\vka}(\eta _{\rm e})\, \cR_{\vkb}(\eta _{\rm e})\,
\cR_{\vkc}(\eta _{\rm e})\rangle ^{(1,2,3)}$ are such that the full
slow-roll bispectrum differs from Eq.~(\ref{eq:bispecloc}).

\par

At this stage, it is worth discussing again our conventions. We have
seen below Eq.~(\ref{eq:newconvention}) that, often in the literature,
the two-point correlation function is defined as $\langle
\bar{\cR}_{\vka} \bar{\cR}_{\vkb}\rangle \equiv
(2\pi)^3P_{\cR}(k_1)\delta^{(3)}\l(\vka+\vkb\r)$, where
$\bar{\cR}_{\vka}\equiv (2\pi)^{3/2}\cR_{\vka}$ and
$P_{\cR}(k_1)\equiv \vert f_{{\bm k}_1}\vert^2\neq
\calP_{\cR}(k_1)$. Then, in order to mimic and/or generalize the
definition of the two-point correlation function, the following
definition of the bispectrum $\cB_{\cR}(\vka,\vkb,\vkc)$ is introduced
\begin{equation}
\label{eq:defBR}
  \langle \bar{\cR}_{\vka} \bar{\cR}_{\vkb} \bar{\cR}_{\vkc} \rangle
  =(2\pi)^3
  \cB_{\cR}(k_1,k_2,k_3)\;
  \delta^{(3)}\l(\vka+\vkb+\vkc\r).
\end{equation}
Notice that we could have also used another definition $\langle
\cR_{\vka}\cR_{\vkb} \cR_{\vkc} \rangle =(2\pi)^3
\cB_{\cR}(k_1,k_2,k_3)\; \delta^{(3)}\l(\vka+\vkb+\vkc\r)$, which
would have resulted in a difference by a factor of $(2\pi)^{9/2}$
[and, by the way, explains the appearance of such a factor in
  Eq.~(\ref{eq:threezeta})]. Here, we do not follow this route and use
the convention~(\ref{eq:defBR}). Then, Eq.~(\ref{eq:bispecloc})
implies that
\begin{eqnarray}
\cB_{\cR}(k_1,k_2,k_3)&=&-\frac65\fnlloc 
\l(\vert f_{{\bm k}_2}\vert^2\vert f_{{\bm k}_3}\vert^2
+{\rm two~permutations}\r) 
\\
\label{eq:bispecRloc}
&=& -\frac65\fnlloc 
\l[P_{\cR}(k_2)P_{\cR}(k_3)
+{\rm two~permutations}\r].
\end{eqnarray}
It is also frequent to define the bispectrum of the Bardeen potential
$\Phi$ rather than the conserved quantity $\cR$. Concretely, the
definition reads\footnote{As already mentioned, our convention for
the Fourier transform is such that
\begin{equation}
\Phi(\eta, {\bm x})=\frac{1}{(2\pi)^{3/2}}
\int {\rm d}{\bm k}\, \Phi_{\bm k}(\eta)\, {\rm e}^{-i{\bm k}\cdot {\bm x}}
\end{equation}
and, following the notation that we have already introduced,
$\bar{\Phi}_{\bm k} =(2\pi)^{3/2}\Phi_{\bm k}$. Moreover, if the
Bardeen potential quantum operator is written as
\begin{equation}
\label{eq:qfieldphi}
\hat{\Phi}(\eta, {\bm x})=\int \frac{{\rm d}^3{\bm k}}{(2\pi)^{3/2}}
\left[a_{\bm k}u_{\bm k}(\eta){\rm e}^{i{\bm k}\cdot {\bm x}}
+a_{\bm k}^{\dagger}u_{\bm k}^*(\eta){\rm e}^{-i{\bm k}\cdot {\bm x}}\right],
\end{equation}
then one has $P_{\Phi}\equiv \vert u_{\bm k}\vert^2$ and
$\calP_{\phi}\equiv k^3\vert u_{\bm k}\vert^2/(2\pi^2)$.}
\begin{equation}
\langle
\bar{\Phi}_{\vka}\bar{\Phi}_{\vkb} \bar{\Phi}_{\vkc} \rangle =(2\pi)^3
\cB_{\Phi}(\vka,\vkb,\vkc)\; \delta^{(3)}\l(\vka+\vkb+\vkc\r).
\end{equation}
Since $\zeta=5\Phi/3=-\cR$, we have
$\cB_{\Phi}=-27\cB_{\cR}/125$. However, since $f_{\bm k}=-(5/3)u_{\bm
  k}$, we also have $P_{\cR}=(25/9)P_{\Phi}$. As a consequence, from 
Eq.~(\ref{eq:bispecRloc}), one obtains that
\begin{equation}
\cB_{\Phi}(k_1,k_2,k_3)= 2\fnlloc 
\l[P_{\Phi}(k_2)P_{\Phi}(k_3)
+{\rm two~permutations}\r],
\end{equation}
which is a formula that often appears in the literature. 

\par

Of course, we can also put the slow-roll bispectrum calculated before
under the form given by Eq.~(\ref{eq:defBR}). For this purpose, let us
write Eqs.~(\ref{eq:term1}), (\ref{eq:term2}), (\ref{eq:term3}) and
(\ref{eq:term7}) as
\begin{equation}
\langle
\cR_{\vka}(\eta _{\rm e})\, \cR_{\vkb}(\eta _{\rm e})\,
\cR_{\vkc}(\eta _{\rm e})\rangle ^{(i)}
\equiv \frac{(2\pi)^3}{(2\pi)^{9/2}}{\cal F}^{(i)}
\f{1}{k_{1}^3\, k_{2}^3\,k_{3}^3\,}\,
\delta^{(3)}(\vka+\vkb+\vkc),
\end{equation}
where the concrete expression of the ${\cal F}^{(i)}$ can be read off
from those equations. Then, the bispectrum for single-field slow-roll
models can be written as
\begin{equation}
\label{eq:bispecsrwithF}
\cB_{\cR}^{\rm sr}(k_1,k_2,k_3)=\f{1}{k_{1}^3\, k_{2}^3\,k_{3}^3\,}
\sum_{i=1,2,3,7}{\cal F}^{(i)}.
\end{equation}
The previous result can also be used to define an effective, scale
dependent, $\fnl$ parameter. If we equate the full bispectrum
$\sum_{i=1,2,3,7} \langle \cR_{\vka}(\eta _{\rm e})\, \cR_{\vkb}(\eta
_{\rm e})\, \cR_{\vkc}(\eta _{\rm e})\rangle ^{(i)}$ to the expression
of Eq.~(\ref{eq:bispecloc}), one obtains
\begin{eqnarray}
\fnl^{\rm sr}(\vka,\vkb,\vkc)
&=&-\frac{10}{3}\; (2\,\pi)^{-4}\;
\sum_{i=1,2,3,7}{\cal F}^{(i)}
\nonumber \\ & & \times 
\l[k_1^{3}\; {\cal P}_{\cR}(k_2)\; {\cal P}_{\cR}(k_3)
+{\rm two~permutations}\r]^{-1}.\label{eq:fnl-d}
\end{eqnarray}
If, for instance, we evaluate this quantity for ${\bm k}_1=-{\bm k}_2$
and a vanishing ${\bm k}_3$ (so that ${\bm k}_1+{\bm k}_2+{\bm k}_3$
is also zero which is mandatory given the presence of the Dirac
function in the above expressions), then the expressions of ${\cal
  F}^{(i)}$ simplify such that one obtains
\begin{equation}
\sum_{i=1,2,3,7}{\cal F}^{(i)}=\frac{H^4k^3}{16\Mp^4\epsilon_1}
\left(\frac12+\frac32+0+\frac{\epsilon_2}{\epsilon_1}\right)
=\frac{H^4k^3}{16\Mp^4\epsilon_1^2}(2\epsilon_1+\epsilon_2),
\end{equation}
which, using Eq.~(\ref{eq:specindices}), can be written
as~\cite{Maldacena:2002vr,2010JCAP...10..020R}
\begin{equation}
\fnl^{\rm sr,sq}=\frac{5}{12}(\nS-1),
\end{equation}
where ``sq'' means ``squeezed'' and refers to the fact that we have
taken the particular configuration ${\bm k}_1=-{\bm k}_2$ and a
vanishing ${\bm k}_3$. Notice that, since we calculate a three-point
correlation function, the sign of $\fnl$ is non trivial.  The sign
that we have obtained results from the choice made in
Eq.~(\ref{eq:deffnl}) and from the fact that we evaluate the
correlator of $\cR$. Finally, very roughly speaking (see
Sec.~\ref{sec:observ} for a more complete discussion) the present
status of the art is such that one can detect Non-Gaussianities if
$\vert \fnl\vert >5$. For slow-roll models, since $\nS\simeq 0.96$,
one obtains $\fnl^{\rm sr,sq}\simeq -1.6\times 10^{-2}$, a number that
is therefore undetectable. This conclusion is in fact valid for any
configuration one may choose. Let us also mention that other
consistency relations for Non-Gaussianities have recently been studied
in Refs.~\cite{Sreenath:2013xra,Sreenath:2014nka,Sreenath:2014nca}.

\par

Clearly, a detection of a non-vanishing three-point correlation
function, given present day technology, would immediately rule out
single field slow-roll models with a standard kinetic term. It is
therefore quite remarkable that Non-Gaussianity has not been detected
so far. Let us also stress that the opposite statement is not
true. The fact that we do not see Non-Gaussianities does not imply
that the more complicated models of inflation are necessarily ruled
out even if some of them do predict large Non-Gaussianities. For the
calculation of the three-point correlation functions of these more
complicated models, we again refer to Ref.~\cite{Byrnes:2014pja}.

\subsection{Inflationary four-point Correlation Functions}
\label{subsec:correl4}

Obviously, the next step is to calculate the four-point correlation
function or
trispectrum~\cite{Seery:2006vu,Byrnes:2006vq,Arroja:2008ga}. Of
course, when we consider higher order correlation functions, the
calculations become more and more complicated. In the previous
sub-section, we calculated the action at third order in the
perturbations in order to derive the inflationary three-point
correlation function. In order to calculate the four-point correlation
function, one therefore needs to evaluate the perturbed action at
fourth order. In order to get an idea of how involved it can be, let
us consider again Eq.~(\ref{eq:deffnl}) but expanded up to third order
\begin{equation}
\label{eq:deffnlgnl}
\cR(\eta, {\bm x})=\cR_{_{\mathrm{G}}}(\eta, {\bm x})
-\frac{3\,\fnl}{5}\, 
\cR_{_{\mathrm{G}}}^2(\eta,{\bm x})
+\frac{9}{25}\gnl\cR_{_{\mathrm{G}}}^3(\eta,{\bm x})\cdots ,
\end{equation}
thus introducing the parameter $\gnl$. Here, we write $\fnl$ in order
to avoid cumbersome notations but it should be clear that
$\fnl=\fnlloc$ (and this will be the case in the rest of this
section). The cube of the curvature perturbation can be expressed as
\begin{equation}
\cR_{_{\mathrm{G}}}^3(\eta,{\bm x})=
\frac{1}{(2\pi)^{3/2}}\int {\rm d}{\bm k}
\, (2\pi)^{-3}\int {\rm d}{\bm p}\, {\rm d}{\bm q}\, 
\cR_{{\bm k}-{\bm p}-{\bm q}}\, \cR_{\bm p}\, \cR_{\bm q}\, 
{\rm e}^{-i {\bm k}\cdot {\bm x}},
\end{equation}
which allows us to identify the Fourier transform of the cube of the
curvature perturbation [as we identified the Fourier transform of the
square of the curvature perturbation in
Eq.~(\ref{eq:fouriersquare})]. Then, the four-point correlator takes
the form
\begin{align}
& \left \langle \cR_{{\bm k}_1}\cR_{{\bm k}_2}
\cR_{{\bm k}_3}\cR_{{\bm k}_4}\right \rangle =
\nonumber \\ 
& \biggl \langle 
\left[\cR_{{\bm k}_1}^{_{\rm G}}
-\frac{3\,\fnl}{5}(2\pi)^{-3/2}\int {\rm d}{{\bm p}_1}\, 
\cR_{{\bm p}_1}^{_{\rm G}}\, \cR_{{\bm k}_1-{\bm p}_1}^{_{\rm G}}
+\frac{9\gnl}{25}(2\pi)^{-3}\int {\rm d}{\bm p}_1{\rm d}{\bm q}_1
\cR_{{\bm k}_1-{\bm p}_1-{\bm q}_1}^{_{\rm G}}
\cR_{{\bm p}_1}^{_{\rm G}}\cR_{{\bm q}_1}^{_{\rm G}}
\right]
\nonumber \\ & 
\left[\cR_{{\bm k}_2}^{_{\rm G}}
-\frac{3\,\fnl}{5}(2\pi)^{-3/2}\int {\rm d}{{\bm p}_2}\, 
\cR_{{\bm p}_2}^{_{\rm G}}\, \cR_{{\bm k}_2-{\bm p}_2}^{_{\rm G}}
+\frac{9\gnl}{25}(2\pi)^{-3}\int {\rm d}{\bm p}_2{\rm d}{\bm q}_2
\cR_{{\bm k}_2-{\bm p}_2-{\bm q}_2}^{_{\rm G}}
\cR_{{\bm p}_2}^{_{\rm G}}\cR_{{\bm q}_2}^{_{\rm G}}
\right]
\nonumber \\ & 
\left[\cR_{{\bm k}_3}^{_{\rm G}}
-\frac{3\,\fnl}{5}(2\pi)^{-3/2}\int {\rm d}{{\bm p}_3}\, 
\cR_{{\bm p}_3}^{_{\rm G}}\, \cR_{{\bm k}_3-{\bm p}_3}^{_{\rm G}}
+\frac{9\gnl}{25}(2\pi)^{-3}\int {\rm d}{\bm p}_3{\rm d}{\bm q}_3
\cR_{{\bm k}_3-{\bm p}_3-{\bm q}_3}^{_{\rm G}}
\cR_{{\bm p}_3}^{_{\rm G}}\cR_{{\bm q}_3}^{_{\rm G}}
\right]
\nonumber \\ &  
\left[\cR_{{\bm k}_4}^{_{\rm G}}
-\frac{3\,\fnl}{5}(2\pi)^{-3/2}\int {\rm d}{{\bm p}_4}\, 
\cR_{{\bm p}_4}^{_{\rm G}}\, \cR_{{\bm k}_4-{\bm p}_4}^{_{\rm G}}
+\frac{9\gnl}{25}(2\pi)^{-3}\int {\rm d}{\bm p}_4{\rm d}{\bm q}_4
\cR_{{\bm k}_4-{\bm p}_4-{\bm q}_4}^{_{\rm G}}
\cR_{{\bm p}_4}^{_{\rm G}}\cR_{{\bm q}_4}^{_{\rm G}}
\right]
\biggr\rangle .
\end{align}
Expanding this expression, one arrives at the following formula
\begin{align}
\label{eq:fourpoint}
& \left \langle \cR_{{\bm k}_1}\cR_{{\bm k}_2}
\cR_{{\bm k}_3}\cR_{{\bm k}_4}\right \rangle =
\left \langle \cR_{{\bm k}_1}^{_{\rm G}}\cR_{{\bm k}_2}^{_{\rm G}}
\cR_{{\bm k}_3}^{_{\rm G}}\cR_{{\bm k}_4}^{_{\rm G}}\right \rangle 
\nonumber \\ & 
-\frac{3\,\fnl}{5}(2\pi)^{-3/2}\biggl(
\int{\rm d}{\bm p}_4
\left \langle 
\cR_{{\bm k}_1}^{_{\rm G}}
\cR_{{\bm k}_2}^{_{\rm G}}
\cR_{{\bm k}_3}^{_{\rm G}}
\cR_{{\bm p}_4}^{_{\rm G}}
\cR_{{\bm k}_4-{\bm p}_4}^{_{\rm G}}
\right\rangle +{\rm three~permutations}\biggr)
\nonumber \\ &
+\frac{9}{25}\fnl^2 (2\pi)^{-3}\biggl(
\int{\rm d}{\bm p}_1
\int{\rm d}{\bm p}_2
\left \langle 
\cR_{{\bm p}_1}^{_{\rm G}}
\cR_{{\bm k}_1-{\bm p}_1}^{_{\rm G}}
\cR_{{\bm p}_2}^{_{\rm G}}
\cR_{{\bm k}_2-{\bm p}_2}^{_{\rm G}}
\cR_{{\bm k}_3}^{_{\rm G}}
\cR_{{\bm k}_4}^{_{\rm G}}
\right\rangle +{\rm five~permutations}\biggr)
\nonumber \\ & 
+\frac{9}{25}\gnl (2\pi)^{-3}\biggl(
\int{\rm d}{\bm p}_4
\int{\rm d}{\bm q}_4
\left \langle 
\cR_{{\bm k}_1}^{_{\rm G}}
\cR_{{\bm k}_2}^{_{\rm G}}
\cR_{{\bm p}_3}^{_{\rm G}}
\cR_{{\bm k}_4-{\bm p}_4-{\bm q}_4}^{_{\rm G}}
\cR_{{\bm p}_4}^{_{\rm G}}
\cR_{{\bm q}_4}^{_{\rm G}}
\right\rangle 
\nonumber \\ &
+ {\rm three~permutations}\biggr)
+\cdots ,
\end{align}
where the dots denote higher order terms. The first term in the above
expansion is non-vanishing but can be expressed as the square of
two-point correlation functions and will be ignored in the
following. The second term is zero since it involves five-point
correlation functions of Gaussian quantities. The two last terms are
the terms of interest. We see that they are given in terms of a
six-point correlation function, a quantity which is not zero for
Gaussian quantities. These terms can be evaluated by means of the
Wick's theorem and lead to the sum of fifteen terms, each of them
being made of the product of three two-point correlation functions.
For the term proportional to $\fnl^2$, among the fifteen only eight of
them actually contribute. An example of a term contributing is given
by
\begin{align}
& \int {\rm d}{\bm p}_1{\rm d}{\bm p}_2
\left \langle 
\cR_{{\bm p}_1}^{_{\rm G}}
\cR_{{\bm p}_2}^{_{\rm G}}
\right\rangle
\left \langle 
\cR_{{\bm k}_1-{\bm p}_1}^{_{\rm G}}
\cR_{{\bm k}_3}^{_{\rm G}}
\right\rangle
\left \langle 
\cR_{{\bm k}_2-{\bm p}_2}^{_{\rm G}}
\cR_{{\bm k}_4}^{_{\rm G}}
\right\rangle
\nonumber \\ &
=
\int {\rm d}{\bm p}_1{\rm d}{\bm p}_2
\frac{(2\pi)^2}{2p_1^3}\calP_{\cR}(p_1)\delta^{(3)}({\bm p}_1+{\bm p}_2)
\frac{(2\pi)^2}{2k_3^3}\calP_{\cR}(\vert {\bm k}_1-{\bm p}_1\vert)
\delta^{(3)}({\bm k}_1-{\bm p}_1+{\bm k}_3)
\nonumber \\ & \times
\frac{(2\pi)^2}{2k_4^3}\calP_{\cR}(\vert {\bm k}_2-{\bm p}_2\vert)
\delta^{(3)}({\bm k}_2-{\bm p}_2+{\bm k}_4)
\nonumber \\ & 
=\frac{(2\pi)^6}{8}\int {\rm d}{\bm p}_1\frac{\calP_{\cR}(p_1)}{p_1^3}
\frac{\calP_{\cR}(\vert {\bm k}_1-{\bm p}_1\vert)}{k_3^3}
\frac{\calP_{\cR}(\vert {\bm k}_2+{\bm p}_1\vert)}{k_4^3}
\delta^{(3)}({\bm k}_1-{\bm p}_1+{\bm k}_3)
\nonumber \\ & \times
\delta^{(3)}({\bm k}_2+{\bm p}_1+{\bm k}_4)
\nonumber \\ &
\label{eq:finalform}
=\frac{(2\pi)^6}{8}
\frac{1}{\vert {\bm k}_1+{\bm k}_3\vert^3k_3^3k_4^3}
\calP_{\cR}(\vert {\bm k}_1+{\bm k}_3\vert)
\calP_{\cR}(k_3)\calP_{\cR}(k_4)
\delta^{(3)}({\bm k}_1+{\bm k}_2+{\bm k}_3+{\bm k}_4).
\end{align}
In fact among the eight terms mentioned above, four are identical to
the one we have just calculated and the remaining four are all given
by Eq.~(\ref{eq:finalform}), but with $\vert {\bm k}_1+{\bm k}_3\vert$
replaced with $\vert {\bm k}_1+{\bm k}_4\vert$. On the other hand, an
example of a non-contributing term is
\begin{align}
\label{eq:ignored1}
& \int {\rm d}{\bm p}_1{\rm d}{\bm p}_2
\left \langle 
\cR_{{\bm p}_1}^{_{\rm G}}
\cR_{{\bm k}_1-{\bm p}_1}^{_{\rm G}}
\right\rangle
\left \langle 
\cR_{{\bm p}_2}^{_{\rm G}}
\cR_{{\bm k}_2-{\bm p}_2}^{_{\rm G}}
\right\rangle
\left \langle 
\cR_{{\bm k}_3}^{_{\rm G}}
\cR_{{\bm k}_4}^{_{\rm G}}
\right\rangle.
\end{align}
We see that the first two-point correlation function appearing in the
above integral will lead to a term proportional to $\delta^{(3)}({\bm
  p}_1+{\bm k}_1-{\bm p}_1)=\delta^{(3)}({\bm k}_1)$, which, in some
sense, is homogeneous. This explains why the term in
Eq.~(\ref{eq:ignored1}) can be ignored.

\par

Let us now come back to Eq.~(\ref{eq:fourpoint}) and consider the term
proportional to $\gnl$. Using again Wick's theorem, this term can be
expressed as the sum of fifteen terms made of the product of three
two-point correlation functions. Among these fifteen terms, only six
participate to the final expression (and they all give the same
contribution). One example is 
\begin{align}
\label{eq:contributeg}
& \int {\rm d}{\bm p}_4{\rm d}{\bm q}_4
\left \langle 
\cR_{{\bm k}_1}^{_{\rm G}}
\cR_{{\bm k}_4-{\bm p}_4-{\bm q}_4}^{_{\rm G}}
\right\rangle
\left \langle 
\cR_{{\bm k}_2}^{_{\rm G}}
\cR_{{\bm p}_4}^{_{\rm G}}
\right\rangle
\left \langle 
\cR_{{\bm k}_3}^{_{\rm G}}
\cR_{{\bm q}_4}^{_{\rm G}}
\right\rangle
\nonumber \\ &
=
\int {\rm d}{\bm p}_4{\rm d}{\bm q}_4
\frac{(2\pi)^2}{2k_1^3}\calP_{\cR}(k_1)\delta^{(3)}({\bm k}_1
+{\bm k}_4-{\bm p}_4-{\bm q}_4)
\frac{(2\pi)^2}{2k_2^3}\calP_{\cR}(k_2)
\delta^{(3)}({\bm k}_2+{\bm p}_4)
\nonumber \\ & \times
\frac{(2\pi)^2}{2k_3^3}\calP_{\cR}(k_3)
\delta^{(3)}({\bm k}_3+{\bm q}_4)
\nonumber \\ & 
=\frac{(2\pi)^6}{8}\frac{1}{k_1^3k_2^3k_3^3}
\calP_{\cR}(k_1)
\calP_{\cR}(k_2)
\calP_{\cR}(k_3)
\delta^{(3)}({\bm k}_1+{\bm k}_2+{\bm k}_3+{\bm k}_4).
\end{align}
Putting everything together, one obtains the following 
expression
\begin{eqnarray}
& & \left \langle \cR_{{\bm k}_1}\cR_{{\bm k}_2}
\cR_{{\bm k}_3}\cR_{{\bm k}_4}\right \rangle =
\frac{9\fnl^2}{25}(2\pi)^{-3}\biggl[4\times 
\frac{(2\pi)^6}{8}
\frac{1}{\vert {\bm k}_1+{\bm k}_3\vert^3k_3^3k_4^3}
\calP_{\cR}(\vert {\bm k}_1+{\bm k}_3\vert)
\calP_{\cR}(k_3)\calP_{\cR}(k_4)
\nonumber \\ 
& & +{\rm eleven~permutations}\biggr]
\delta^{(3)}({\bm k}_1+{\bm k}_2+{\bm k}_3+{\bm k}_4)
+\frac{9\gnl}{25}(2\pi)^{-3}\biggl[6\times
\frac{(2\pi)^6}{8}\frac{1}{k_1^3k_2^3k_3^3}
\nonumber \\ & & \times
\calP_{\cR}(k_1)
\calP_{\cR}(k_2)
\calP_{\cR}(k_3)
+
{\rm three~permutations}\biggr]
\delta^{(3)}({\bm k}_1+{\bm k}_2+{\bm k}_3+{\bm k}_4)
\\ 
\label{eq:finalfour}
& &=\frac{36\fnl^2}{25}(2\pi)^{-3}\biggl(\vert f_{{\bm  k}_1+{\bm k}_3}\vert^2
\vert f_{{\bm k}_3}\vert^2\vert f_{{\bm k}_4}\vert^2 +{\rm eleven~permutations}
\biggr)\delta^{(3)}({\bm k}_1+{\bm k}_2+{\bm k}_3+{\bm k}_4)
\nonumber \\ & &
+\frac{54\gnl}{25}(2\pi)^{-3}\biggl(\vert f_{{\bm  k}_1}\vert^2
\vert f_{{\bm k}_3}\vert^2\vert f_{{\bm k}_3}\vert^2 +{\rm three~permutations}
\biggr)\delta^{(3)}({\bm k}_1+{\bm k}_2+{\bm k}_3+{\bm k}_4).\nonumber \\
\end{eqnarray}
The fact that we have eleven permutation in the first term comes from
the fact that we had six terms and that each of these terms separates in
two groups. At the end, this gives twelve terms. Usually, the
definition of the trispectrum is given in terms of $\bar{\cR}_{\bm k}$
(see the above discussions about conventions) and reads
\begin{eqnarray}
\left \langle \bar{\cR}_{{\bm k}_1}\bar{\cR}_{{\bm k}_2}
\bar{\cR}_{{\bm k}_3}\bar{\cR}_{{\bm k}_4}\right \rangle =
(2\pi)^3{\cal T}_{\cR}(k_1,k_2,k_3,k_4)
\delta^{(3)}({\bm k}_1+{\bm k}_2+{\bm k}_3+{\bm k}_4),
\end{eqnarray}
with
\begin{eqnarray}
\label{eq:triinf}
{\cal T}_{\cR}(k_1,k_2,k_3,k_4)&=&
\tau_{_{\rm NL}}\biggl(\vert f_{{\bm  k}_1+{\bm k}_3}\vert^2
\vert f_{{\bm k}_3}\vert^2\vert f_{{\bm k}_4}\vert^2 +{\rm eleven~permutations}
\biggr)
\nonumber \\ &+&
\frac{54\gnl}{25}\biggl(\vert f_{{\bm  k}_1}\vert^2
\vert f_{{\bm k}_3}\vert^2\vert f_{{\bm k}_3}\vert^2 +{\rm three~permutations}
\biggr).
\end{eqnarray}
One can check that our result~(\ref{eq:finalfour}) matches exactly
this form provided that
\begin{equation}
\tau_{_{\rm NL}}=\frac{36\fnl^2}{25}.
\end{equation}
This equation is called the Suyama-Yamaguchi consistency
relation~\cite{Suyama:2010uj} (more precisely, it is in fact a
particular case of $\tau_{_{\rm NL}}\geq 36f_{_{\rm NL}}^2/25$). The
above equation indicates that the tri-spectrum is expected to be
quadratic in the slow-roll parameters and, hence, even harder to
detect than the three-point correlation function. Of course, it should
be stressed again that the scale dependence of Eq.~(\ref{eq:triinf})
is not what would emerge from an exact calculation starting from the
perturbed action at fourth order. In Sec.~\ref{sec:observ}, we will
discuss the constraints put by the Planck experiment on the
tri-spectrum.

\subsection{Adiabatic and Isocurvature Perturbations}
\label{subsec:iso}

Another important consequence that follows from the Planck data is
that the perturbations are adiabatic. Before discussing in more detail
in Sec.~\ref{sec:observ} how this conclusion is reached, we now
explain what it means and what it implies for inflation.

\par

The post inflationary Universe is made of four fluids: photons,
neutrinos, baryons and cold dark matter (we are ignoring dark
energy). In order to calculate the CMB anisotropies, one needs to
integrate the equations governing the behavior of these four
fluids. But we also need to specify initial conditions, just after
inflation, at the onset of the radiation dominated era. Different
initial conditions will lead to different subsequent evolutions and,
therefore, to different CMB patterns. Adiabaticity refers to a
situation where one has~\cite{Bucher:1999re}
\begin{equation}
\label{eq:adia}
\delta_{\rm cdm}=\delta_{\rm b}=\frac34 \delta_{\gamma}
=\frac34\delta _{\nu},
\end{equation}
where $\delta_X\equiv \delta \rho_X/\rho_X$ is the density contrast
(``cdm'' stands for cold dark matter, ``b'' for baryons, $\gamma$ for
photons and $\nu$ for neutrinos). It may be surprising that CMB data
single out particular initial conditions and it is interesting to
discuss why the conditions~(\ref{eq:adia}) play an important
role. Equally important is the question of what they can teach us
about inflation: after all, these initial conditions are the results
of what happened during inflation. As a consequence, they certainly
tell us something about the type of inflationary expansion that took
place in the early Universe.

\par

Let us start by giving the equations controlling the evolution of the
four fluids mentioned before. Each fluid is characterized by its
density contrast $\delta_X$ and by its velocity $v_X$. From energy
conservation, one can derive the following equations
\begin{eqnarray}
\label{eq:conc}
\left(\delta_{\rm c}-3\Psi\right)'-k^2v_{\rm c} 
=\Delta _{\rm c}'-k^2v_{\rm c} &=& 0, \\
\label{eq:conb}
\left(\delta_{\rm b}-3\Psi\right)'-k^2v_{\rm b} 
=\Delta_{\rm b}'-k^2v_{\rm b} 
&=& 0, \\
\label{eq:congamma}
\left(\delta_{\gamma}-4\Psi\right)'-\frac43k^2v_{\gamma} 
=\Delta_{\gamma}'-\frac43k^2v_{\gamma} 
&=& 0, \\
\label{eq:connu}
\left(\delta_{\nu}-4\Psi\right)'-\frac43k^2v_{\nu} 
=\Delta_{\nu}'-\frac43k^2v_{\nu} 
&=& 0,
\end{eqnarray}
where $\Psi$ is the second Bardeen potential already considered before
(but, in the present context, we no longer necessarily have
$\Psi=\Phi$) and where the quantities $\Delta_X$ are defined by the
above equations. The space component of the conservation equation
gives an equation for the velocities. For cold dark matter, one
obtains
\begin{equation}
\label{eq:convc}
v_{\rm c}'+{\cal H}v_{\rm c}+\Phi =0,
\end{equation}
where $\Phi$ is the other Bardeen potential. In the early Universe,
baryons and photons are tightly coupled. This means that $v_{\rm
  b}=v_{\rm \gamma}\equiv v_{{\rm b}\gamma}$. The corresponding
equation of motion reads
\begin{equation}
\label{eq:convbgamma}
v_{{\rm b}\gamma}'+\frac{R}{1+R}{\cal H}v_{{\rm b}\gamma}+\Phi
+\frac14\frac{\delta_{\gamma}}{1+R}
+\frac{4\eta_{{\rm b}\gamma}}{3a\rho_{\gamma}}
\frac{R}{1+R}k^2v_{{\rm b}\gamma}=0,
\end{equation}
where $\eta_{{\rm b}\gamma}$ is the viscosity (or anisotropic stress)
of the fluid made of baryons and photons and $R$ is three quarters of
the baryon to photon energy density ratio, namely $R\equiv 3\rho_{\rm
  b}/(4\rho_{\gamma})$. Finally, the conservation equation for the
neutrinos can be written as
\begin{equation}
\label{eq:convnu}
v_{\nu}'+\Phi+\frac14 \delta _{\nu}+\frac{\eta_{\nu}}{a\rho_{\nu}}k^2v_{\nu}=0,
\end{equation}
where $\eta_{\nu}$ is the neutrinos viscosity (notice that the
viscosity does not appear in the time component of the conservation
equations). Since the above formulas contain the two Bardeen
potentials, they must be supplemented by additional equations
governing the behavior of $\Phi$ and $\Psi$. These are of course the
perturbed Einstein equations. By combining the time-time and
time-space Einstein equations, one arrives at
\begin{equation}
\label{eq:barone}
-\frac{k^2}{{\cal H}^2}\Psi-\frac92\Psi \sum _{X}\Omega_X(1+w_X)
=\frac32\sum_X\Omega_X\Delta_X-\frac92{\cal H}
\sum_X\Omega_X(1+w_X)v_X,
\end{equation}
where the sum runs over the four species mentioned above, where $w_X$
is the equation of state parameter of the fluid $X$ and $\calH\equiv
a'/a$. Finally the space space component of the Einstein equations
(with $i\neq j$) leads to
\begin{equation}
\label{eq:bartwo}
\frac{k^2}{{\cal H}^2}\left(\Phi-\Psi\right)=\frac{6k^2}{a\rho_{\rm cri}}
\left(\eta_{{\rm b}\gamma}k^2v_{{\rm b}\gamma}+\eta_{\nu}k^2v_{\nu}\right),
\end{equation}
where we remind that $\rho_{\rm cri}$ is the critical energy
density. At this stage we have all the equations necessary to
understand the behavior of the four fluids: we have ten quantities
(namely four $\delta_X$, four $v_X$, $\Psi$ and $\Phi$) and ten
equations, namely Eqs.~(\ref{eq:conc}), (\ref{eq:conb}),
(\ref{eq:congamma}), (\ref{eq:connu}), (\ref{eq:convc}),
(\ref{eq:convbgamma}), (\ref{eq:convnu}), (\ref{eq:barone})
and~(\ref{eq:bartwo}) (the tenth equation is simply $v_{\rm
  b}=v_{\gamma}$). The only thing which remains to be done is to
specify the initial conditions. Integrating this system of ten
equations analytically is not possible (even if linear). This has to
be done numerically. However, since we are mainly interested in the
behavior of the system on large scales, the problem gets
simplified. Indeed, let us introduce the quantity, introduced by
Bardeen, Steinhardt and Turner, $\zeta _{_{\rm BST}}$ defined
by~\cite{Bardeen:1983qw,Martin:1997zd}
\begin{equation}
\label{eq:defzetabst}
  \zeta _{_{\rm BST}}=-\Psi-\frac{{\cal H}}{\rho '}\delta \rho 
=\sum_X\frac{\rho_X'}{\rho'}\zeta_X\, ,
\end{equation}
where $\rho=\sum_X\rho_X$ is the total energy density and $\zeta_X$
can be expressed as
\begin{equation}
\zeta_X=-\Psi+\frac{\delta_X}{3\left(1+w_X\right)}.
\end{equation}
From Eqs.~(\ref{eq:conc}), (\ref{eq:conb}), (\ref{eq:congamma}) and
(\ref{eq:connu}), we see that, on large scales (where the terms
$\propto k^2v_X$ go to zero), each $\zeta_X$ is conserved, namely
$\zeta'_X=0$. Now, we understand the particular role of the
conditions~(\ref{eq:adia}). Indeed, they amount to simply choose
\begin{equation}
\label{eq:adiazeta}
\zeta_{\rm cdm}=\zeta_{\rm b}=\zeta_{\gamma}=\zeta_{\nu}\equiv \zeta_{\rm adia}.
\end{equation} 
and, in this case, we have
\begin{equation}
\zeta_{_{\rm BST}}=\zeta _{\rm adia}\sum_X\frac{\rho_X'}{\rho'}=\zeta_{\rm adia},
\end{equation}
which is a constant. Therefore, for adiabatic initial conditions, the
quantity $\zeta_{_{\rm BST}}$ is conserved on large scales. Another
way to see the same thing is to differentiate $\zeta _{_{\rm BST}}$
(using the expression of $\delta \rho '$ obtained from energy
conservation). Then, one arrives at the following equation
\begin{equation}
\zeta _{_{\rm BST}}'=-\frac{{\cal H}}{\rho +p}\delta p_{\rm nad}
-\frac13 \partial _i\partial ^iv^{\rm (gi)}\, ,
\end{equation}
which shows that, on large scales, the conservation of $\zeta_{_{\rm
    BST}}$ is controlled by the non-adiabatic pressure [here,
  $v^{({\rm gi})}$ is the scalar component of the gauge-invariant
  velocity]. This quantity is defined by the following expression
\begin{equation}
\label{defpnonadia}
\delta p_{\rm nad}=\delta p-c_{_{\rm S}}^2\delta \rho \, ,
\end{equation}
where $\delta \rho $, $\delta p$ are the total perturbed energy
density and pressure, respectively. The quantity $c_{_{\rm S}}^2\equiv
p'/\rho '$ is the (total) sound velocity. In the case where one has
two fluids (in order to keep things simple), expressing the perturbed
energy density and the perturbed pressure explicitly, one arrives at
\begin{eqnarray}
\delta p_{\rm nad} &=& \left(\delta p_1-c_{_{\rm S 1} }^2\delta \rho _1\right) 
+\left(\delta p_2-c_{_{\rm S 2} }^2\delta \rho _2\right)
\nonumber \\ & &
+\left(c_{_{\rm S1} }^2-c_{_{\rm S2} }^2\right)
\frac{\left(\rho _1+p_1\right)\left(\rho _2+p_2\right)}{\rho +p}
S_{12}\, ,
\end{eqnarray}
where $S_{12}$ is given by
\begin{equation}
\label{eq:defS12}
S_{12}=\frac{\delta \rho _1}{\rho _1+p_1}
-\frac{\delta \rho _2}{\rho _2+p_2}=3\left(\zeta_1-\zeta_2\right)\, .
\end{equation}
and where $c_{_{\rm S i}}\equiv p_i'/\rho_i'$. The non-adiabatic
pressure contains two contributions. The terms $\delta p_i-c_{_{\rm S
    i} }^2\delta \rho _i$ originate from intrinsic entropy
perturbations (if any) of the fluids under consideration while the
term proportional to $S_{12}$ represents the entropy of mixing. Let us
summarize: for adiabatic perturbations, $\zeta_{_{\rm BST}}$ is a
conserved quantity. For non adiabatic perturbations, this quantity can
evolve even on large scales and this evolution is given by
Eq.~(\ref{eq:defzetabst}).

\par

Let us also remark that one can work in terms of the quantity $\zeta $
defined by~\cite{Martin:1997zd} and already introduced before
\begin{equation}
\zeta =\Phi +\frac23\frac{{\cal H}^{-1}\Phi '+\Phi
}{1+\omega }\, .
\end{equation}
If one has $\Psi=\Phi$, then
\begin{equation}
\zeta _{_{\rm BST}}=-\zeta -\frac{k^2}{3\epsilon_1 {\cal H}^2}\Phi \,
,
\end{equation}
and, in the standard situation, when there is no entropy
perturbations, the quantities $\zeta $ and $\zeta _{_{\rm BST}}$ are
both conserved on super-Hubble scales. Notice that, strictly speaking,
$\zeta$ stays constant only in absence of shear viscosity.

\par

Let us now try to understand how the presence or the absence of
adiabatic perturbations can affect CMB anisotropies. On large scales,
the temperature fluctuations can be expressed as
\begin{equation}
\label{eq:sw}
\frac{\delta T}{T}\simeq \frac14 \delta _{\gamma}\vert _{\rm lss}
+\Phi\vert _{\rm lss},
\end{equation}
where ``lss'' means ``last scattering surface'' and indicates when the
radiation density contrast and the Bardeen potential must be
evaluated. Since last scatterings occur during the matter dominated
era, using the time-time component of the perturbed Einstein equation,
one obtains $-2\Phi\vert_{\rm lss}\simeq R_{\rm cdm}\delta_{\rm
  cdm}\vert_{\rm lss}+R_{\rm b}\delta_{\rm b}\vert_{\rm lss}$ where
$R_{\rm cdm}\equiv \rho_{\rm cdm}/(\rho_{\rm cdm}+\rho_{\rm b})$ and
$R_{\rm b}\equiv \rho_{\rm b}/(\rho_{\rm cdm}+\rho_{\rm b})$. It is
conventional to measure the non-adiabatic perturbation with respect to
photons. Therefore, one introduces the notation $S_X\equiv
S_{X\gamma}\equiv 3(\zeta_X-\zeta_{\gamma})$. Then, one obtains
\begin{equation}
\label{eq:philssone}
  \Phi \vert_{\rm lss}=-\frac35\zeta_{\gamma}-\frac15 R_{\rm
    cdm}S_{\rm cdm}
-\frac15 R_{\rm b}S_{\rm b},
\end{equation}
where we recall that $R_{\rm b}$ and $R_{\rm cdm}$ are evaluated at
last scattering. Notice also that, in principle, we do not need a
subscript ``lss'' for $\zeta_X$ or $S_X$ because they are constant (in
time) quantities since $\zeta_X$ is conserved. In particular, they
should be viewed as the value of $\zeta_X$ at the onset of the
radiation dominated era, just after inflation and, therefore, $S_X$
could also be written as $S_X^{\rm ini}$ in order to emphasize this
point. To calculate the temperature anisotropies, we use
Eq.~(\ref{eq:sw}) and write $\delta T/T=\delta_{\gamma}/4\vert_{\rm
  lss}+\Phi_{\rm lss}=\zeta_{\gamma}+2\Phi\vert_{\rm lss}$. As a
result, one obtains
\begin{equation}
\label{eq:dt1}
\frac{\delta T}{T}=
-\frac15\zeta_{\gamma}-\frac25 R_{\rm
    cdm}S_{\rm cdm}
-\frac25 R_{\rm b}S_{\rm b}.
\end{equation}
Finally, during the Radiation Dominated (RD) era, one can write
\begin{equation}
\label{eq:zetard}
\zeta_{_{\rm RD}}=R_{\gamma}\zeta_{\gamma}+R_{\nu}\zeta_{\nu}
=\zeta_{\gamma}+R_{\nu}\frac{S_{\nu}}{3},
\end{equation}
with $R_{\gamma}\equiv \rho_{\gamma}/(\rho_{\gamma}+\rho_{\nu})$ and
$R_{\nu}\equiv \rho_{\nu} /(\rho_{\gamma}+\rho_{\nu})$, these
quantities being evaluated during the radiation dominated era. Using
Eq.~(\ref{eq:zetard}) to obtain an expression of $\zeta_{\gamma}$ and
using this expression in Eq.~(\ref{eq:dt1}), it follows that
\begin{equation}
\label{eq:dTnonadia}
\frac{\delta T}{T}=
-\frac15\zeta_{_{\rm RD}}-\frac25 R_{\rm
    cdm}S_{\rm cdm}
-\frac25 R_{\rm b}S_{\rm b}+\frac{1}{15}R_{\nu}S_{\nu},
\end{equation}
which coincides with Eq.~(7) of Ref.~\cite{Gordon:2002gv}. The term
$\zeta_{_{\rm RD}}/5$ represents the adiabatic contribution. In fact,
one can also define an effective isocurvature mode taking into account
both cold dark matter and baryons entropy fluctuations by defining
\begin{equation}
\label{eq:Scdmeff}
S_{\rm cdm}^{\rm eff}\equiv S_{\rm cdm}+\frac{R_{\rm b}}{R_{\rm cdm}}S_{\rm b},
\end{equation}
such that Eq.~(\ref{eq:dTnonadia}) now reads
\begin{equation}
\label{eq:dTnonadiafinal}
\frac{\delta T}{T}=
-\frac15\zeta_{_{\rm RD}}-\frac25 R_{\rm
    cdm}S_{\rm cdm}^{\rm eff}+\frac{1}{15}R_{\nu}S_{\nu}.
\end{equation}
We therefore have two adiabatic modes that are, as will be seen in
Sec.~\ref{sec:observ}, denoted by the Planck collaboration CDI (for
the effective cold dark matter) and NDI (for neutrinos). In fact,
there is a third mode, NVI, related to neutrinos velocity. Since the
expression of the temperature is modified by the presence of
isocurvature modes, the temperature multipole moments will also be
affected, for concrete and quantitative results see for instance
Ref.~\cite{Langlois:2000ar}. As a consequence, when compared to the
CMB data, one can put constraints on their amplitude.


As will be discussed in Sec.~\ref{sec:observ}, so far, CMB
measurements are consistent with adiabaticity. This gives non trivial
information about inflation. Indeed, if non adiabatic pertubations
were observed it would mean that inflation can not be driven by a
single scalar field. As for Non-Gaussianities, this would have implied
that single-field slow-roll inflation with a standard kinetic term
were ruled out. This class of models has therefore passed another
non-trivial test. Of course, this is the situation now and this could
very well change in the future. In that case, what would be the
implications for inflation? A natural explanation would be to have
multiple field inflation and we now explain in detail why using a
simple example~\cite{Polarski:1992dq,Peter:1994dx,
  Polarski:1994rz,Langlois:1999dw,Langlois:2000ar,Gordon:2000hv,
  Amendola:2001ni,Bartolo:2001rt,
  Wands:2002bn,Gordon:2002gv,Byrnes:2006fr,Wands:2007bd,Choi:2008et}.

\par

Assume that, instead of having one field, we now have a collection of
fields that all play a role during inflation. For simplicity, and
because we want to be explicit, let us consider the case where we have
two fields, $\phi_{\rm h}$ and $\phi_{\ell}$, and where the potential
is quadratic for each field, without interaction term, namely
$V=V_{\rm h}+V_{\ell}\equiv m_{\rm h}^2\phi_{\rm
  h}^2/2+m_{\ell}^2\phi_{\ell}/2$. Then, the equations of motion for
the background are given by
\begin{align}
& H^2 = \frac{\kappa}{3}\left(\frac12\dot{\phi}^2_{\rm h}
+\frac12\dot{\phi}^2_{\ell}
+\frac12 m_{\rm h}^2\phi_{\rm h}^2
+\frac12 m_{\ell }^2\phi_{\ell}^2
\right), \\
& \ddot{\phi}_{\rm h} +3H\dot{\phi}_{\rm h}+m_{\rm h}^2\phi_{\rm h} = 0, \\
& \ddot{\phi}_{\ell } +3H\dot{\phi}_{\ell}+m_{\ell}^2\phi_{\ell} = 0,
\end{align}
where, as is standard in the literature, we have used notations that
make obvious the fact that one field is heavy and the other light,
meaning that $R\equiv m_{\rm h}/m_{\ell}>1$ [not to be confused with
  the $R$ introduced in Eq.~(\ref{eq:convbgamma})]. These equations
cannot be solved exactly but one can use the slow-roll
approximation. The first Hubble flow parameter is given by, see
Eq.~(\ref{eq:eps1})
\begin{equation}
\epsilon_1=\frac{\kappa}{2H^2}\left(\dot{\phi}_{\rm h}^2
+ \dot{\phi}_{\ell}^2\right), 
\end{equation}
and $\epsilon_1\ll 1$ implies that $\kappa \dot{\phi}_{\rm
  h}^2/(2H^2)\ll 1$ and $\kappa \dot{\phi}_{\ell }^2/(2H^2)\ll
1$. These two conditions are similar to what would be obtained in the
single-field case. This means that, as usual, the kinetic term can be
neglected in the Friedmann equation. On the other hand, the second
Hubble flow parameter can be written as
\begin{equation}
\epsilon_2=2\epsilon_1+\frac{2}{H}\frac{\ddot{\phi}_{\rm h}
\dot{\phi}_{\rm h}+\ddot{\phi}_{\ell}
\dot{\phi}_{\ell}}{\dot{\phi}_{\rm h}^2+\dot{\phi}_{\ell}^2}.
\end{equation}
In the single-field case, this relation reduces to $\epsilon_2
=2\epsilon_1+2\ddot{\phi}/(H\dot{\phi})$ and the acceleration in the
Klein-Gordon equation can also be neglected since $\epsilon_2\ll 1$
implies that $\ddot{\phi}/(H\dot{\phi})\ll 1$. However, in the
two-field case, the properties $\ddot{\phi_{\rm h}}/(H\dot{\phi_{\rm
    h}})\ll 1$ and $\ddot{\phi_{\ell }}/(H\dot{\phi_{\ell }})\ll 1$
cannot be deduced from $\epsilon_2\ll 1$. As a consequence, neglecting
the acceleration in the Klein-Gordon equations for the heavy and light
fields is in fact an additional assumption that we will make in the
following. Then, using that $\dot{H}=-\kappa\left(\dot{\phi}_{\rm
  h}^2+\dot{\phi}_{\ell}^2\right)/2$ (which, by the way, shows that
the Hubble parameter always decreases) and the slow-roll Klein-Gordon
equation to relate the first time derivative of the fields to the
derivative of the potential, one obtains
\begin{equation}
\label{eq:eps1double}
\epsilon_1
\simeq 2\Mp^2\frac{R^4\phi_{\rm h}^2+\phi_{\ell}^2}
{\left(R^2\phi_{\rm h}^2+\phi_{\ell}^2\right)^2}.
\end{equation}
If the heavy field dominates, $R\phi_{\rm h}\gg \phi_{\ell}$ or,
equivalently, $m_{\rm h}\phi_{\rm h}\gg m_{\ell}\phi_{\ell}$, then
$\epsilon_1\simeq 2\Mp^2/\phi_{\rm h}^2$. As expected, this expression
is similar to that one would obtain in single Large Field Inflation
(LFI). And if the light field dominates, \ie if $\phi_{\ell}\gg
R^2\phi_{\rm h}$, then $\epsilon_1\simeq
2\Mp^2/\phi_{\ell}^2$. Therefore, we will assume the following initial
conditions which guarantee that slow-roll is valid
\begin{equation}
\label{eq:icdouble}
\phi_{\rm h}\gg \sqrt{2}\Mp, \quad \phi_{\ell}\gg \sqrt{2}\Mp, 
\quad R\phi_{\rm h}\gg \phi_{\ell}.
\end{equation}
The second condition is a priori less obvious so let us discuss it a
little bit more. The domination of the heavy field comes to an end
when $R\phi_{\rm h}=\phi_{\ell}$. At this transition, the first Hubble
flow parameter is given by $\epsilon_{1 {\rm
    t}}=2\Mp^2(1+R^2)/\phi_{\ell {\rm t}}^2$. So inflation does not
stop provided $\phi_{\ell {\rm t}}\gg \sqrt{2}\sqrt{1+R^2}\Mp\sim
\sqrt{2}R\Mp$. Since the light field is almost constant during the
phase dominated by the heavy field (see below), this justifies our
initial condition. But it is also possible to consider a situation
where inflation stops at the transition, namely $\phi_{\ell {\rm
    t}}<\sqrt{2}R\Mp$~\cite{Polarski:1992dq,Peter:1994dx,
  Polarski:1994rz}. After the transition, $\epsilon_1\simeq
2\Mp^2/\phi_{\ell }^2$ and if one wants inflation to start again, one
needs the condition~(\ref{eq:icdouble}) for the light field. 

\par

As long as the slow-roll approximation is valid, the equations of
motion can be integrated and the solution for the field vacuum
expectation values reads~\cite{Polarski:1992dq,Peter:1994dx,
  Polarski:1994rz}
\begin{equation}
\label{eq:fieldsrdouble}
\phi_{\rm h}=\sqrt{\frac{4s}{\kappa}}\sin \left[\theta(s)\right], \quad
\phi_{\ell }=\sqrt{\frac{4s}{\kappa}}\cos \left[\theta(s)\right]
\end{equation}
while the Hubble parameter is given by
\begin{equation}
\label{eq:frieddouble}
H^2(s)=\frac{2s}{3}m_{\ell}^2
\left[1+\left(R^2-1\right)\sin ^2\theta\right].
\end{equation}
This is a parametric representation of the solution in terms of the
variable $s$ defined by $s=-\ln (a/a_{\rm end})$, with 
\begin{equation}
\label{eq:s}
s=s_0\frac{\left(\sin \theta \right)^{2m_{\ell}^2/\left(m_{\rm h}^2-m_{\ell}^2\right)}}
{\left(\cos \theta \right)^{2m_{\rm h}^2/\left(m_{\rm h}^2-m_{\ell}^2\right)}}.
\end{equation}
The initial phase, dominated by the heavy field, corresponds to
$\theta \rightarrow \pi/2$, $s\rightarrow \infty$,
$\epsilon_1\rightarrow 0$ and $\phi_{\rm h}/\phi_{\ell}=\tan
\theta\rightarrow \infty$. As already mentioned, this happens when
$m_{\rm h}\phi_{\rm h}>m_{\ell}\phi_{\ell}$ or $\theta <\theta_{\rm
  t}$ where the ``transition angle'' $\theta_{\rm t}$ is given by
$\tan \theta_{\rm t}\equiv R^{-1}$. The fact that $R>1$ implies that
$\theta_{\rm t}<\pi/4$. If $R\gg 1$ then $\theta_{\rm t}\simeq R^{-1}$
is a small angle. In that case (\ie $R\gg 1$), in the regime where
$\theta\gg \theta _{\rm t}$ (\ie $\theta $ is not a small angle), we
have the following behavior for $s$: $s\simeq
s_0\cos^{-2}\theta$. This implies that the heavy and light fields are
given by
\begin{equation}
\label{eq:heavyphase}
\phi_{\rm h}\simeq \sqrt{\frac{4s_0}{\kappa}} \tan \theta \gg \Mp, \quad 
\phi_{\ell}\simeq \sqrt{\frac{4s_0}{\kappa}}.
\end{equation}
In this regime, the heavy field is super-Planckian and the model is
effectively equivalent to large field inflation (LFI). This is
confirmed by writing the Friedmann equation~(\ref{eq:frieddouble})
using Eq.~(\ref{eq:heavyphase})
\begin{equation}
H^2\simeq \frac{2s}{3}m_{\ell}^2R^2\sin^2 \theta
=\frac{2s_0}{3}m_{\rm h}^2\tan ^2\theta
=\frac{\kappa}{6}m_{\rm h}^2\phi_{\rm h}^2,
\end{equation}
which is exactly the Friedmann equation for LFI. On the other hand, as
announced above, the light field is frozen and its back-reaction is
negligible. This provides an interpretation for the parameter $s_0$:
it is nothing but the vacuum expectation value of the frozen light
field. Let us also notice that the condition $\phi_{\ell}\gg
\sqrt{2}\Mp$ translates into $s_0\gg 1/2$. Moreover the condition for
avoiding an interruption of inflation reads $s_0\gg R^2/2$.

\par

Then, the next question is to calculate the behavior of the two scalar
fields after the transition. The light field now drives the expansion
of space-time. The situation is a little subtle because one can still
have $\dot{H}\ll H^2$ but $3H\dot{\phi}_{\rm h}\neq -m_{\rm
  h}^2\phi_{\rm h}$. In other words, the background still inflates but
the heavy field, that has become a test field, is not necessarily in
slow-roll. In that case, this, in principle, invalidates
Eqs.~(\ref{eq:fieldsrdouble}), (\ref{eq:frieddouble}) and~(\ref{eq:s})
since they all assume $\dot{H}\ll H^2$ and the two fields in
slow-roll: in other words, having the kinetic terms negligible in the
Friedmann equation and only one field in slow-roll is not sufficient
to derive Eqs.~(\ref{eq:fieldsrdouble}), (\ref{eq:frieddouble})
and~(\ref{eq:s}). In that case, we need to return to the exact
Klein-Gordon equation for the heavy field. If we write $\phi_{\rm
  h}=a^{-3/2}f_{\rm h}$, then it takes the form
\begin{equation}
\ddot{f}_{\rm h}-\left[\frac32\left(\dot{H}+\frac32H^2\right)-m_{\rm h}^2
\right]f_{\rm h}=0.
\end{equation}
Since the background is still in slow-roll, one can neglect the term
$\dot{H}$ in the above equation. Then, we see that the behavior of
the field depends on the ratio $H/m_{\rm h}$. Since $H$ is decreasing,
the term proportional to the mass necessarily becomes dominant at some
time and then the field oscillates, namely
\begin{equation}
\phi_{\rm h}\simeq a^{-3/2}\cos \left(m_{\rm h}t\right).
\end{equation}
The frequency of the oscillations is given by the mass of the
field. The amplitude of the oscillations decreases as $\propto
a^{-3/2}$ and, therefore, the heavy field becomes negligible very
rapidly. During the oscillations of the heavy field, inflation
continues driven by the light field. It comes to an end when the
vacuum expectation of the light field becomes sub-Planckian. 

\par

Having described the behavior of the background, we can now turn to
the perturbations. They are described by the Bardeen potential already
introduced before, $\Phi$, and the two perturbed scalar fields $\delta
\phi_{\rm h}$ and $\delta \phi_{\ell}$. The corresponding equations of
motion read
\begin{align}
\dot{\Phi}+H\Phi & =\frac{\kappa}{2}\left(
\dot{\phi}_{\rm h}\delta \phi_{\rm h}+\dot{\phi}_{\ell}
\delta \phi_{\ell}\right), \\
\ddot{\delta \phi}_{\rm h}+3H\dot{\delta \phi}_{\rm h}
+\left(\frac{k^2}{a^2}+m_{\rm h}^2\right)\delta \phi_{\rm h}
&= 4\dot{\phi}_{\ell }\dot{\Phi}-2m_{\ell }^2\Phi\phi_{\ell}, \\
\ddot{\delta \phi}_{\ell}+3H\dot{\delta \phi}_{\ell}
+\left(\frac{k^2}{a^2}+m_{\ell}^2\right)\delta \phi_{\ell}
&= 4\dot{\phi}_{\ell}\dot{\Phi}-2m_{\ell}^2\Phi\phi_{\ell}.
\end{align}
Unfortunately, this system of equations cannot be solved
analytically. However, on large scales, namely for wavelengths larger
than the Hubble radius, the expression of the growing mode of the
Bardeen potential and of the two perturbed scalar fields can be
established. They read~\cite{Polarski:1992dq,Peter:1994dx,
  Polarski:1994rz}
\begin{eqnarray}
\Phi &=& -C_1\frac{\dot{H}}{H^2}-H\frac{{\rm d}}{{\rm d}t}
\left(\frac{d_{\rm h}V_{\rm h}+d_{\ell}V_{\ell}}{V_{\rm
    h}+V_{\ell}}\right), \\ 
\label{eq:perth}
\frac{\delta \phi_{\rm h}}{\dot{\phi}_{\rm
    h}} &=& \frac{C_1}{H}-2H\left(\frac{d_{\rm h}V_{\rm
    h}+d_{\ell}V_{\ell}}{V_{\rm h}+V_{\ell}} -d_{\rm h}\right), \\
\label{eq:pertl}
\frac{\delta \phi_{\ell}}{\dot{\phi}_{\ell }} 
&=& \frac{C_1}{H}-2H\left(\frac{d_{\rm h}V_{\rm
    h}+d_{\ell}V_{\ell}}{V_{\rm h}+V_{\ell}} -d_{\ell }\right),
\end{eqnarray}
where $C_1(k)$, $d_{\rm h}(k)$ and $d_{\ell}(k)$ are integration
constants. At this point, the following remark is in order. We have
seen that, in the theory of cosmic inflation, the source of the
perturbations are the quantum vacuum fluctuations. This of course
remains true in a model where we have several scalar fields. This
means that the quantities $\Phi$, $\delta \phi_{\rm h}$ and $\delta
\phi_{\ell}$ are in fact quantum operators. A convenient way to
describe this situation without introducing all the machinery of
quantum field theory is simply to write that the amplitude of the
perturbed fields at Hubble radius crossing are given by $\delta
\phi_{\rm h}=H/\sqrt{2k^3}e_{\rm h}({\bm k})$ and $\delta \phi_{\ell
}=H/\sqrt{2k^3}e_{\ell }({\bm k})$, where $e_{\rm h}({\bm k})$ and
$e_{\ell}({\bm k})$ are two independent Gaussian stochastic processes
satisfying $\langle e_{\rm h}({\bm k})\rangle =\langle e_{\ell}({\bm
  k})\rangle =0$ and $\langle e_{\rm h}({\bm k})e_{\rm h}({\bm
  k}')\rangle =\delta ^{(3)}({\bm k}-{\bm k}')$, $\langle e_{\ell
}({\bm k})e_{\ell }({\bm k}')\rangle =\delta ^{(3)}({\bm k}-{\bm
  k}')$, $\langle e_{\rm h}({\bm k})e_{\ell }({\bm k}')\rangle
=0$. This parametrization raises in fact non trivial questions such
as the quantum-to-classical transition of quantum cosmological
perturbations but, in this review, we will not discuss these
issues~\cite{Grishchuk:1997pk,Polarski:1995jg,Martin:2012pea}.

\par

Then, let us simplify the expression of the perturbed heavy scalar
field, see Eq.~(\ref{eq:perth}), by using the explicit form of the
potential. One arrives at
\begin{eqnarray}
\frac{\delta \phi_{\rm h}}{\dot{\phi}_{\rm h}}&=& 
=\frac{C_1}{H}-2H\left(d_{\ell }-d_{\rm h}\right)
\frac{V_{\ell}}{V_{\rm h}+V_{\ell}}\\
\label{eq:contrastfieldh}
&=& \frac{C_1}{H}-2H
\left(d_{\ell}-d_{\rm h}\right)\frac{m_{\ell}^2\phi_{\ell}^2}
{m_{\rm h}^2\phi_{\rm h}^2+m_{\ell}^2\phi_{\ell}^2}\\
&\simeq & \frac{C_1}{H}+2HC_3,
\end{eqnarray}
where $C_3\equiv d_{\rm h}-d_{\ell}$ and where, in the last equality,
we have assumed that the light field was dominant (namely the second
phase of inflation). Then, one can use the slow-roll relation
$3H\dot{\phi}_{\rm h}\simeq -m_{\rm h}^2\phi_{\rm h}$ and obtains
\begin{equation}
\delta \phi_{\rm h}\simeq -\frac{C_1}{3}\frac{m_{\rm h}^2}{H^2}\phi_{\rm h}
-\frac{2}{3}C_3m_{\rm h}^2\phi_{\rm h}\simeq 
-\frac{2}{3}C_3m_{\rm h}^2\phi_{\rm h},
\end{equation}
where, in the last equality, we have used the fact that, before the
onset of oscillations, $H\gg m_{\rm h}$. Then, as already mentioned
above, the field starts oscillating and the slow-roll approximation is
no longer valid. As a consequence, the above equations can no longer
be used. During the oscillations, one has equipartition between the
kinetic and potential energy. As a consequence, $\langle \rho_{\rm
  h}\rangle \simeq m^2\langle \phi_{\rm h}^2\rangle$. This implies
that $\delta \rho_{\rm h}\simeq m_{\rm h}^22\phi_{\rm h}\delta
\phi_{\rm h}$ and, therefore,
\begin{equation}
\frac{\delta \rho_{\rm h}}{\rho_{\rm h}}
\simeq 2\frac{\delta \phi_{\rm h}}{\phi_{\rm h}}.
\end{equation}
But, in fact, the perturbed Klein-Gordon equation for large scales
modes, if one neglects its right hand side, is the same as the
background Klein-Gordon equation provided the potential is quadratic
in the field (which is precisely the case in the present
situation). As a consequence, $\delta \phi_{\rm h}$ is in fact always
proportional to $\phi _{\rm h}$, the slow-roll approximation being
satisfied or not. In other words, $\delta \phi_{\rm h}/\phi_{\rm h}$
and hence $\delta \rho_{\rm h}/\rho_{\rm h}$ are constant. So if we
assume that the heavy field decays into cold dark matter after its
oscillations, one has~\cite{Polarski:1992dq,Peter:1994dx,
  Polarski:1994rz}
\begin{equation}
\delta _{\rm cdm}=\frac{\delta \rho_{\rm h}}{\rho_{\rm h}}\biggl
\vert_{\rm end\, osci}=\frac{\delta \rho_{\rm h}}{\rho_{\rm h}}\biggl
\vert_{\rm start\, osci}=2\frac{\delta \phi_{\rm h}}{\phi_{\rm h}}
\biggl\vert_{\rm start\, osci}=-\frac{4}{3}C_3(k)m_{\rm h}^2.
\end{equation}
We conclude that, if one is able to express the constant $C_3(k)$,
then one can establish the expression of the cold dark matter density
contrast. But this is in fact an easy task. Indeed, reproducing the
same calculation as the one which led to
Eq.~(\ref{eq:contrastfieldh}), one obtains
\begin{equation}
\frac{\delta \phi_{\ell}}{\dot{\phi}_{\ell}}
=\frac{C_1}{H}-2HC_3\frac{m_{\rm h}^2\phi_{\rm h}^2}{m_{\rm h}^2\phi_{\rm h}^2
+m_{\ell}^2\phi_{\ell}^2}.
\end{equation}
Then, using this formula and Eq.~(\ref{eq:contrastfieldh}), one can
eliminate $C_1(k)$ and find an expression for
$C_3(k)$. Straightforward manipulations lead to
\begin{equation}
C_3(k)=-\frac{1}{2H}\left(\frac{\delta \phi_{\ell}}{\dot{\phi}_{\ell}}
-\frac{\delta \phi_{\rm h}}{\dot{\phi}_{\rm h}}\right).
\end{equation}
The next step is to replace the derivatives of the fields by their
slow-roll expressions and $\delta \phi_{\rm h}$ by $He_{\rm
  h}/\sqrt{2k^3}$ (and a similar expression for $\delta \phi_{\ell}$)
as was discussed after Eq.~(\ref{eq:pertl}). One arrives at
\begin{equation}
C_3(k)=\frac{3H}{2m_{\rm h}^2}
\frac{1}{\sqrt{2k^3}}
\left(\frac{m_{\rm h}^2}{m_{\ell}^2}\phi_{\ell}^{-1}e_{\ell}
-\phi_{\rm h}^{-1}e_{\rm h}\right).
\end{equation}
Finally, one can estimate the entropy perturbation. According to the
definitions introduced before, see Eq.~(\ref{eq:defS12}), one has
$S_{\rm cdm}\equiv S_{{\rm cdm}\gamma}\equiv 3\left(\zeta_{\rm
  cdm}-\zeta_\gamma\right)=\delta _{\rm cdm}-4\delta_{\gamma}/3\simeq
\delta_{\rm cdm}$ because $\delta \rho_{\rm cdm}\sim \delta
\rho_{\gamma}$ and, during the radiation dominated era,
$\rho_{\gamma}\gg \rho_{\rm cdm}$. As a consequence, one has
\begin{equation}
\label{eq:Scdm}
S_{{\rm cdm},{\bm k}}\simeq -\sqrt{\frac{2}{k^3}}H
\left[R^2\phi_{\ell}^{-1}e_{\ell}({\bm k})
-\phi_{\rm h}^{-1}e_{\rm h}({\bm k})\right],
\end{equation}
where the quantities $H$, $\phi_{\rm h}$ and $\phi_{\ell}$ should be
viewed here as scale dependent quantities since they are expressed at
Hubble radius crossing. In fact, as will be discussed in
Sec.~\ref{sec:observ}, their scale dependence permits the calculation
of the isocurvature perturbations power spectrum. With these
equations, one can now predict the CMB temperature anisotropies by
using Eq.~(\ref{eq:dTnonadia}). But, in fact, the most important
conclusion is of course that $S_{{\rm cdm},{\bm k}}\neq 0$. This means
that, in a model of inflation with more than one field, isocurvature
perturbations can be produced. This justifies our claim that, if
non-adiabatic perturbations are observed in the future, a natural
explanation will be to consider that several scalar fields play a role
during inflation.

\par

In fact, there is even more. Indeed, during the radiation dominated
era, the adiabatic perturbations can be written as $\zeta_{_{\rm RD}}
=3\Phi/2=C_1(k)$. The constant $C_1(k)$ can also be evaluated easily
using the solutions~(\ref{eq:perth}) and~(\ref{eq:pertl}). The
corresponding expression reads
\begin{equation}
\label{eq:zetadoubleinf}
\zeta_{_{\rm RD}}=-\frac{\kappa}{2}\frac{H}{\sqrt{2k^3}}
\left[\phi_{\rm h}e_{\rm h}({\bm k})+\phi_{\ell}e_{\ell}({\bm
  k})\right].
\end{equation}
In particular, one has introduced before the power spectrum of the
conserved quantity $\zeta _{\bm k}$ according to $\langle \zeta_{{\bm
    k}_1}\zeta_{{\bm k}_2}^*\rangle =2\pi^2/k_1^3\calP(k_1)\delta
^{(3)}({\bm k}_1-{\bm k}_2)$\footnote{Notice that this expression is
  consistent with the definition given above, in the text between
  Eq.~(\ref{eq:qfieldzeta}) and Eq.~(\ref{eq:meanzetasquare}), namely
  $\langle \zeta_{{\bm k}_1}\zeta_{{\bm k}_2}\rangle
  =2\pi^2/k_1^3\calP(k_1)\delta ^{(3)}({\bm k}_1+{\bm k}_2)$ because
  $\zeta_{-{\bm k}_2}=\zeta_{{\bm k}_2}^*$.}. In the same manner, one
can define the power spectrum of the non-adiabatic perturbations by
\begin{equation}
\label{eq:defSS}
\langle S_{{\rm cdm},{\bm k}_1}S_{{\rm cdm},{\bm k}_2}^*\rangle
\equiv \frac{2\pi^2}{k_1^3}\calP_{S_{\rm cdm}}(k_1)
\delta ^{(3)}\left({\bm k}_1-{\bm k}_2\right).
\end{equation}
But the most important aspect of the above calculations is that
adiabatic and isocurvature perturbations turn out to be
correlated~\cite{Langlois:1999dw,Langlois:2000ar}. This means that the
correlator $\langle \zeta_{{\bm k}_1} S_{{\rm cdm},{\bm
    k}_2}^*\rangle$ is non-vanishing. This correlator can be expressed
as
\begin{equation}
\Re \langle \zeta_{{\bm k}_1} S_{{\rm cdm},{\bm k}_2}^*\rangle
\equiv\frac{2\pi^2}{k_1^3}{\cal C}_{\zeta, S_{\rm cdm}}
\delta ^{(3)}\left({\bm k}_1-{\bm k}_2\right).
\end{equation}
This is because the expressions of $\zeta_{\bm k}$, see
Eq.~(\ref{eq:zetadoubleinf}), and $S_{{\rm cdm},{\bm k}}$, see
Eq.~(\ref{eq:Scdm}), both depend on $e_{\rm h}$ and $e_{\ell}$. From
the above definition, let us also notice that one can define a
correlation spectrum by
\begin{equation}
\label{eq:defcorr}
\calP_{\zeta, S_{\rm cdm}}\equiv \frac{{\cal C}_{\zeta, S_{\rm cdm}}}
{\sqrt{\calP_{\zeta}}\sqrt{\calP_{S_{\rm cdm}}}}.
\end{equation}
Let us stress that, when one constrains the amplitude of isocurvature
modes using the CMB data, it is of course important to take into
account the fact that adiabatic and isocurvature perturbations can be
correlated. As we will see in the next section, this was done in the
analysis of the Planck data.

\section{Inflation after Planck}
\label{sec:observ}

We have previously studied the predictions of inflation for different
cosmological observables. In this section, we review what is
experimentally known about these observables and discuss the
corresponding implications for cosmic inflation.

\begin{figure}
\begin{center}
\includegraphics[width=15cm]{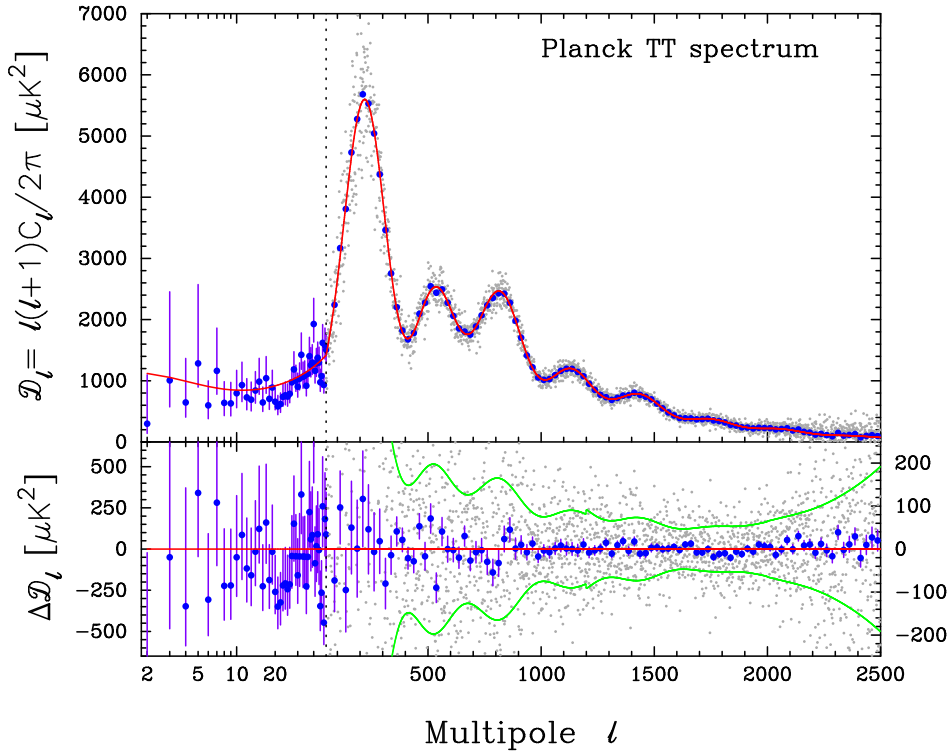}
\end{center}
\caption{Temperature anisotropy multipole moments obtained from the
  Planck $2013$ data versus the angular scale $\ell$ (notice that, for
  $\ell \leq 49$, the scale is logarithmic). The gray points denote
  the value of the multipole $C_{\ell}$ for each $\ell$ while the blue
  points represent the value of $C_{\ell}$ averaged in bands of width
  $\Delta \ell \simeq 31$. The red solid line shows the prediction of
  the best fit six-parameters $\Lambda$CDM model. The error bars
  correspond to $\pm 1\sigma$ uncertainties. The lower panel shows the
  residual signal once the best fit model has been subtracted. Figure
  taken from Ref.~\cite{Ade:2013zuv}.}
\label{fig:TTplanck13}
\end{figure}

\begin{figure}
\begin{center}
\includegraphics[width=15cm]{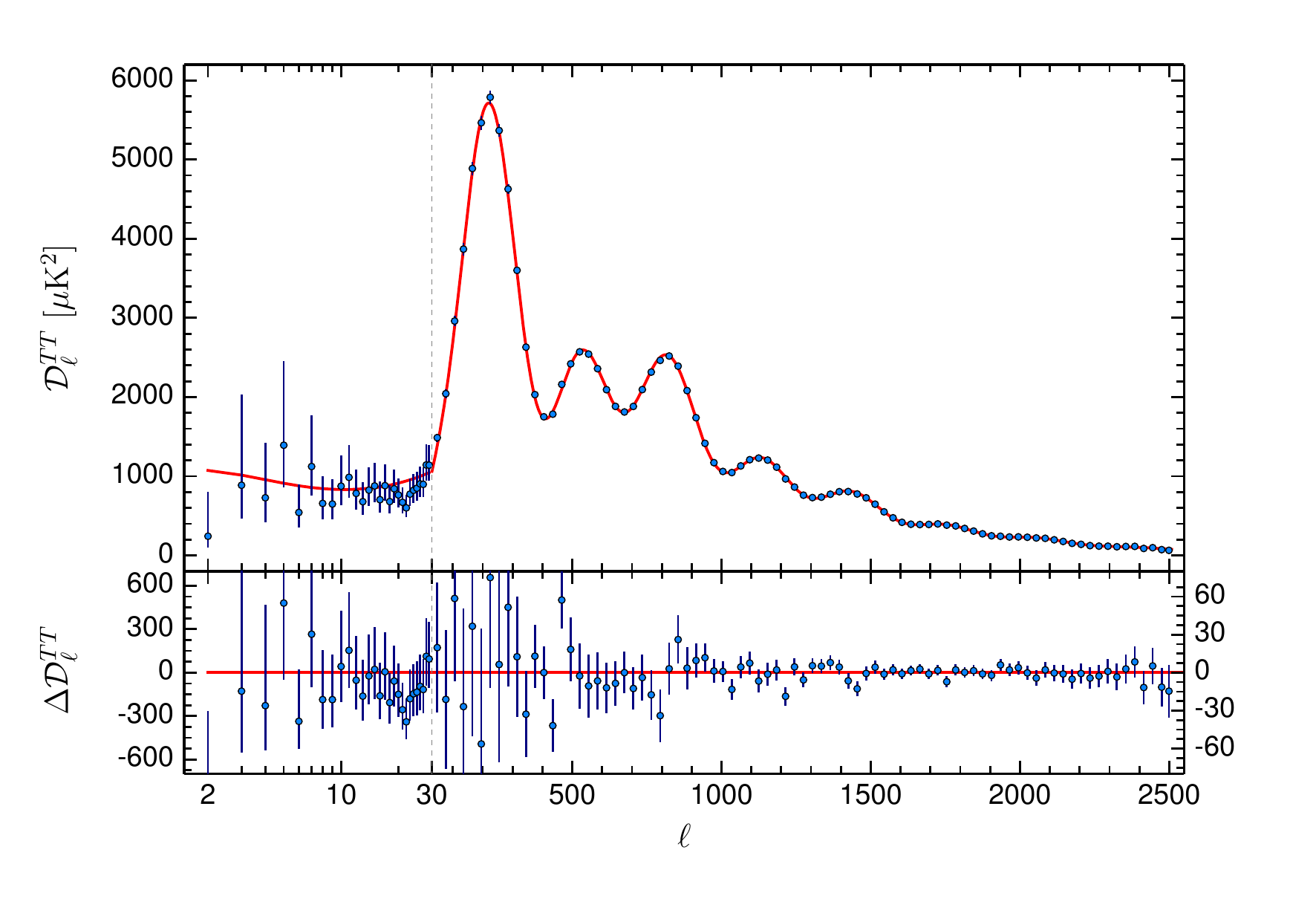}
\end{center}
\caption{Same as Fig.~\ref{fig:TTplanck13} but with the Planck $2015$
  data. Notice that the quantity ${\cal D}_{\ell}$ is defined by
  ${\cal D}_{\ell}=\ell(\ell+1)C_{\ell}/(2\pi)$. This plot should be
  compared to Fig.~\ref{fig:TTplanck13}. Figure taken from
  Ref.~\cite{Planck:2015xua}.}
\label{fig:TTplanck2015}
\end{figure}

The Planck CMB data have been released for the first time in
$2013$~\cite{Ade:2013zuv} and, more recently, in $2015$, new
measurements have been made public~\cite{Planck:2015xua}. Planck
$2013$ has measured the CMB temperature anisotropies and the
corresponding multipole moments $C_{\ell }^{_{\rm TT}}$ are
represented in Fig.~\ref{fig:TTplanck13}. Let us remind that these
quantities are defined as follows. After foregrounds subtraction, the
Planck measurements can be used to construct a map of the CMB
temperature anisotropy, namely
\begin{equation}
\frac{\delta T}{T}({\bm e})=\sum_{\ell m}a_{\ell m}Y_{\ell m}({\bm e}),
\end{equation}
where $Y_{\ell m}$ are the spherical harmonics and where the vector
${\bm e}$ specifies a direction in the sky. In practice, $\delta T/T$
can be expressed as
\begin{equation}
\label{eq:clform}
\frac{\delta T}{T}({\bm e})=
\int \frac{{\rm d}{\bm k}}{(2\pi)^{3/2}}
\left[F({\bm k})+G({\bm k})\frac{\partial}{\partial \eta_0}\right]
{\rm e}^{i d_{\rm A}{\bm k}\cdot 
{\bm e}/a(\eta_{\rm lss})}\, ,
\end{equation}
where $d_{\rm A}=a(\eta_{\rm lss})r_0+a(\eta_{\rm
  lss})\left(\eta_0-\eta_{\rm lss}\right)$ ($r_0$ being Earth's radial
coordinate and $\eta_0$ denoting the present time) is the angular
distance to the surface of last scattering and the quantity ${\bm
  k}/a(\eta_{\rm lss})$ represents the physical wavenumber of the
Fourier mode under consideration at the time of recombination. The
quantities $F({\bm k})$ and $G({\bm k})$ encode the behavior of
cosmological perturbations and are called ``form factors'' in
Ref.~\cite{Weinberg:2002kg}. Already at this stage, we see that the
configuration where the wavelengths of the perturbations become equal
to the angular distance of the last scattering surface plays a
preferred role. Then, the two-point correlation function in real space
can be written as
\begin{equation}
\left \langle \frac{\delta T}{T}({\bm e}_1) \frac{\delta T}{T}({\bm
  e}_2) \right \rangle=
\frac{1}{4\pi}\sum_{\ell=0}^{+\infty}(2\ell+1)C_{\ell }^{_{\rm
    TT}}P_{\ell }(\cos \theta),
\end{equation}
where $\theta $ is the angle between the two vectors ${\bm e}_1$ and
${\bm e}_2$. This expression defines the multipole moments
$C_{\ell}^{_{\rm TT}}$.

\par

\begin{figure}
\begin{center}
\includegraphics[width=13cm]{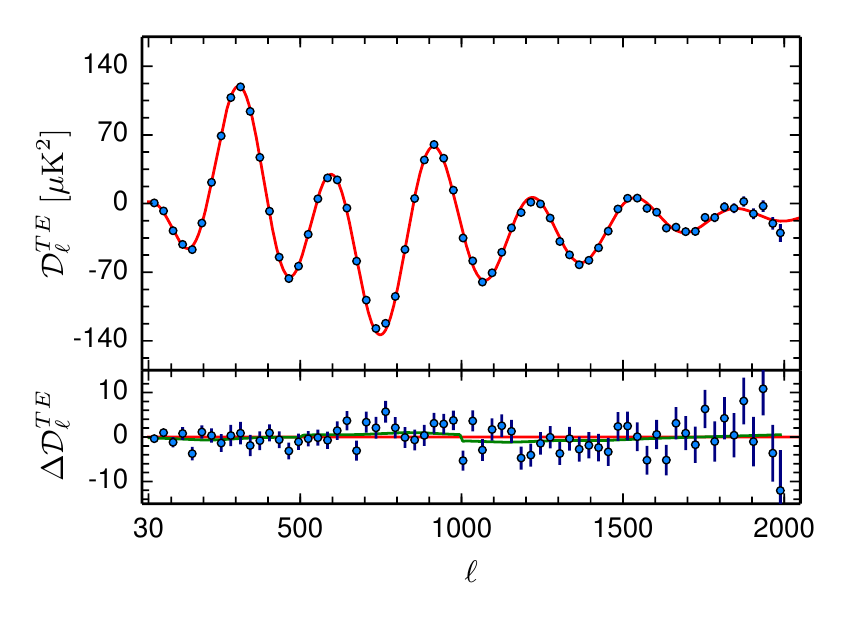}
\end{center}
\caption{Multipole moments corresponding to the correlation between
  temperature and $E$-mode polarization anisotropies obtained from
  Planck $2015$. The red solid line corresponds to the best fit
  obtained with temperature measurements only. The lower panel shows
  the residual with respect to this best fit. Figure taken from
  Ref.~\cite{Planck:2015xua}.}
\label{fig:TEplanck2015}
\end{figure}

\begin{figure}
\begin{center}
\includegraphics[width=13cm]{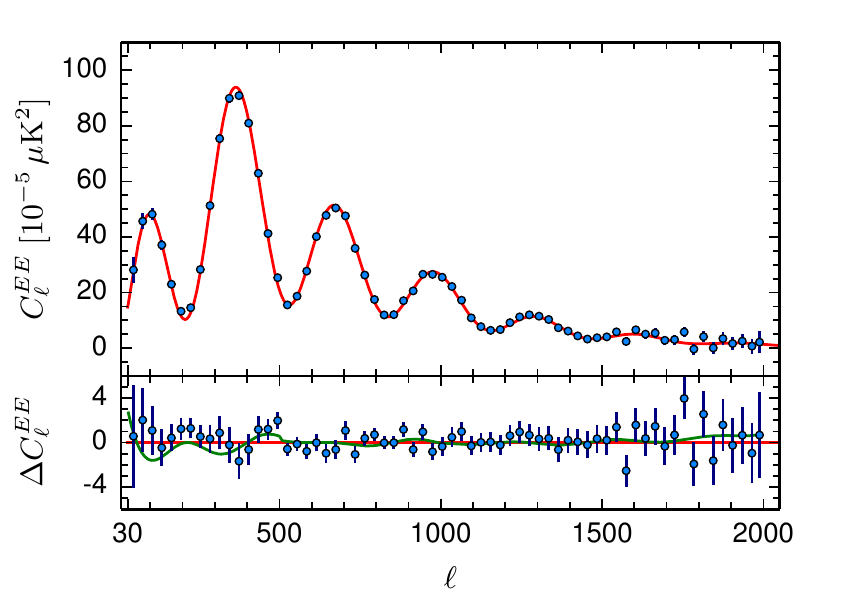}
\end{center}
\caption{Same as in Fig.~\ref{fig:TEplanck2015} but for the $E$-mode
  power spectrum obtained from Planck $2015$. Figure taken from
  Ref.~\cite{Planck:2015xua}.}
\label{fig:EEplanck2015}
\end{figure}

The big novelty of the Planck $2015$
data~\cite{Planck:2015xua,Ade:2015oja} is that they not only lead to a
more accurate measurements of the $C_{\ell}^{_{\rm TT}}$, see
Fig.~\ref{fig:TTplanck2015}, but they also provide measurements of the
$E$-mode CMB polarization. One can then define quantities similar to
the $C_{\ell }^{_{\rm TT}}$ for the correlation between temperature
and $E$-mode polarization fluctuations and for the $E$-mode power
spectrum. The corresponding multipole moments $C_{\ell }^{_{\rm TE}}$
and $C_{\ell}^{_{\rm EE}}$ are represented in
Figs.~\ref{fig:TEplanck2015} and~\ref{fig:EEplanck2015}.

Before focusing on the consequences of these data for inflation, let
us briefly discuss their implications for the standard model of
Cosmology. It is important to understand that the constraints on the
cosmological parameters can depend on the model analyzed and on the
data used to perform the analysis. In $2013$, Planck used the
temperature anisotropy measurement plus the WMAP polarization
measurement on large scales ($\ell \leq 23$), the corresponding
likelihood function being denoted PlanckTT+WP. In $2015$, at least
five different likelihoods have been used: PlanckTT utilizes
temperature data only and is an hybrid, meaning that the temperature
likelihood is not the same for low multipoles ($\ell \leq 30$) and
high multipoles; PlanckTT+lowP makes use of PlanckTT and low-$\ell$
polarization data; PlanckTE+lowP corresponds to the TE likelihood at
$\ell \geq 30$ plus low-$\ell$ polarization data only; PlanckTT,TE,EE
+lowP makes use of the TT, TE and EE likelihoods at $\ell \geq 30$ and
of the temperature and polarization data at small scales. Depending on
which likelihood is used, the constraints on cosmological parameters
can slightly change.

\par

The theoretical framework used to analyze the data is the flat [\ie
  ${\cal K}=0$ in Eq.~(\ref{eq:fried})] $\Lambda$CDM model. In order
to specify it, we need to know the energy budget of the Universe, \ie
the photon energy density $\rho_{\gamma}$, the neutrino energy density
$\rho_{\nu}$, the baryons energy density $\rho_{\rm b}$, the cold dark
matter energy density $\rho_{\rm c}$ and the dark energy density
$\rho_{\Lambda}$ (here assumed to be a cosmological constant). Then,
in principle, on can calculate the behavior of the scale factor $a(t)$
since we know $\rho\equiv\sum_i\rho_i$ in the right hand side of the
Friedmann equation~(\ref{eq:fried}). Of course we also need the Hubble
rate today, $H_0$ or $h\equiv H_0/(100\, \mbox{km}\times \mbox{s}^{-1}
\times \mbox{Mpc}^{-1})$ which is, therefore, another free
parameter. We also need the power spectrum of scalar fluctuations
assumed to be of the power-law form $P(k)\propto A_{_{\rm
    S}}k^{n_{_{\mathrm S}}-1}$ where $A_{_{\rm S}}$ is the amplitude
of the fluctuations and $\nS$ the spectral index. Here, gravitational
waves are supposed to be absent, $r=0$. So, in this simple framework,
the perturbations are characterized by two numbers. In the case of
inflation, we need three parameters, the amplitude of scalar
fluctuations and the two first slow-roll parameters. We notice
(again!) that the parametrization used here is different from what we
generically obtain from inflation where the power spectrum is not of
the power-law form and where $r$ is necessarily non-vanishing (but can
be very small). Finally, we need a parameter describing reionization
and we take the optical depth $\tau$. The interpretation of this
parameter is as follows. After recombination, the photons are supposed
to propagate freely from the surface of last scattering to
us. However, at the epoch of the formation of the first stars,
estimated to be $z_{\rm re}\sim 10$, the Universe is ionized again. As
a consequence, some of the CMB photons scatter off free electrons
again. The probability to ``avoid'' this additional scattering is
${\rm e}^{-\tau}$ where
\begin{equation}
\tau\equiv \sigma_{_{\rm T}} \int _{t_{\rm re}}^{t_{\rm now}}n_{\rm e}{\rm d}t,
\end{equation}
is the optical depth. In this expression, $\sigma_{_{\rm T}}$ is the
Thomson cross-section and $n_{\rm e}$ is the number density of free
electrons. When additional scattering occur, the direction of the
photon change randomly and this washes out the CMB anisotropy on small
angular scales, $\ell>\ell _{\rm re}$, namely $a_{\ell m}\rightarrow
a_{\ell m}{\rm e}^{-\tau}$. For $\ell <\ell_{\rm re}$, the CMB
anisotropies are not changed. The value of $\ell_{\rm re}$ clearly
depends on $z_{\rm re}$. The previous considerations imply that, on
small scales, the amplitude of the fluctuations becomes $A_{_{\rm
    S}}{\rm e}^{-2\tau}$ and there is therefore a partial degeneracy
between $A_{_{\rm S}}$ and $\tau$.

\par

In the Planck papers, one of the free parameters is in fact taken to
be $\theta_{_{\rm MC}}$, where the subscript ``MC'' reminds that this
quantity is used in \COSMOMC. By definition, it is equal to be
$\theta_{_{\rm MC}} \equiv 100 (r_{_{\rm S}}/d_{\rm A})\vert_{\rm
  approx}$. Here, $r_{_{\rm S}}$ is the sound horizon at last
scattering, namely
\begin{equation}
r_{_{\rm S}}=\int _{0}^{\eta_{\rm lss}}c_{_{\rm S}}{\rm d}\eta,
\end{equation}
where $c_{_{\rm S}}$ is the sound speed of the baryon-photon fluid, \ie
\begin{equation}
c_{_{\rm S}}^2=\frac{\delta p_{{\rm b}-\gamma}}{\delta \rho_{{\rm b}-\gamma}}
=\frac{\delta p_{\gamma}}{\delta \rho_{\rm b}+\delta \rho_{\gamma}}
=\frac13\frac{4\rho_{\gamma}}{4\rho_{\gamma}+3\rho_{\rm b}}=
\frac13 \frac{1}{1+R},
\end{equation}
where $R=3\rho_{\rm b}/(4\rho_{\gamma})$. The quantity $d_{\rm A}$ is,
as already mentioned, the angular distance to the last scattering
surface and naturally appears in the expression of the multipole
moments, see Eq.~(\ref{eq:clform}). Therefore, $r_{_{\rm S}}/d_{\rm
  A}$ is in fact the angular size of the sound horizon. $\theta_{_{\rm
    MC}}$ is defined approximately because its value is calculated at
a redshift which is given by a fitting formula~\cite{Hu:1995en}
\begin{equation}
z_{\rm lss}=1048\left[1+0.00124\left(\Omega _{\rm b}h^2\right)^{-0.738}
\right]\left[1+g_1\left(\Omega_{\rm m}h^2\right)^{g_2}\right],
\end{equation}
where the function $g_1$ and $g_2$ can be expressed as
\begin{eqnarray}
g_1 &=& 0.0783\left(\Omega _{\rm b}h^2\right)^{-0.238}
\left[1+39.5 \left(\Omega _{\rm m}h^2\right)^{0.763}\right]^{-1}, \\
g_2&=& 0.560\left[1+21.1\left(\Omega _{\rm b}h^2\right)^{1.81}\right]^{-1}.
\end{eqnarray}
with $\Omega_{\rm m}=\Omega _{\rm b}+\Omega_{\rm c}$. In practice,
instead of including $h$ in the list of free parameters, we consider
$\theta_{_{\rm MC}}$.

\par

We conclude that, a priori, we have a $9$ parameters: $h$ or
$\theta_{_{\rm MC}}$, $\rho_{\gamma}$, $\rho_{\nu}$, $\rho_{\rm b}$,
$\rho_{\rm c}$, $\rho_{\Lambda}$, $A_{_{\rm S}}$, $\nS$ and
$\tau$. However, the photon energy density is not a free parameter
because it is given by $\pi^2T_0^4/15$ where $T_0=2.7255\pm
0.00006~\mbox{K}$ is the CMB temperature. In the same way, the
neutrino energy density is fixed since $\rho_{\nu}=N_{\rm
  eff}(7/8)(4/11)^{4/3}\rho_{\gamma}$ with $N_{\rm eff}=3$. Moreover,
the fact that the spatial sections are assumed to be flat means that,
say $\rho_{\Lambda}$, can be deduced from the knowledge of the other
parameters. Therefore, the ``base'' model used in the Planck articles
is in fact a six-parameter scenario and it is sufficient to fit the
CMB data.

\par

Planck $2013$ (\ie PlanckTT+WP using the terminology introduced
before) found the following results ($68\%$ confidence
limits)~\cite{Ade:2013zuv}
\begin{eqnarray}
\Omega_{\rm b}h^2 &=& 0.022032\pm 0.00028, \quad 
\Omega_{\rm c}h^2=0.1199\pm 0.0027, \\
100\theta_{_{\rm MC}} &=& 1.04131\pm 0.00063, \quad 
\tau=0.089^{+0.012}_{-0.014}, \\
\nS &=& 0.9603\pm 0.0073, \quad \ln\left(10^{10}A_{_{\rm S}}\right)
=3.089^{+0.024}_{-0.027}.
\end{eqnarray}
On the other hand, Planck $2015$ with PlanckTT, TE, EE+lowP (as
already mentioned, using the other likelihoods described before would
lead to slightly different numbers) gives~\cite{Planck:2015xua}
\begin{eqnarray}
\Omega_{\rm b}h^2 &=& 0.02225\pm 0.00016, \quad 
\Omega_{\rm c}h^2=0.1198\pm 0.0015, \\
100\theta_{_{\rm MC}} &=& 1.04077\pm 0.00032, \quad 
\tau=0.079\pm0.017, \\
\nS &=& 0.9645\pm 0.0049, \quad \ln\left(10^{10}A_{_{\rm S}}\right)
=3.094\pm 0.0049.
\end{eqnarray}
The consistency between Planck $2013$ and Planck $2015$ is evidently
very good. 

\par

More involved data analysis can be carried out by opening the
parameter space (for instance by considering gravitational waves, a
running for the scalar power spectrum, a time-dependent dark energy
equation of state etc ...) and/or adding more data sets. In the
following, we will describe the corresponding results for the
observables that are especially relevant for inflation.

\subsection{Spatial Curvature}
\label{subsec:curvplanck}

As discussed in Sec.~\ref{subsec:hepinf}, see Eq.~(\ref{eq:omegak}),
maybe the most important prediction of inflation is that our Universe
should be spatially flat (although there are contrived inflationary
models for which this is not true~\cite{Linde:1995rv}). Therefore, one
can follow the strategy described above and relax the assumption that
the curvature of spacelike sections is flat. Then, the Planck $2013$
data plus the WMAP data on large scale polarization imply
that~\cite{Ade:2013zuv,Ade:2013uln}
\begin{equation}
\Omega_{\cal K}=-0.058^{+0.046}_{-0.026}.
\end{equation}
If, in addition, Baryonic Acoustic Oscillations (BAO) data are
included, one obtains $\Omega_{\cal K}=-0.004\pm 0.0036$. 

\par

The Planck $2015$~\cite{Planck:2015xua} results have confirmed and
tightened this conclusion. Indeed, at $95\%$ confidence level,
PlanckTT,TE,EE+lowP leads to $\Omega _{\cal
  K}=-0.040^{+0.038}_{-0.041}$. If lensing data and BAO are taken into
account, one arrives at the impressive following result
\begin{equation}
\Omega_{\cal K}=0.000\pm 0.005.
\end{equation}
Therefore, we live in a spatially flat Universe in agreement with one
of the most basic prediction of inflation.

\par

As already mentioned, when one relaxes the assumption that the
Universe is spatially flat, this introduces a new parameter and,
therefore, we are no longer in the framework of the six-parameters
$\Lambda$CDM base model considered before. As a consequence, a priori,
the constraints on the other parameters may change. This is in
particular the case of the spectral index $\nS$ and its significant
deviation from the scale-invariant case which is very important for
inflation. However, Ref.~\cite{Ade:2015oja} has shown that in the
framework where $\Omega _{\cal K}\neq 0$ and where tensor modes are
present, the constraint on $\nS$ becomes
\begin{equation}
\nS=0.969\pm 0.005,
\end{equation}
using PlanckTT,TE,EE+lowP. The conclusion that the scale invariant case
is ruled out seems therefore robust. In fact, Ref.~\cite{Ade:2015oja}
has shown that this conclusion is valid for other type of extensions
such as different relativistic degrees of freedom (the parameter
$N_{\rm eff}$ defined above, running, dark energy equation of state
etc \dots). This is of course crucial for inflation.

\subsection{Isocurvature Perturbations}
\label{subsec:isoplanck}

Let us now investigate the Planck constraints on isocurvature
perturbations~\cite{Ade:2013zuv,Ade:2013uln,Ade:2015oja}. We have
discussed before, in Sec.~\ref{subsec:iso}, two types of isocurvature
perturbations. Firstly, there is the effective mode taking into
account cold dark matter and baryons entropy fluctuations, see
Eq.~(\ref{eq:Scdmeff}) denoted, as already mentioned, CDI in the
Planck papers. Secondly, there is also the Neutrino Density
Isocurvature (NDI) mode and the Neutrino Velocity Isocurvature (NVI)
mode. Each mode is characterized by its power spectrum as in
Eq.~(\ref{eq:defSS}) and each cross term can also be described by the
correlation spectrum as in Eq.~(\ref{eq:defcorr}). Therefore, the most
general situation can be parametrized by the $4\times 4$ matrix
$\calP_{ab}(k)$ where $a=\zeta, S_{\rm CDI}, S_{\rm NDI}, S_{\rm NVI}$
with the convention that $\calP_{\zeta \zeta}\equiv \calP_{\zeta}$ and
similar expressions for the diagonal terms. Of course, this matrix is
symmetrical.

\begin{figure}
\begin{center}
\includegraphics[width=10cm]{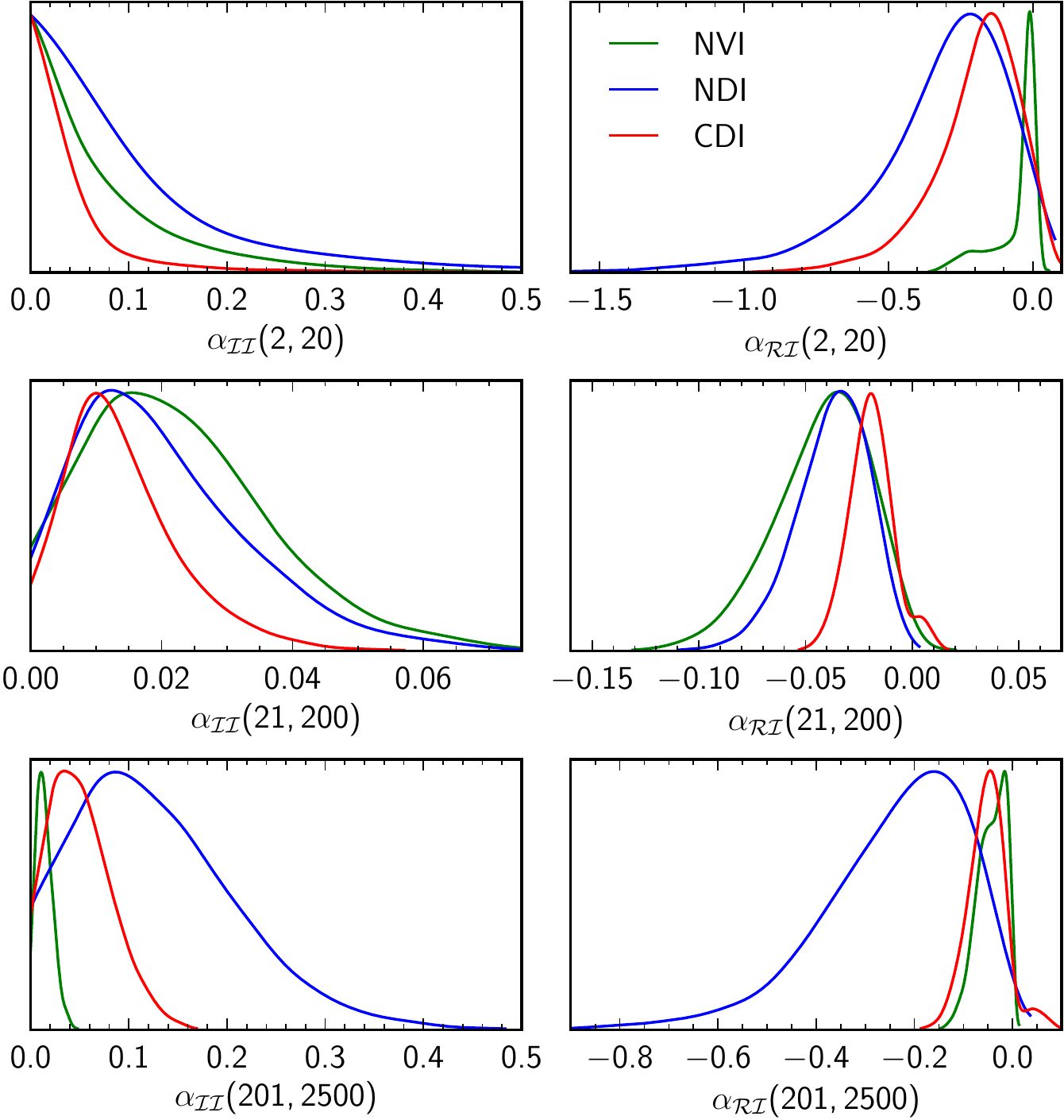}
\end{center}
\caption{Posterior distributions for the quantities
  $\alpha_{ab}\left(\ell_{\rm min},\ell_{\rm max}\right)$ introduced
  in Eq.~(\ref{eq:defalpha}), inferred from the Planck $2013$ data. No
  statistically significant deviation from adiabaticity is
  found. Notice that ${\cal R}=-\zeta$ is used in this plot. Figure
  taken from Ref.~\cite{Ade:2013uln}.}
\label{fig:iso1d}
\end{figure}

\begin{figure}
\begin{center}
\includegraphics[width=10cm]{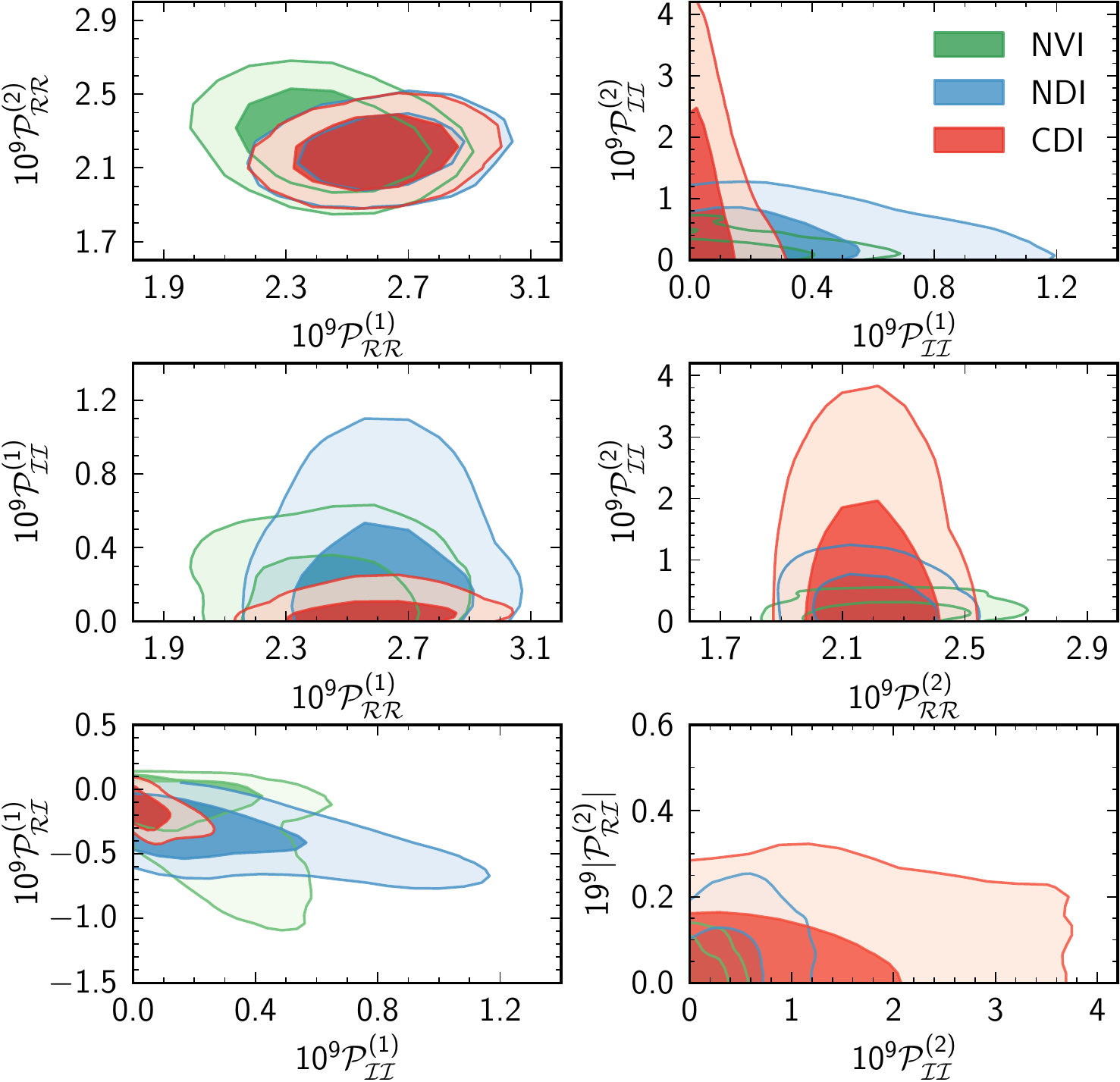}
\end{center}
\caption{Two-dimensional distributions for the quantities
  $\calP_{ab}(k_{1,2})\equiv \calP_{ab}^{(1,2)}$ inferred from the
  Planck $2013$ data for $a=\calR$, CDI (red), NDI (blue) and NVI
  (green). Again, amplitude of isocurvature spectra and correlation
  spectra are all consistent with adiabaticity. In these plots, notice
  that $\calP_{ab}$ has indices $a=\calR$, ${\cal I}$ with ${\cal
    I}\equiv \mbox{CDI}, \mbox{NDI}, \mbox{NVI}$. Figure taken from
  Ref.~\cite{Ade:2013uln}.}
\label{fig:iso2d}
\end{figure}

Usually, only a $2\times 2$ matrix is analyzed and a power law is
assumed for each of the power spectra with independent spectral
index. But this is not the route followed by the Planck team. Instead,
they have assumed the following phenomenological form for
$\calP_{ab}(k)$
\begin{equation}
\calP_{ab}(k)=\exp \left[
\left(\frac{\ln k-\ln k_2}{\ln k_1-\ln k_2}\right)
\ln \calP_{ab}(k_1)
+
\left(\frac{\ln k-\ln k_1}{\ln k_2-\ln k_2}\right)
\ln \calP_{ab}(k_2)\right],
\end{equation}
where the two scales $k_1$ and $k_2$ are chosen to be $k_1=2\times
10^{-3}\mbox{Mpc}^{-1}$ and $k_2=0.1\mbox{Mpc}^{-1}$ so that the
entire Planck window is spanned. The positive definiteness of the
matrix requires $\left(\calP_{ab}\right)^2\leq \calP_{aa}\calP_{bb}$.

\par

Then, the following quantities are defined
\begin{equation}
\label{eq:defalpha}
\alpha_{ab}\left(\ell_{\rm min},\ell_{\rm max}\right)
=\frac{\left(\Delta T\right)^2_{ab}
\left(\ell_{\rm min},\ell_{\rm max}\right)}
{\left(\Delta T\right)^2_{\rm tot}
\left(\ell_{\rm min},\ell_{\rm max}\right)},
\end{equation}
where 
\begin{equation}
\left(\Delta T\right)^2_{ab}
\left(\ell_{\rm min},\ell_{\rm max}\right)=\sum_{\ell =\ell_{\rm min}}
^{\ell=\ell_{\rm max}}\left(2\ell+1\right)C_{ab,\ell}^{\rm TT}.
\end{equation}
In this expression, the quantity $C_{ab,\ell}^{\rm TT}$ represents the
multipole moments calculated with the primordial spectrum taken to be
$\calP_{ab}$. $\left(\Delta T\right)^2_{\rm tot}$ is just the sum of
all contributions. So, in the standard situation, there is just one
contribution and the multipole moments are computed with
$\calP_{\zeta\zeta}=\calP_{\zeta}$. If one has $\alpha_{\zeta
  \zeta}=1$, this means that the perturbations are fully adiabatic.

\par

In Figs.~\ref{fig:iso1d} and~\ref{fig:iso2d}, we have respectively
represented the one-dimensional posterior distribution of
$\alpha_{ab}\left(\ell_{\rm min},\ell_{\rm max}\right)$ and the
two-dimensional distribution for the power spectra
$\calP_{ab}(k_{1,2})$ for the three modes, CDI, NDI and NVI obtained
from Planck $2013$. The conclusion is clear: there is no statistically
significant deviation from pure adiabaticity.

\begin{figure}
\begin{center}
\includegraphics[width=9cm]{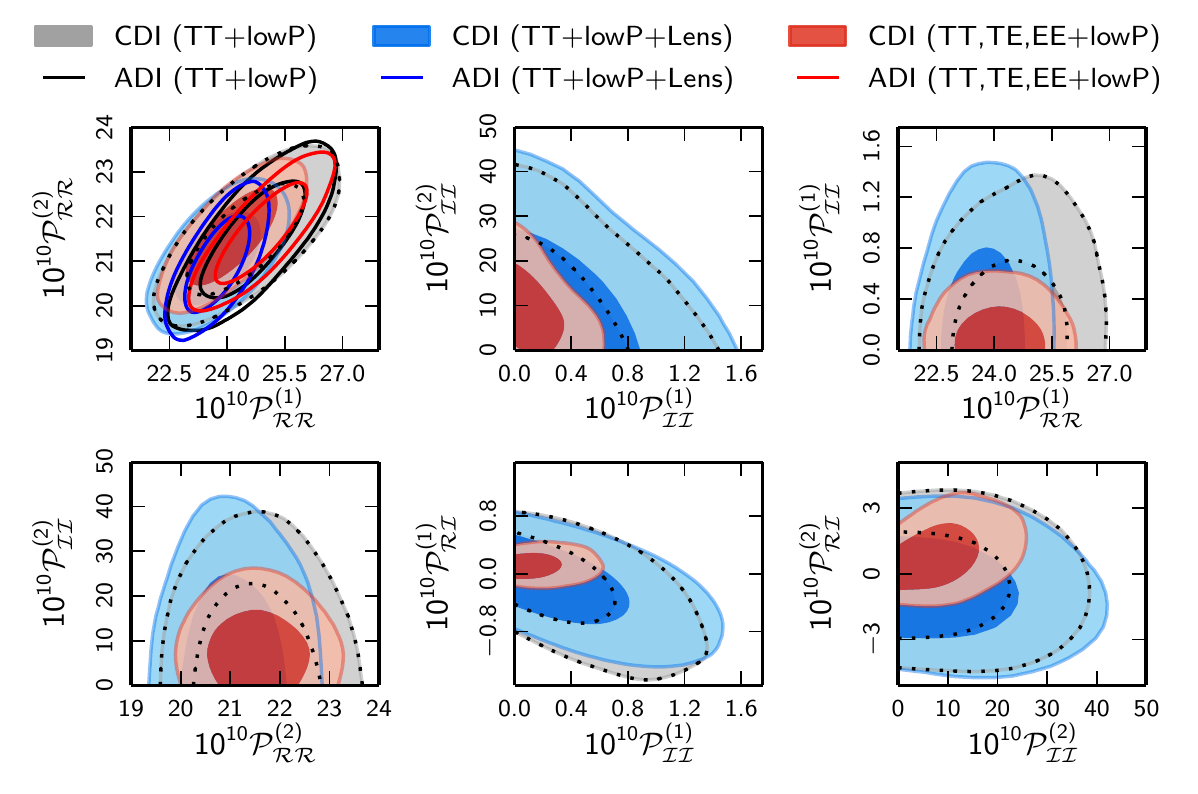}
\includegraphics[width=9cm]{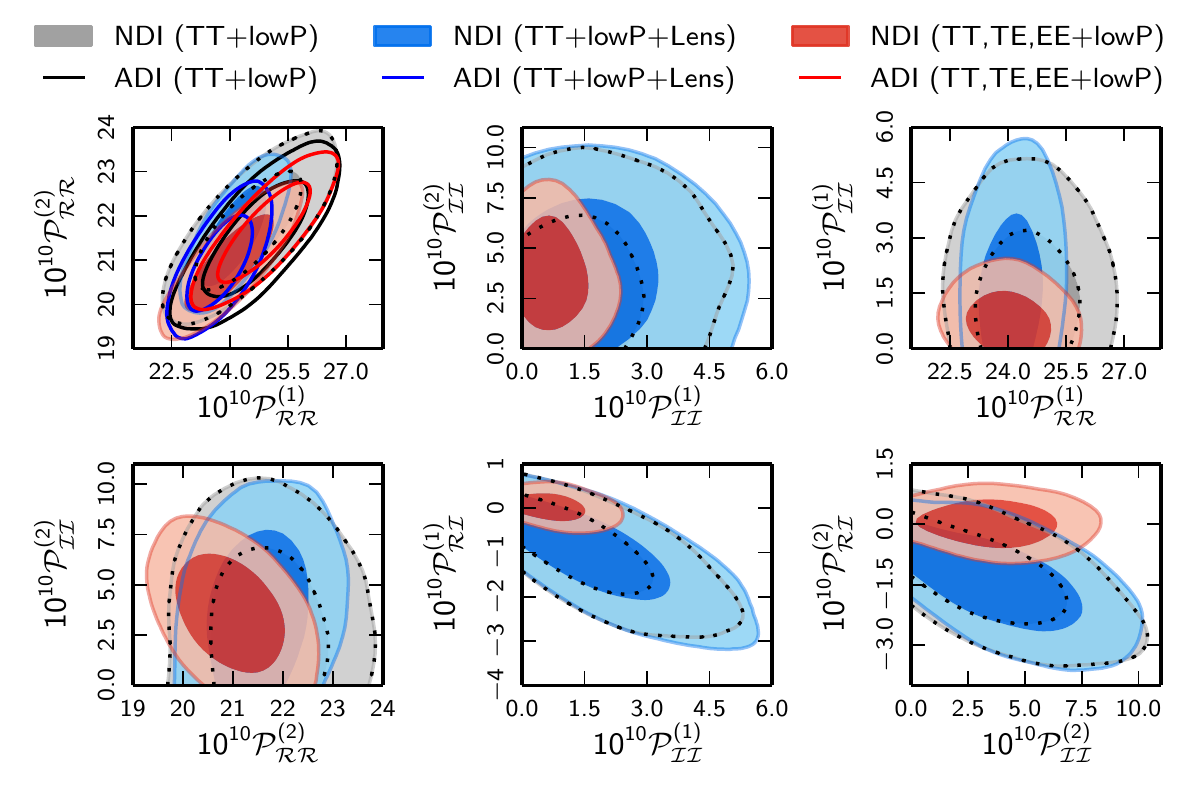}
\includegraphics[width=9cm]{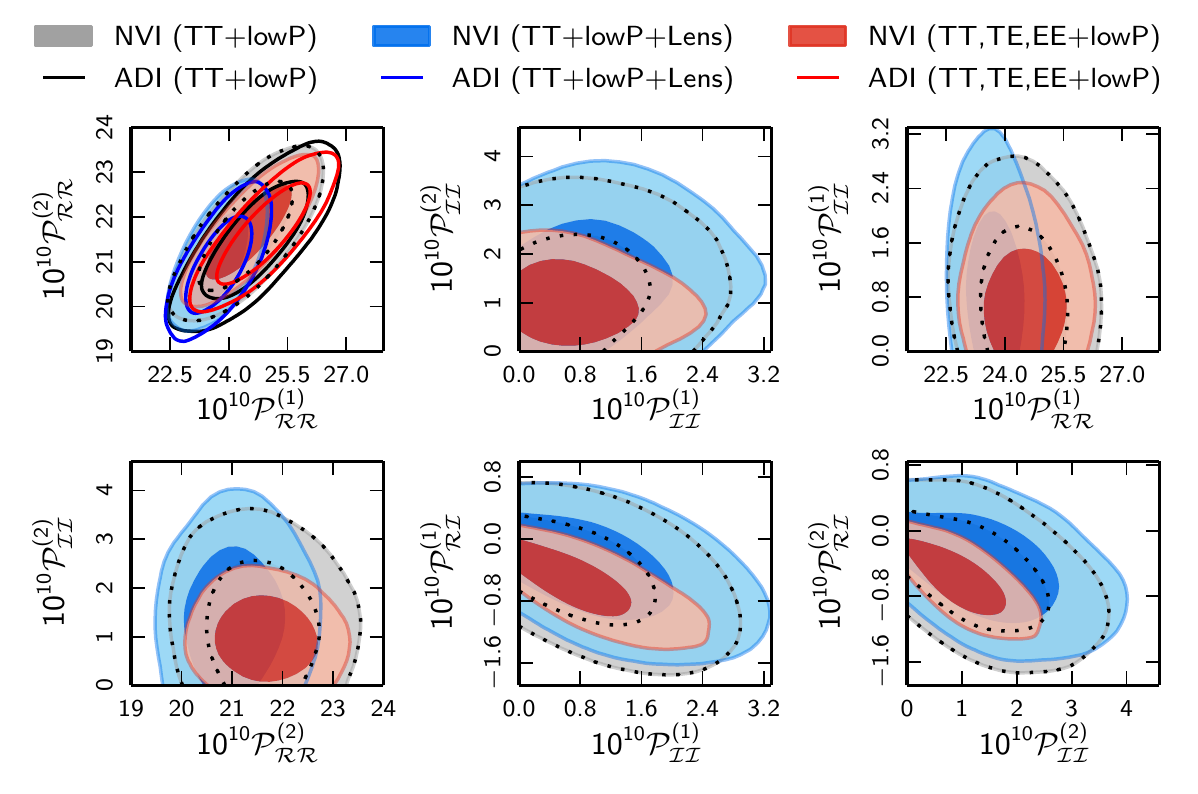}
\end{center}
\caption{Two-dimensional distributions for the quantities
  $\calP_{ab}(k_{1,2})\equiv \calP_{ab}^{(1,2)}$ inferred from the
  Planck $2015$ for different choices of likelihoods indicated by
  different colors (gray, blue and red). This plot should be compared
  to Fig.~\ref{fig:iso2d}. The six upper plots correspond to a
  situation where we have a mixture of adiabatic (denoted ADI) and CDI
  modes, the six middle plots to a situation where we have ADI and NDI
  and the bottom six plots to a case where one has ADI and NVI. Figure
  taken from Ref.~\cite{Ade:2015oja}.}
\label{fig:iso2d2015}
\end{figure}

\begin{figure}
\begin{center}
\includegraphics[width=12cm]{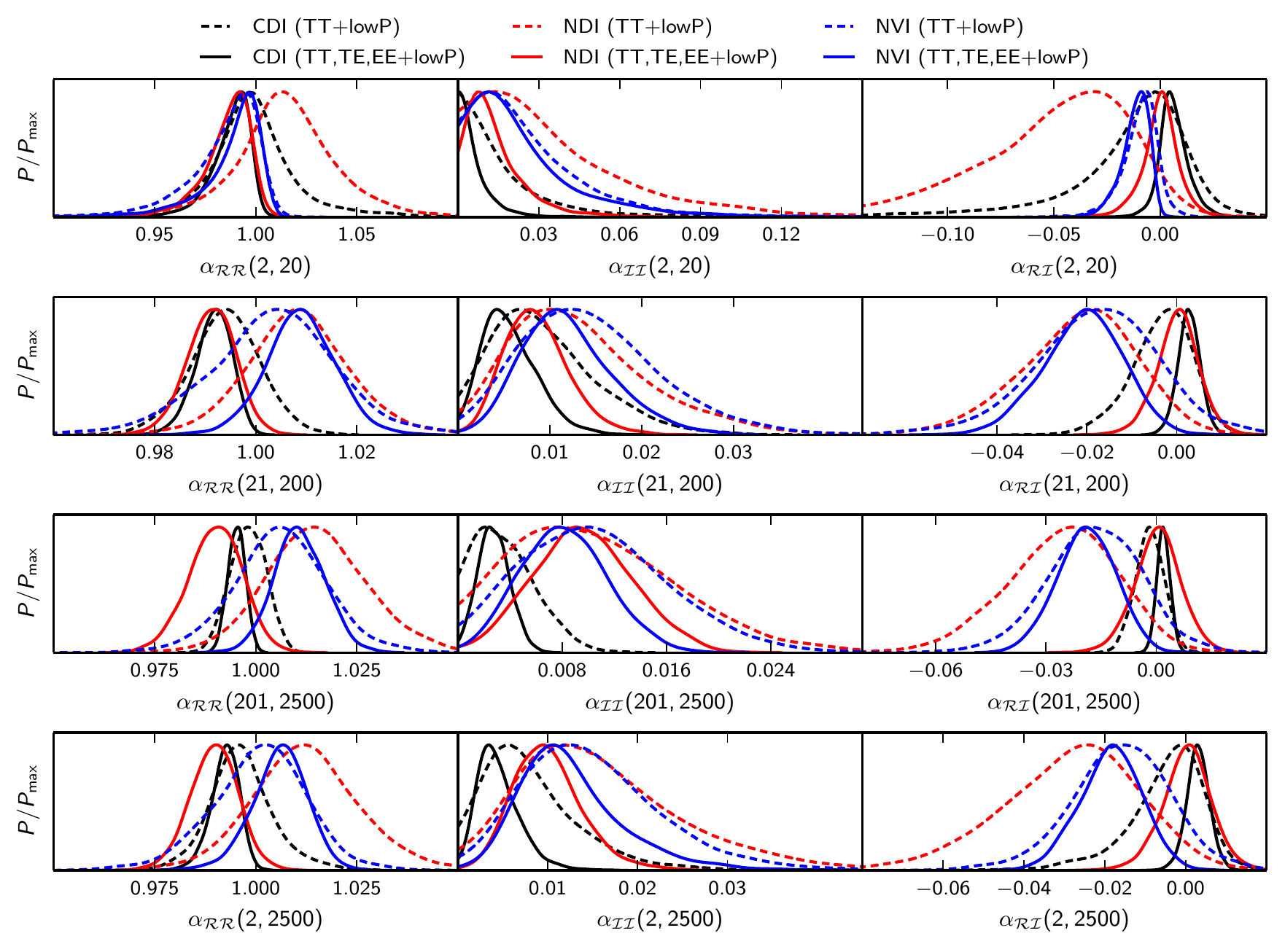}
\end{center}
\caption{Posterior distributions for the quantities
  $\alpha_{ab}\left(\ell_{\rm min},\ell_{\rm max}\right)$ introduced
  in Eq.~(\ref{eq:defalpha}), inferred from the Planck $2015$
  data. This plot should be compared to Fig.~\ref{fig:iso1d}. Figure
  taken from Ref.~\cite{Ade:2015oja}.}
\label{fig:iso1d2015}
\end{figure}

In Ref.~\cite{Ade:2015oja}, the constraints on isocurvature modes
implied by the Planck $2015$ data have been derived. This work is
particularly interesting since one expects the polarization data to
have a good constraining power on the amplitude of the isocurvature
modes. In this analysis, uniform priors for $\calP_{\zeta \zeta}(k_1)$
and $\calP_{\zeta \zeta}(k_2)$ are assumed in the range
$\left[10^{-9},10^{-8}\right]$. For the power spectrum of the
isocurvature power spectra, the same choice is made in the range
$\left[0,10^{-8}\right]$. Finally, the adiabatic-isocurvature
correlation function at $k=k_1$ is taken in the range
$\left[-10^{-8},10^8\right]$. The same quantity, but at $k=k_2$, is
fixed through an assumption about the correlation spectrum, see
Eq.~(\ref{eq:defcorr}). Ref.~\cite{Ade:2015oja} restricts itself to
scale independent correlation spectrum,
\begin{equation}
\cos \Delta_{ab}=\frac{\calP_{ab}}{\sqrt{\calP_{aa}\calP_{bb}}}
\end{equation}
in the range $\left[-1,1\right]$. Writing the above equation at
$k=k_1$ and $k=k_2$ and requiring that the value be the same (since
the correlation spectrum is scale-independent) allows us to derive the
parameter $\calP_{ab}(k_2)$.

\par

The constraints on $\calP_{ab}(k_1)$ and $\calP_{ab}(k_2)$ obtained
from Planck $2015$ are represented in Fig.~\ref{fig:iso2d2015}. The
constraints on the quantities $\alpha_{ab}(\ell_{\rm min},\ell_{\rm
  max})$ are displayed in Fig.~\ref{fig:iso1d2015}. The conclusions
obtained from Planck $2013$ are confirmed and even tightened. No
isocurvature mode is detected and the primordial fluctuations are
fully compatible with exact adiabaticity. This has of course very
important implications for inflation. As explained before, this is a
non trivial test for single-field slow-roll models. The Planck $2013$
data were compatible with this simple class of models and did not
require to introduce additional fields. The results of Planck $2015$
do not modify this claim. As we are going to see in the next section,
this is also the conclusion reached by the Planck measurements of
Non-Gaussianities.

\subsection{Non-Gaussianties}
\label{subsec:ngplanck}

\begin{figure}
\begin{center}
\includegraphics[width=10cm]{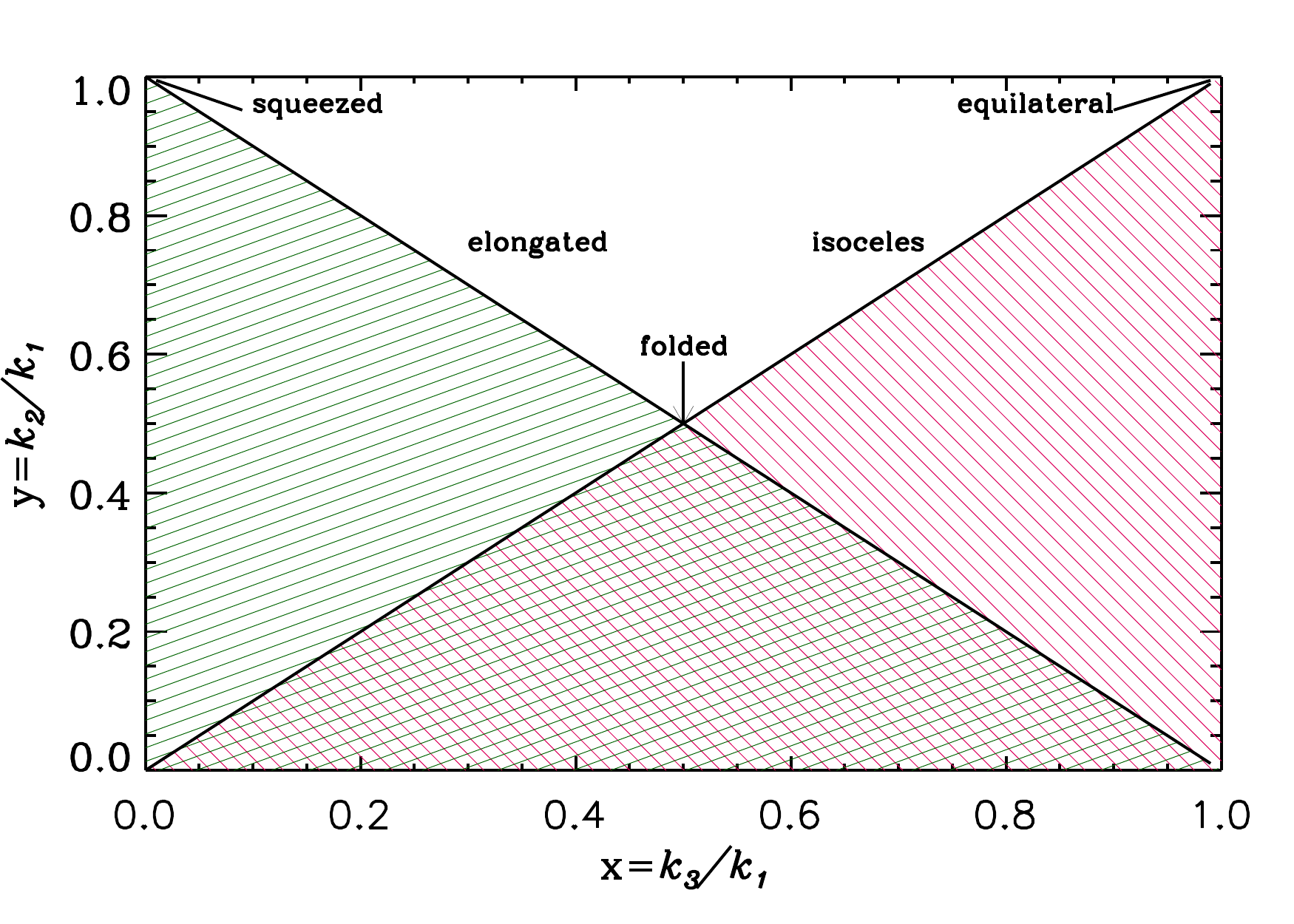}
\end{center}
\caption{Visualization of the bispectrum shape. From the fact that the
  three vectors $\vka$, $\vkb$ and $\vkc$ form a triangle, it is
  possible to faithfully represent the bispectrum in the white
  triangle. Then, different configurations correspond to vertices or
  edges of that triangle.}
\label{fig:triangle}
\end{figure}

Let us now turn to the constraints on primordial Non-Gaussianity, see
Ref.~\cite{Ade:2013ydc}. Before discussing what was measured by the
Planck satellite, it is interesting to review how the results are
sometimes presented in the literature~\cite{2010CQGra..27l4010K}. In
order to visualize the bispectrum, it is convenient to plot the
quantity ${\cal B}_{\cR}(k_1,k_2,k_3)(k_1k_2k_3)^2$ in terms of the
ratios $x\equiv k_3/k_1$ and $y\equiv k_2/k_1$ with the conditions
that $\vka+\vkb+\vkc=0$ and $k_1\geq k_2\geq k_3$. This immediately
implies that $0\leq x\leq 1$ and $0\leq y\leq 1$ and, therefore, the
visualization can be restricted to this square, see
Fig.~\ref{fig:triangle}. The fact that $k_2\geq k_3$ means that $y>x$
and, as a consequence, the red hatched region is in fact
forbidden. Then, since the three vectors $\vka$, $\vkb$ and $\vkc$
form a triangle, every edge length is smaller than the sum of the
length of the two other edges. This means that $y>1-x$ and the green
hatched region is also forbidden. The conditions $k_2<k_1+k_3$ (namely
$y<1+x$) and $k_3<k_1+k_2$ (namely $y>x-1$) lead to new constraints
but outside the square $[0,1]\times [0,1]$ and, therefore, are not
interesting for us. The previous considerations show that it is
sufficient to plot ${\cal B}_{\cR}(k_1,k_2,k_3)(k_1k_2k_3)^2$ in the
white, non hatched, region in Fig.~\ref{fig:triangle} in order to have
a complete representation of the bispectrum.

\par

It is common practice to single out particular configurations. The
squeezed triangle corresponds to $k_1\sim k_2\gg k_3$ which means
$x\sim 0$ and $y\sim 1$. The equilateral configuration is given by
$k_1\sim k_2 \sim k_3$ or $x\sim y \sim 1$. The folded case is defined
by $k_1\simeq 2k_2\simeq 2 k_3$ or $x\sim y\sim 1/2$. These three
configurations correspond to three vertices of the white triangle in
Fig.~\ref{fig:triangle}. Another configuration is the elongated one
for which $k_1\sim k_2+k_3$ or $x+y\sim 1$ and is therefore
represented by a line in Fig.~\ref{fig:triangle}. The same is true for
the isosceles triangle $k_1>k_2\sim k_3$ or $x\sim y$.
 
\begin{figure}
\begin{center}
\includegraphics[width=7.1cm]{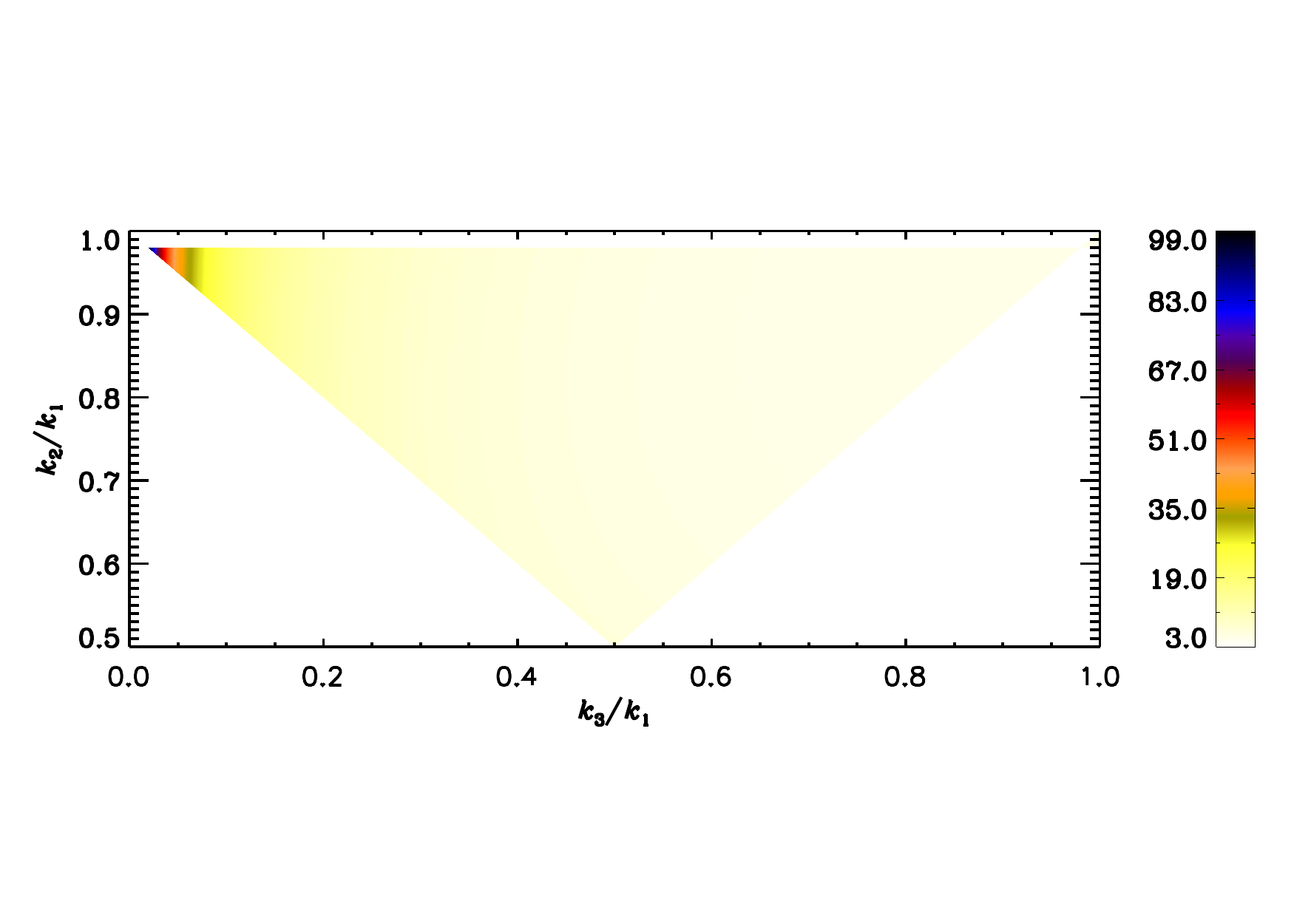}
\includegraphics[width=7.1cm]{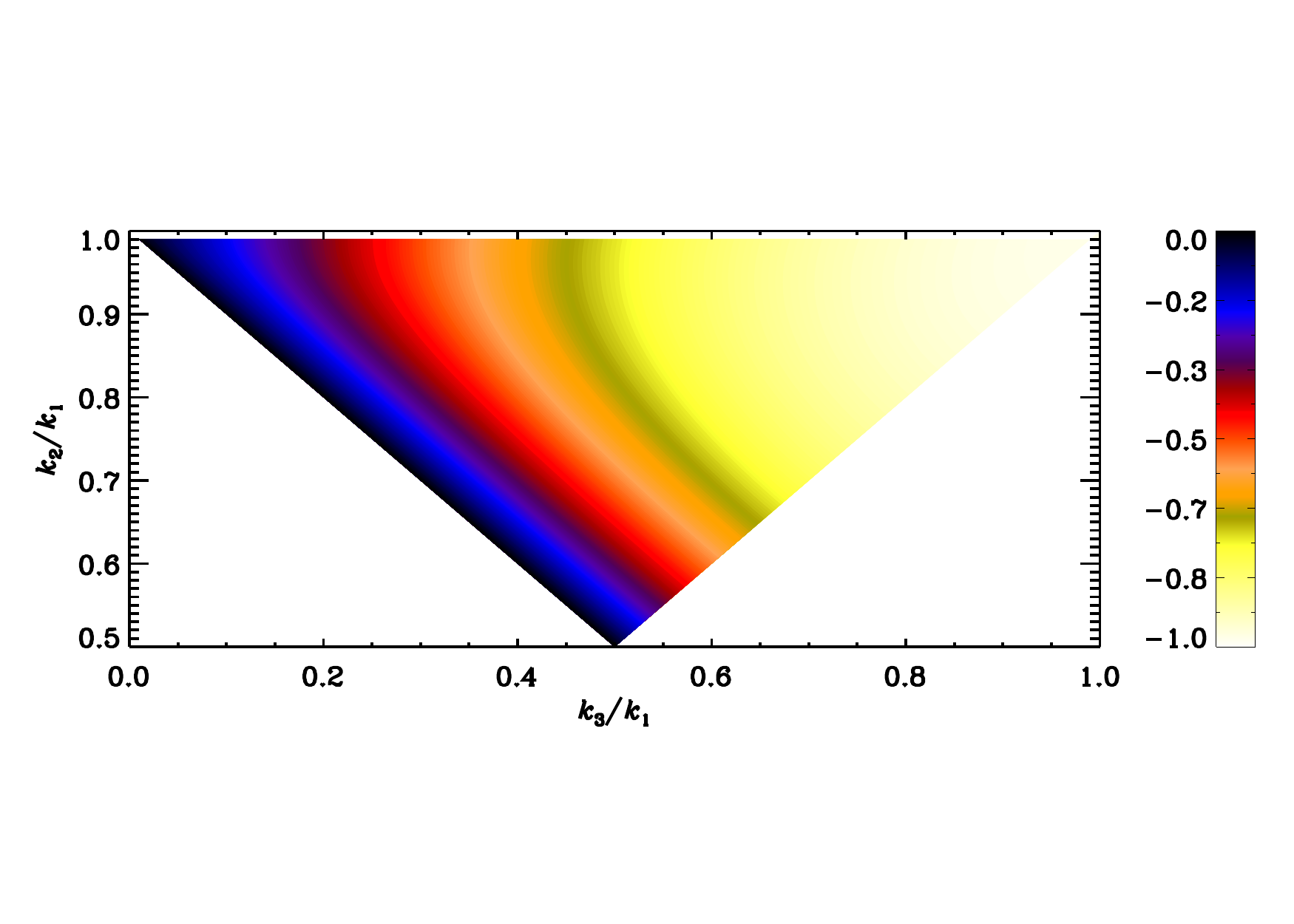}
\includegraphics[width=7.1cm]{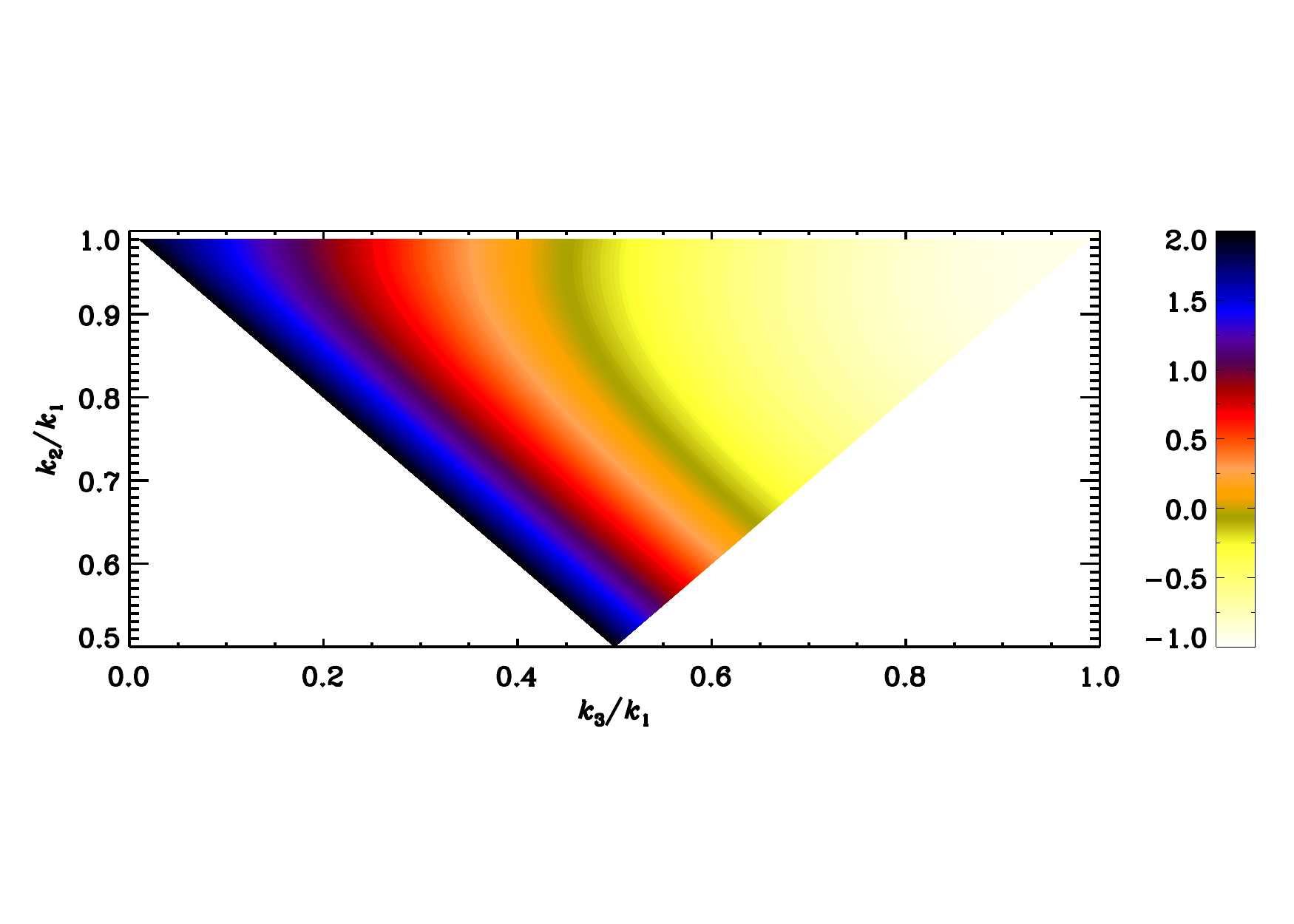}
\includegraphics[width=7.1cm]{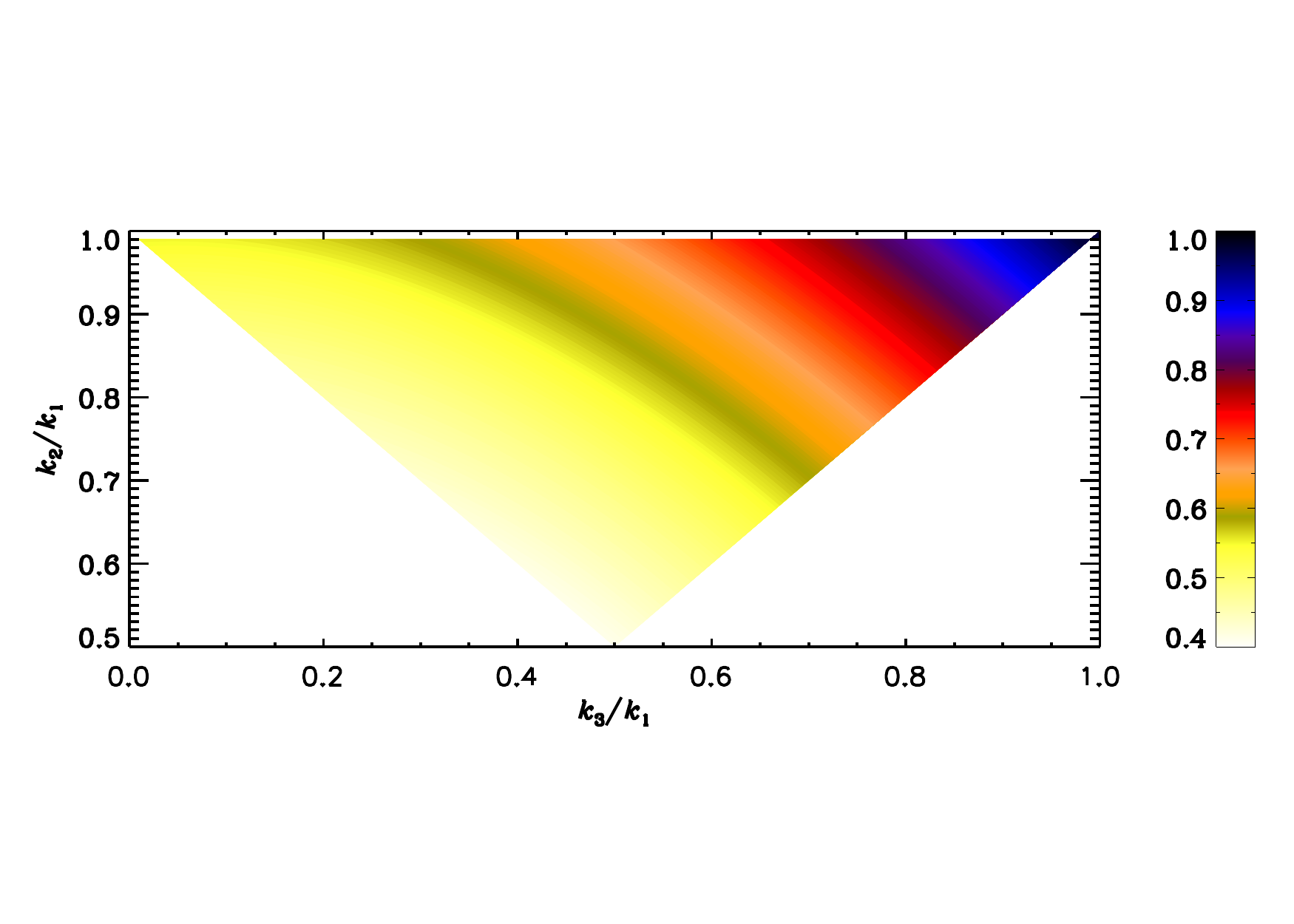}
\end{center}
\caption{Bispectrum for different shape configuration: local (top left
  panel), equilateral (top right panel), orthogonal (bottom left
  panel). The bottom right panel represents the slow-roll prediction
  computed for a model where $V=m^2\phi^2/2$. Notice that the absolute
  normalization in these figures is irrelevant.}
\label{fig:bispec}
\end{figure}

Let us now study how the local bispectrum looks like in this
representation. From Eq.~(\ref{eq:bispecRloc}), one can write
\begin{eqnarray}
{\cal B}_{\cR}(k_1,k_2,k_3) &=& -\frac{6\fnlloc}{5}
(2\pi^2)^2\frac{A_{_{\rm S}}^2}{k_1^6}\frac{1}{x^3y^3}(1+y^3+x^3).
\end{eqnarray}
where $\calP(k)=A_{_{\rm S}}(k/k_*)^{\nS-1}$ and, for simplicity, we
have taken $\nS=1$ [strictly speaking, one should consider the power
  spectrum of Eq.~(\ref{spectrumsr}) but, in fact, this does not
  change significantly the result of this calculation]. It follows,
since ${\cal B}_{\cR}(k_1,k_2,k_3)(k_1k_2k_3)^2=k_1^6y^2x^2{\cal
  B}_{\cR}$ that
\begin{equation}
{\cal B}_{\cR}(k_1,k_2,k_3)(k_1k_2k_3)^2=-\frac{6\fnlloc A_{_{\rm S}}^2}{10}
(2\pi^2)^2\frac{1}{xy}(1+y^3+x^3).
\end{equation}
In Fig.~\ref{fig:bispec} (top left panel) we have represented this
function (without the overall factor in the above expression). As it
is clear from the plot (and also from the analytical expression), the
local shape peaks at the squeezed triangle. The local shape has been
constrained by the Planck 2013 data and one obtains~\cite{Ade:2013ydc}
$\fnlloc=2.7\pm 5.8$ at $68\% {\rm CL}$. The Planck $2015$
data~\cite{Ade:2015ava} with temperature only implies $\fnlloc=2.5\pm
5.7$ and including polarization data, one arrives at
\begin{equation}
\fnlloc=0.8\pm 5,
\end{equation}
thus tightening the conclusion that the perturbations are Gaussian.

\par

Another shape that was studied by the Planck team is the equilateral
one. It is defined by
\begin{eqnarray}
\label{eq:defequibi}
\cB_{\cR}(k_1,k_2,k_3)&=&\frac{18}{5}\fnleq(2\pi^2)^2A_{_{\rm S}}^2
\Biggl[\frac{1}{k_1^3k_2^3}
+\frac{1}{k_2^3k_3^3}
+\frac{1}{k_1^3k_3^3}
+\frac{2}{(k_1k_2k_3)^2}
-\frac{1}{k_1k_2^2k_3^3}
\nonumber \\ & &
-\frac{1}{k_1k_3^2k_2^3}
-\frac{1}{k_2k_1^2k_3^3}
-\frac{1}{k_2k_3^2k_1^3}
-\frac{1}{k_3k_1^2k_2^3}
-\frac{1}{k_3k_2^2k_1^3}\Biggr],
\end{eqnarray}
where, again, one has taken $\nS=1$ for simplicity. One can re-express the
bispectrum in terms of our variables $x$ and $y$ and then multiply by
$(k_1k_2k_3)^2$. This gives
\begin{eqnarray}
\cB_{\cR}(k_1,k_2,k_3)(k_1k_2k_3)^2&=&
\frac{18}{5}\fnleq(2\pi^2)^2A_{_{\rm S}}^2
\frac{1}{xy}\bigl(x^3+1+y^3+2xy-y-x-y^2
\nonumber \\ & & 
-y^2x-x^2-x^2y\bigr).
\end{eqnarray}
The corresponding bispectrum has been represented in
Fig.~\ref{fig:bispec} (top right panel). The coefficient $\fnleq$ has
been constrained by Planck 2013 which finds~\cite{Ade:2013ydc}
$\fnleq=-42\pm 75$, a value compatible with zero. With the Planck
$2015$ data~\cite{Ade:2015ava} (temperature only), one obtains that
$\fnleq=-16\pm 70$ and, including polarization,
\begin{equation}
\fnleq=-4\pm 43.
\end{equation}

\par

Finally, the last shape studied by Planck is the orthogonal one for
which the bispectrum can be expressed as
\begin{eqnarray}
\cB_{\cR}(k_1,k_2,k_3)&=&\frac{18}{5}\fnlortho(2\pi^2)^2A_{_{\rm S}}^2
\Biggl[\frac{3}{k_1^3k_2^3}
+\frac{3}{k_2^3k_3^3}
+\frac{3}{k_1^3k_3^3}
+\frac{8}{(k_1k_2k_3)^2}
-\frac{3}{k_1k_2^2k_3^3}
\nonumber \\ & &
-\frac{3}{k_1k_3^2k_2^3}
-\frac{3}{k_2k_1^2k_3^3}
-\frac{3}{k_2k_3^2k_1^3}
-\frac{3}{k_3k_1^2k_2^3}
-\frac{3}{k_3k_2^2k_1^3}\Biggr],
\end{eqnarray}
which leads to
\begin{eqnarray}
\cB_{\cR}(k_1,k_2,k_3)(k_1k_2k_3)^2&=&
\frac{18}{5}\fnlortho(2\pi^2)^2A_{_{\rm S}}^2
\frac{1}{xy}\bigl(3x^3+3+3y^3+8xy-3y-3x-3y^2
\nonumber \\ & & 
-3y^2x-3x^2-3x^2y\bigr), 
\end{eqnarray}
and is plotted in Fig.~\ref{fig:bispec} (bottom left panel). The
coefficient $\fnlortho$ has been measured by Planck 2013 and the
result reads~\cite{Ade:2013ydc}: $\fnlortho=-25\pm 39$. This
conclusion is confirmed by the Planck $2015$
measurements~\cite{Ade:2015ava}, namely $\fnlortho=-34\pm 33$
(temperature only). If polarization data are included, then one finds
\begin{equation}
\fnlortho=-26\pm 21.
\end{equation}
Once again, the measured value is compatible with Gaussian primordial
fluctuations.

\par

It is also also interesting to represent explicitly the slow-roll
result using the same visualization tools. This bispectrum was derived
in Eq.~(\ref{eq:bispecsrwithF}). Expressed in terms of $x$ and $y$,
each term ${\cal F}^{(i)}$ reads
\begin{eqnarray}
\label{eq:f1}
{\cal F}^{(1)} &=& \frac{H^4}{16\Mp^4\epsilon_1}k_1^3
\Biggl[\left(1+\frac{1}{1+x+y}\right)\frac{x^2y^2}{1+x+y}
+\left(1+\frac{y}{1+x+y}\right)\frac{x^2}{1+x+y}
\nonumber \\ & & +
\left(1+\frac{x}{1+x+y}\right)\frac{y^2}{1+x+y}\Biggr],\\
\label{eq:f2}
{\cal F}^{(2)} &=& \frac{H^4}{16\Mp^4\epsilon_1}\times -\frac{k_1^3}{2}
(1+y^2+x^2)\Biggl[-(1+x+y)+\frac{y+x+xy}{1+x+y}
\nonumber \\ & &
+\frac{xy}{(1+x+y)^2}\Biggr]\\
\label{eq:f3}
{\cal F}^{(3)} &=& -\frac{H^4}{16\Mp^4\epsilon_1}k_1^3
\Biggl[\frac12 (-1+x^2-y^2)\frac{x^2}{1+x+y}\Biggl(
2+\frac{1+y}{1+x+y}\Biggr)
\nonumber \\ & &
+\frac12 (-1-x^2+y^2)\frac{y^2}{1+x+y}\Biggl(
2+\frac{1+x}{1+x+y}\Biggr)
\nonumber \\ & &
+\frac12 (-1-x^2-y^2)\frac{1}{1+x+y}\Biggl(
2+\frac{x+y}{1+x+y}\Biggr)\Biggr]\\
\label{eq:f7}
{\cal F}^{(7)} &=& \frac{H^4}{16\Mp^4\epsilon_1}k_1^3
\frac{\epsilon_2}{2\epsilon_1}(1+x^3+y^3).
\end{eqnarray}
If we write ${\cal F}^{(i)}\equiv
H^4/(16\Mp^4\epsilon_1)k_1^3f^{(i)}=(2\pi)^2\epsilon_1A_{_{\rm
    S}}^2k_1^3f^{(i)}$, then we see from the above
equations~(\ref{eq:f1}), (\ref{eq:f2}), (\ref{eq:f3})
and~(\ref{eq:f7}) that the functions $f^{(i)}$ only depend on $x$ and
$y$. In particular, this definition factors out the term $k_1^3$. As a
consequence, using Eq.~(\ref{eq:bispecsrwithF}), the quantity
$\cB_{\cR}^{\rm sr}(k_1,k_2,k_3) (k_1k_2k_3)^2$ can be written
as~\cite{Hazra:2012yn}
\begin{equation}
\label{eq:finalBrsr}
\cB_{\cR}^{\rm sr}(k_1,k_2,k_3)(k_1k_2k_3)^2=\left(2\pi^2\right)^2
\epsilon_1\frac{A_{_{\rm S}}^2}{xy}\sum_{i=1,2,3,7}f^{(i)}(x,y).
\end{equation}
This bispectrum is represented in Fig.~\ref{fig:bispec} (bottom right
panel). We notice that it is similar (up to a sign) to the equilateral
shape~(\ref{eq:defequibi}).

\par

Finally, Planck 2013 has also measured the four point correlation
function for the local configuration. The corresponding constrain on
the $\tau_{_{\rm NL}}$ reads~\cite{Ade:2013ydc}
\begin{equation}
\tau_{_{\rm NL}}<2800, 
\end{equation}
at $95\% $ confidence level, that is to say, a result compatible with
Gaussianity. A recent analysis~\cite{Feng:2015pva} has confirmed this
conclusion. Ref.~\cite{Feng:2015pva} has indeed found $\tau_{_{\rm
    NL}}=0.3\pm0.9\times 10^{4}$ and $g_{_{\rm NL}}=-1.2\pm 2.8\times
10^{5}$. Finally, Planck $2015$~\cite{Ade:2015ava} obtained $g_{_{\rm
    NL}}=(-9.0\pm7.7)\times 10^4$ at $68\%$ confidence level.

\par

We conclude this section on Non-Gaussianity measurements as we
concluded the section on isocurvature modes: the fact that we do not
detect a signal beyond the vanilla situation is another non-trivial
test for single-field slow-roll inflation with a minimal kinetic
term. In the remaining part of this review, we therefore focus on this
class of models and derive the corresponding implications that can
inferred from the Planck data.

\subsection{Slow-Roll Inflation}
\label{subsec:sr}

\begin{figure}
\begin{center}
\includegraphics[width=10cm]{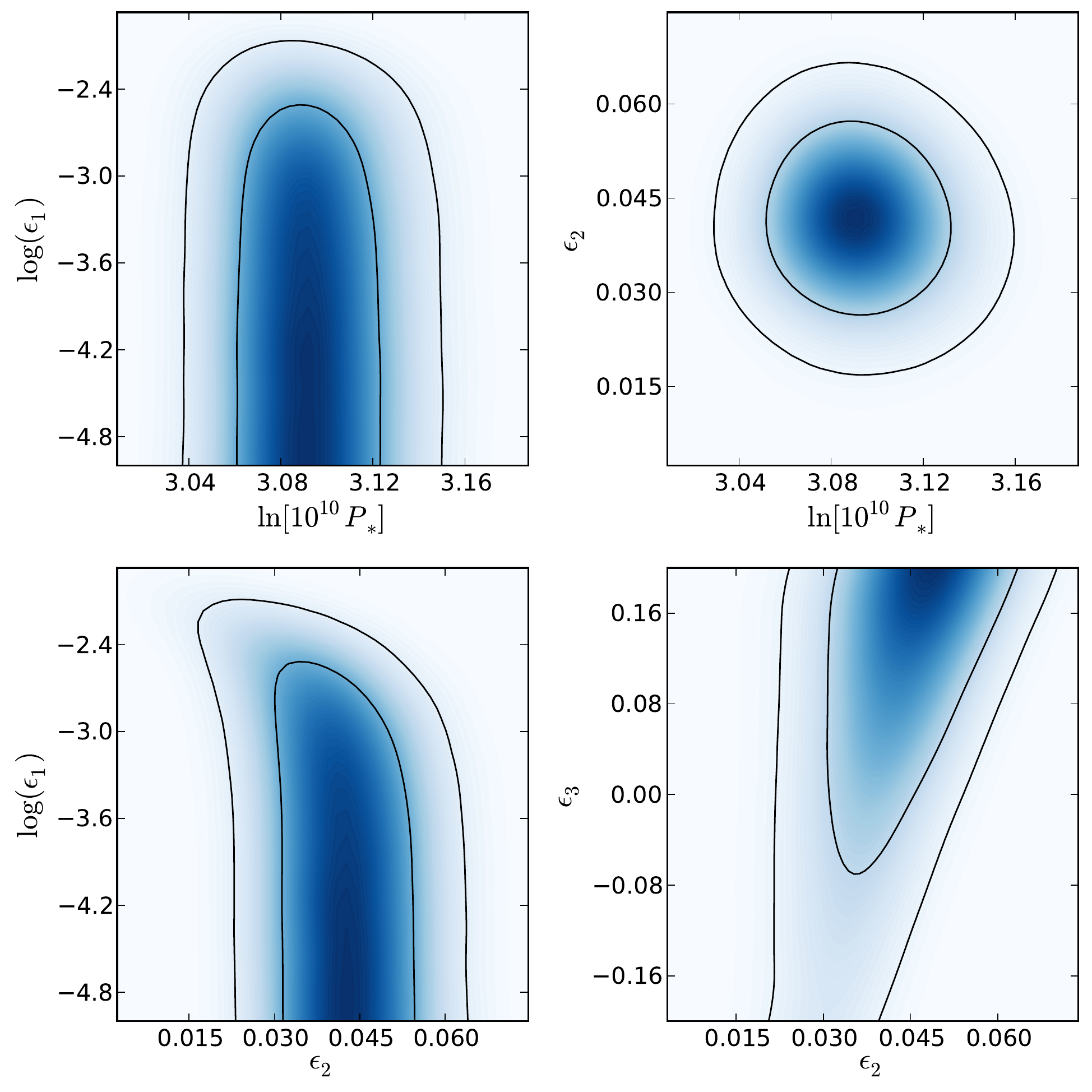}
\end{center}
\caption{Two dimensional posterior distributions of the parameters
  $\epsilon_1$, $\epsilon_2$, $\epsilon_3$ and $P_*$ obtained from the
  Planck 2013 data. The pivot scale is chosen to be $k_*=0.05
  \Mpc^{-1}$ and the priors are taken to be as follows: a Jeffreys'
  prior (\ie a flat prior on the logarithm of the corresponding
  quantity) for $P_*$ such that $\ln (10^{10}P_*) \in [2.7,4.2]$, a
  Jeffreys' prior for $\epsilon_1$ such that $\log(\epsilon_1)\in
  [-5,-0.7]$ (the choice of the upper bound ensures that
  $\epsilon_1<0.2$ and, therefore, that the slow-roll approximation is
  valid) and flat priors for $\epsilon_2 $ and $\epsilon_3$ such that
  $\epsilon_2\in [-0.2,0.2]$ and $\epsilon_3\in [-0.2,0.2]$. Figure
  taken from Ref.~\cite{Martin:2013nzq}.}
\label{fig:sr2nd}
\end{figure}

We have seen in Sec.~\ref{subsec:correl2} that the power spectra of
scalar and tensor perturbations can be expressed in terms of the
slow-roll parameters. Since the CMB measurements constrain the power
spectra, they also constrain the slow-roll
parameters~\cite{Martin:2014lra}. In Fig.~\ref{fig:sr2nd}, we show the
two dimensional marginalized posterior distributions for the
parameters $\epsilon_1$, $\epsilon_2$, $\epsilon_3$ and $P_*$, where
this last quantity represents the overall normalization of the power
spectrum\footnote{That is to say, we have re-written
  Eq.~(\ref{spectrumsr}) as
\begin{equation}
\calP_\zeta=P_*\left[1+\frac{a_1}{a_0}\ln \left(\frac{k}{k_*}\right)
+\frac{a_2}{a_0}\ln ^2\left(\frac{k}{k_*}\right)+\cdots \right],
\end{equation} which defines the quantity $P_*$.} obtained from 
the Planck 2013 data. We see that $P_*$ and $\epsilon_2$ are well
constrained while there only exists an upper bound on $\epsilon_1$ and
almost no constraints on $\epsilon_3$.  Explicitly, one has $3.035
\lesssim \ln \left(10^{10}P_*\right) \lesssim 3.15$,
$\log\left(\epsilon_1\right)\lesssim -2.01$ and $0.023\lesssim
\epsilon_2\lesssim 0.063$ at the two sigma level. Planck
$2015$~\cite{Ade:2015oja} has also analyzed this question and found
$\epsilon_1<0.0068$ and $\epsilon_2=0.029^{+0.008}_{-0.007}$ using
PlanckTT+lowP and restricting the hierarchy at first order in
slow-roll. When high-$\ell$ polarization data are included in the
analysis, one finds $\epsilon_1<0.0066$ and
$\epsilon_2=0.030^{+0.007}_{-0.006}$.

\par

\begin{figure}
\begin{center}
\includegraphics[width=5cm]{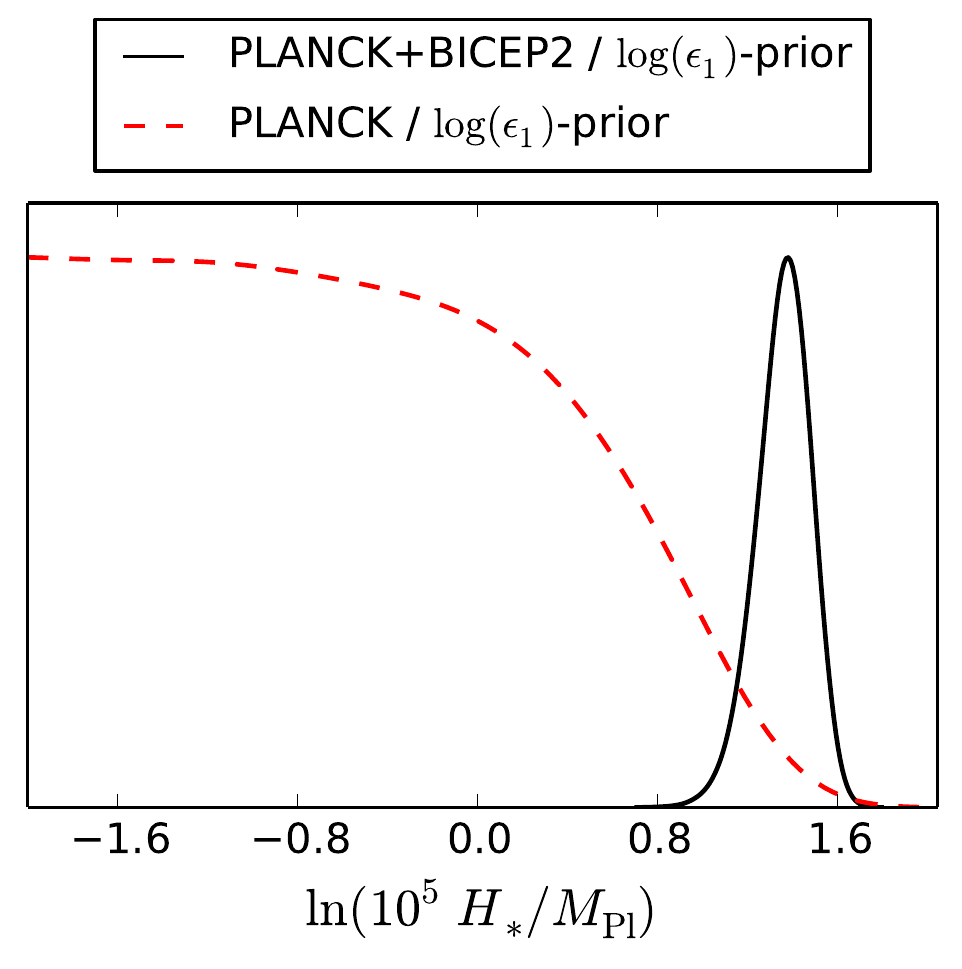}
\includegraphics[width=5cm]{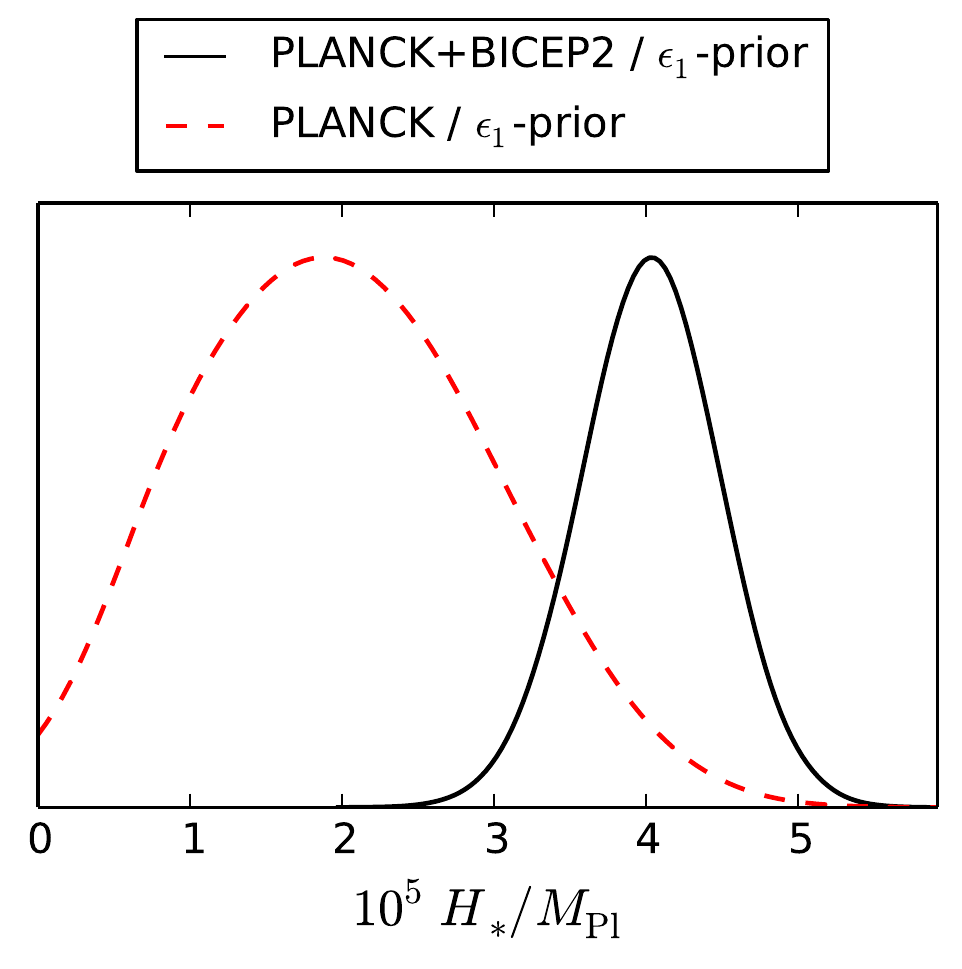}
\caption{Marginalized posterior distribution for the inflationary
  Hubble parameter at the time of pivot crossing with a Jeffreys'
  prior (left panel) and a flat panel (right panel) on $\epsilon_1$
  (left panel). The dashed red line represents the distribution
  obtained from the Planck 2013 data while the solid black line
  corresponds to the case where the Planck 2013 data are combined with
  the BICEP2 measurement (here, interpretated as a detection of
  gravity waves) and illustrates how a detection of primordial
  gravitational waves could allow us to determine the energy scale of
  inflation. Figure taken from Ref.~\cite{Martin:2014lra}.}
\label{fig:Hstar}
\end{center}
\end{figure}

Let us now discuss the physical information on inflation that can be
inferred from the above results. Firstly, from Eqs.~(\ref{spectrumsr})
and~(\ref{eq:scalaramp}), one has at next-to-leading order on
slow-roll
\begin{equation}
P_*=\calP_{\zeta 0}a_0=\frac{H_*^2}{8\pi^2\epsilon_{1*}\Mp^2}
\left[1-2\left(C+1\right)\epsilon_1-C\epsilon_2\right],
\end{equation}
from which we deduce that, at second order in slow-roll, the Hubble
parameter during inflation can be expressed as~\cite{Martin:2014lra}
\begin{equation}
\begin{aligned}
\dfrac{\Hstar^2}{\Mpl^2} & = 8 \pi^2 \epsstar{1} \Pstar\left[1+ 2(1+C)
  \epsstar{1} + C \epsstar{2} \right].
\end{aligned}
\end{equation}
Since we know the posterior of $P_*$ and $\epsilon_1$, one can derive
the corresponding one for $H_*$. The result is represented in
Fig.~\ref{fig:Hstar}, see the red dashed curves. Clearly, the fact
that we only have an upper bound on $\epsilon_1$ implies that we also
only have an upper bound on $H_*$. With the Jeffreys' prior on
$\epsilon_1$ (see the left panel in Fig.~\ref{fig:Hstar}), one obtains
$\ln\left(10^5 \dfrac{\Hstar}{\Mpl} \right) \lesssim 1.6$, that is to
say
\begin{equation}
H_* \lesssim 1.2\times 10^{14} \GeV.
\label{eq:lnH}
\end{equation}
One obtains a similar number if a flat prior on $\epsilon_1$ is
assumed (see the right panel in Fig.~\ref{fig:Hstar}). Those values
can be expressed into gravitating energy scales through
\begin{equation}
\rho_*^{1/4} = 3^{1/4} \sqrt{\Hstar \Mpl}\lesssim 2.2\times 10^{16}\GeV,
\label{eq:V}
\end{equation}
where this value assumes a Jeffreys' prior on $\epsilon_1$ (again, a
similar result is obtained with a flat prior). If primordial gravity
waves are detected, then this would fix the value of $r$ and, hence,
the energy scale of inflation. This is illustrated in
Fig.~\ref{fig:Hstar} where we have also plotted the posterior
distribution of $H_*$ obtained when the BICEP2
results~\cite{Ade:2014xna} are taken into account (assuming, for the
sake of illustration, that they correspond to a detection of gravity
waves).

Secondly, let us now study what the constraints on the slow-roll
parameters mean for the shape of the inflation
potential~\cite{Martin:2014lra}. From Eq.~(\ref{eq:epsfirst}), we see that
this gives an upper bound on the first derivative of the inflation
potential, namely
\begin{equation}
\left \vert V_\phi \right \vert\lesssim 0.14 \frac{V}{\Mp}.
\end{equation}
Using PlanckTT+lowP, the recent Planck $2015$ data~\cite{Ade:2015oja}
implies that $\vert V_{\phi}\vert \lesssim 0.116
\left(V/\Mp\right)$. On the other hand, the second Hubble flow
parameter gives information about the second derivative of the
inflaton potential. From Eq.~(\ref{eq:eps2}), one sees that
\begin{equation}
\Mp^2\frac{V_{\phi \phi}}{V}=2\epsilon_1-\frac{\epsilon_2}{2}.
\end{equation}
From this expression, we also obtain bounds on the second derivative
of the potential. Indeed, one has $\Mp^2 V_{\phi \phi}/V>-\epsilon_{2
  {\rm sup}}$ and $\Mp^2 V_{\phi \phi}/V<2\epsilon_{1{\rm
    sup}}-\epsilon_{2{\rm min}}/2$. Explicitly, one has
\begin{equation}
-0.03\lesssim \Mp^2\frac{V_{\phi \phi}}{V}\lesssim 0.008.
\end{equation}
Planck $2015$~\cite{Ade:2015oja}, using PlanckTT+lowP, finds the
following value $\Mp^2V_{\phi \phi}/V=-0.01^{+0.005}_{-0.009}$ at
$95\%$ confidence level.

\begin{figure}
\begin{center}
\includegraphics[width=10cm]{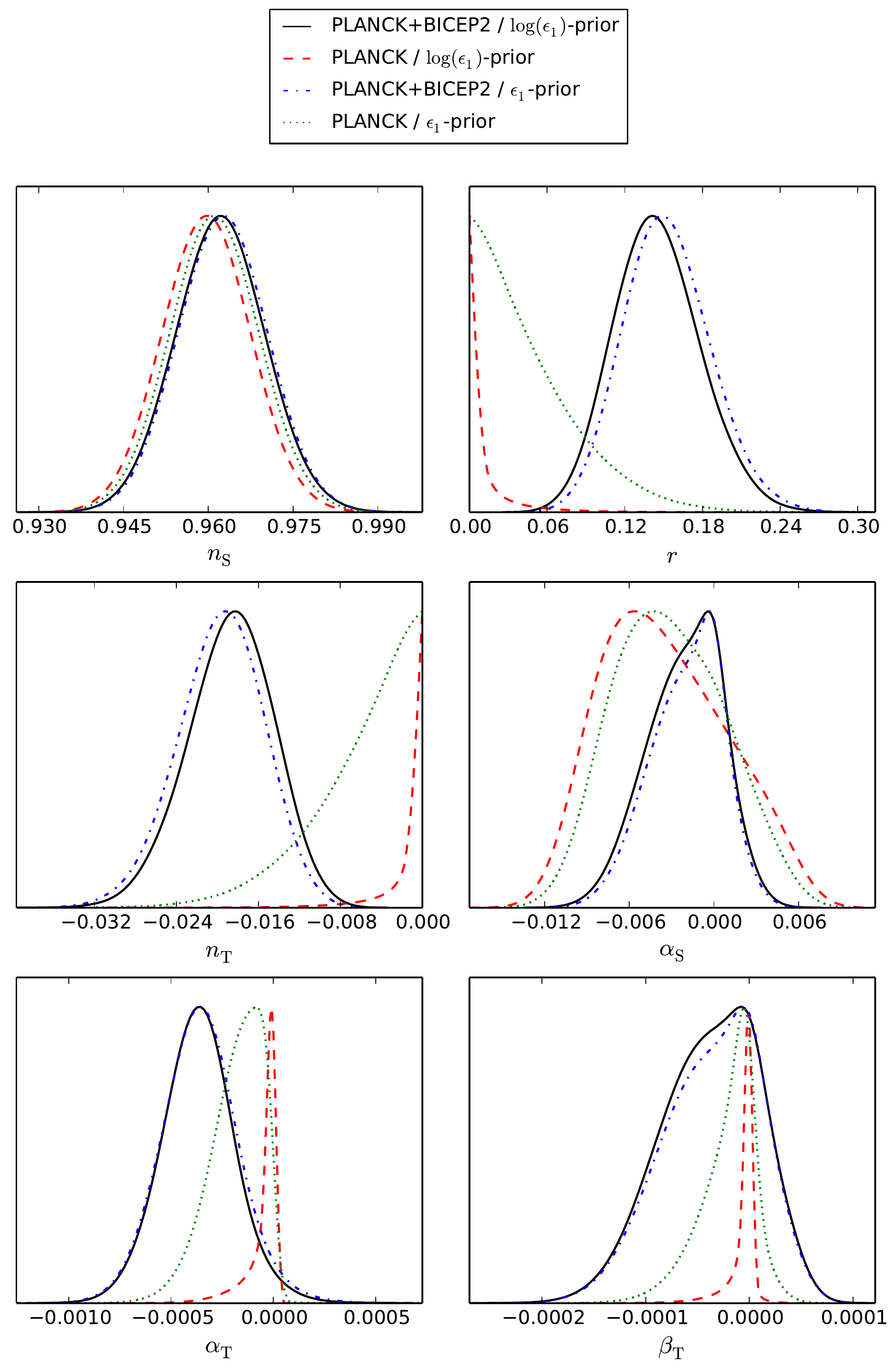}
\caption{Marginalized posterior distributions for the derived power
  law parameters $\nS$, $r$, $\nT$, $\alphaS$, $\alphaT$ and $\betaT$
  obtained by importance sampling from the distributions of the second
  order slow-roll parameters. The dashed red line and the dotted green
  lines are the distributions obtained from Planck 2013. We have also
  represented the results obtained by combining Planck 2013 with
  BICEP2 (see the solid black line and the dotted dashed blue
  line). The most striking feature which would follow from a detection
  of gravity waves (here illustrated by including the BICEP2 results
  taken at face value) is of course that $r$, and therefore $\nT$
  using the slow-roll consistency relations~(\ref{eq:consistency}),
  would now be measured. Figure taken from
  Ref.~\cite{Martin:2014lra}.}
\label{fig:pl}
\end{center}
\end{figure}

Thirdly, although the shape of the power spectrum is entirely
characterized by Eq.~(\ref{spectrumsr}), it is also interesting to
derive constraints on the so-called power-law
parameters~\cite{Martin:2014lra}. These parameters are in fact simple
combinations of the Hubble flow parameters, as exemplified by
Eqs.~(\ref{eq:specindices}). We now investigate this question in more
detail. In Eqs.~(\ref{eq:specindices}), we gave the spectral indices
at first order in slow-roll. At second order, they read
\begin{equation}
\begin{aligned}
\nS & =1 -
(2\epsstar{1} + \epsstar{2}) - 2\epsstar{1}^2 -
(3+2C)\epsstar{1}\epsstar{2} -C \epsstar{2}\epsstar{3}, \\
\nT & = -2\epsstar{1}-2\epsstar{1}^2-2(1+C)\epsstar{1}\epsstar{2},
\end{aligned}
\label{eq:nst}
\end{equation}
while the tensor-to-scalar ratio can be expressed as [see also
Eq.~(\ref{eq:defr})]
\begin{equation}
r  = 16\epsstar{1} (1+C\epsstar{2}).
\label{eq:r}
\end{equation}
One can also define the runnings for scalar and tensor and, in the
slow-roll approximation, they are second-order quantities and their
expressions read
\begin{equation}
\alphaS = -2\epsstar{1}\epsstar{2}-\epsstar{2}\epsstar{3},\qquad
\alphaT  =-2\epsstar{1}\epsstar{2}, 
\label{eq:alphast}
\end{equation}
Finally, let us mention that the running of the running for the tensor
mode is also completely specified by the first three Hubble flow
functions and is given by
\begin{equation}
\betaT = - 2 \epsstar{1} \epsstar{2} \left( \epsstar{2} 
+ \epsstar{3} \right).
\end{equation}
One sees that, in general, one has six independent quantities, namely
$r$, $\nS$, $\nT$, $\alphaS$, $\alphaT$ and $\betaT$. However, the
predictions of slow-roll inflation can be expressed in terms of three
Hubble flow parameters (at least at this order), $\epsilon_1$,
$\epsilon_2$ and $\epsilon_3$. This implies that all the parameters
describing the tensor sector can, in fact, be expressed in terms of
those characterizing the scalar sector. Explicitly, these so-called
consistency relations can be expressed as
\begin{equation}
\begin{aligned}
\nT & \simeq -\frac{r}{8} , \\
\alphaT & \simeq \frac{r}{8}\left[\frac{r}{8} +
  \left(\nS-1\right)\right], \\
\betaT & \simeq \frac{r}{8}\left[\frac{r}{8} +
  \left(\nS-1\right)\right]\left(1-\nS-\frac{r}{4}\right)
+\frac{r}{8}\alphaS.
\end{aligned}
\label{eq:consistency}
\end{equation}
As before, since we know the posterior distributions of the slow-roll
parameters for the Planck 2013 data, we can infer those of the
power-law parameters. They are represented in Fig.~\ref{fig:pl}, see
the red dashed and dotted green lines. We see that the scalar spectral
index $\nS$ is very well constrained and is around $\nS\simeq 0.96$
(the Planck $2013$ value, with WMAP large-angle polarization, reads
$\nS=0.9603\pm 0.0073$). On the other hand, we only have an upper
bound on the tensor-to-scalar ratio which is of course expected since
$r\propto \epsilon_1$. At two sigmas, one obtains $\log(r)\lesssim
-0.88$ which gives
\begin{equation}
\label{eq:rsrupper}
r\lesssim 0.13.
\end{equation}
Notice that this result is obtained assuming a Jeffeys' prior on
$\epsilon_1$. If, instead, a flat prior is chosen, one has
$\log(r)\lesssim -0.64$, leading to $r\lesssim 0.23$. This is because
a flat prior has the tendency to favor large values of $r$ compared to
what is obtained with a Jeffrey's prior. In Fig.~\ref{fig:pl}, we have
also represented the results obtained by combining Planck 2013 and
BICEP2 assuming that this last signal is due to primordial gravity
waves. Of course, in that case $r$ is determined and, as a
consequence, the tensor spectral index is also fixed. It is now known
that the BICEP2 signal can be entirely explained by dust
contamination~\cite{Ade:2014gaa} but, nevertheless, it is interesting
to see what would be the implications for inflation of a detection of
primordial gravity waves.

\par

Recently, Planck $2015$~\cite{Ade:2015oja} has also put constraints on
$r$. As usual, these constraints depend on the data sets used and on
the assumptions made about the theoretical frameworks. Here we just
quote two numbers. Using PlanckTT,TE,EE+lowP and considering that $r$
is the only extra parameter beyond the base $\Lambda$CDM model, one
obtains
\begin{equation}
r_{0.002}<0.1,
\end{equation}
at $95\%$ confidence limit. If instead PlanckTT+lowP+WP (we remind
that WP means the polarization data on large scales measured by WMAP),
this number becomes $r_{0.002}<0.09$. Here, the subscript ``$0.002$''
indicates that the pivot scale is taken to be $0.002\,
\mbox{Mpc}^{-1}$.

\begin{figure}
\begin{center}
\includegraphics[width=10cm]{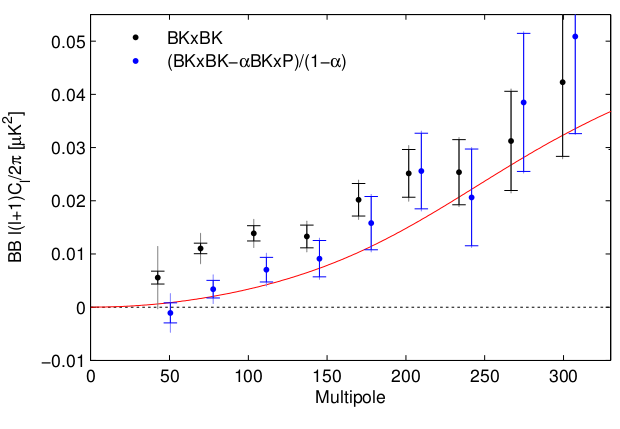}
\caption{$B$-mode CMB polarization multipole moments compared to the
  lensing signal (solid red curve). The black dots represent the
  values of $C_{\ell}^{\mathrm{BB}}$ obtained from the BICEP2/Keck
  array map. On recognizes the bump in the range $\ell \sim [50,120]$
  that deviates from the lensing curve and that was interpreted as a
  detection of primordial gravity waves with $r\sim 0.16$ in
  Ref.~\cite{Ade:2014xna}. The blue dots correspond to the same
  multipole moments but after subtraction of the dust contribution,
  estimated from the cross-spectrum with the Planck $353$ GHz
  channel. Clearly, the new data are in good agreement with what is
  expected from lensing. Figure taken from
  Ref.~\cite{BICEP2/Keck:2015tva}.}
\label{fig:bcp/Planck}
\end{center}
\end{figure}

Very recently, a joint analysis by the BICEP2/Keck Array team and the
Planck collaboration was released~\cite{BICEP2/Keck:2015tva}. The
results are presented in Fig.~\ref{fig:bcp/Planck}. In
Ref.~\cite{Ade:2014xna}, BICEP2 announced the detection of primordial
gravity waves at a level corresponding to a tensor-to-scalar ratio of
$r\sim 0.16$. The reason for this claim can be seen in
Fig.~\ref{fig:bcp/Planck}. In this plot, the red solid curve is the
signal due to the weak lensing of $E$-mode that produces $B$-modes on
small angular scales. This contribution is necessarily present in the
standard model of Cosmology and its amplitude can be inferred
unambiguously once we know the value of the cosmological
parameters. The black dots represent the signal measured by BICEP2 and
Keck Array. As is well visible, in the range $\ell \sim [50, 120]$,
there is an excess of power with respect to the red solid line and,
hence, there must be another source of $B$-modes. BICEP2 interpreted
this excess as a contribution coming from primordial gravity
waves. However, there is another known source of contamination:
dust. BICEP2 could not measure accurately the dust contribution
because it operates at a single frequency only. The BICEP2 team
therefore used theoretical models available at that time to remove the
dust signal. On the other hand, Planck measures the CMB at different
frequencies and, as a consequence, can estimate with good precision
the dust contribution. It is therefore clear that a joint analysis
between the two teams is the best way to use at the same time the good
sensitivity of BICEP2/Keck Array and the good control of the dust
signal of the Planck team. The result of this analysis are the blue
dots in Fig.~\ref{fig:bcp/Planck}. We see that the bump has
disappeared which means that the excess of power observed was probably
entirely due to dust contamination and not to primordial gravity
waves. The signal is now compatible with lensing. The new analysis
suggests that the best value of $r$ is now $r\sim 0.05$ but with very
low significance and $r\sim 0$ cannot be excluded. In other words,
there is no longer a detection of primordial gravity waves. In
addition, one obtains a new upper limit which is now $r<0.12$ (at
$95\%$ confidence limit) instead of $r<0.11$ from the Planck $2013$
data. Notice that we obtained before $r<0.13$ from the Planck $2013$
data, see Eq.~(\ref{eq:rsrupper}), and not $r<0.11$, but this is just
due to some differences between our analysis and the Planck one
(essentially, different priors).

\par

In Ref.~\cite{Martin:2014lra}, it was demonstrated that the sets of
inflationary models preferred by Planck alone and BICEP2 alone are
almost disjoint, indicating a clear tension between the two data
sets. Using a Bayesian measure of compatibility between BICEP2 and
Planck, it was indeed shown that, for models favored by Planck $2013$
the two data sets tended to be incompatible, whereas there was a
moderate evidence of compatibility for the BICEP2 preferred
models. This means that the three assumptions (i) slow-roll inflation
is the correct description of the early Universe (ii) Planck $2013$
data accurately measure CMB temperature anisotropies and (iii) BICEP2
measurement is due to primordial gravity waves are mutually
exclusive. In other words, if one has the theoretical prejudice that
slow-roll inflation did occur in the early Universe, then
Ref.~\cite{Martin:2014lra} already proved that the value $r\sim 0.16$
was likely to be overestimated. In some sense, the fact that dust
contamination can explain the BICEP2 signal reinforces our trust in
inflation!

\par

Let us now turn to the scalar running $\alphaS$. At $95\%$ confidence
level, one finds
\begin{equation}
-0.012\lesssim \alphaS\lesssim 0.006,
\end{equation}
that is to say a value consistent with no running. Finally, one
notices that the quantities $\alphaT$ and $\betaT$ are
well-constrained. It is easy to understand why on the example of
$\alphaT$. One has $\alphaT=(r/8)^2+(\nS-1)(r/8)$, see
Eq.~(\ref{eq:consistency}). In this equation $\nS-1$ is known, one can
take $\nS-1\simeq -0.04$ which means that $\alphaT\simeq
(r/8)^2-0.04(r/8)$. This parabola has a minimum at $r/8\simeq 0.02$
which corresponds to $\alphaT\simeq -9\times 10^{-5}$. The maximum is
for $r\simeq 0.13$ and gives $\alphaT\simeq -0.0004$. We therefore
expect $-0.0004\lesssim \alphaT <-9\times 10^{-5}$ and which (roughly
speaking) explains why the distribution of $\alphaT$ in
Fig.~\ref{fig:pl} is peaked (see the red dashed line).

\par

A last remark is in order at this point. Very often, as already
pointed out, the power spectrum is parametrized as
\begin{equation}
\calP_{\zeta}(k)=A_{_{\rm S}}\left(\frac{k}{k_*}
\right)^{\nS-1+\alphaS/2 \ln (k/k_*)+\cdots },
\end{equation}
and a similar expression for the tensors. Clearly, this is not exactly
what inflation predicts since not expanding
$\nS-1=-2\epsilon_1-\epsilon_2$ (if one works at first order in
slow-roll) in the above formula means in fact keeping an infinite
number of higher order corrections which is clearly inconsistent since
$\nS$ is determined at a fixed order. Of course, since $\nS-1$ is
small, for all practical purposes, this does not impact a lot the
final results.

\subsection{Model Comparison}
\label{subsec:compar}

Let us now turn to model
comparison~\cite{Martin:2010hh,Martin:2014vha,Martin:2013nzq,
  Martin:2014lra}. We would like to determine the models of inflation
that perform the best being given the current CMB data. From a
statistical point of view, this question is subtle. Indeed, suppose we
have two models: $\calM_1$ characterized by one parameter
$\theta_{11}$ and $\calM_2$ characterized by two parameters
$\theta_{21}$, $\theta_{22}$. What does it mean to claim that model
$\calM_1$ is better than model $\calM_2$ (or the opposite)?  Naively,
one could compare the likelihoods of the two models for the values of
the parameters leading to the best fits. But model $\calM_2$ has one
extra parameter and, therefore, one expects this model to
automatically improve the fit. Therefore, in some sense, it would be
``unfair'' to claim that $\calM_2$ is better than $\calM_1$ since it
is ``more complicated''. Moreover, suppose that only for, say,
$\theta_{21}\in \left[10^{-20}, 10^{-19}\right]$ does $\calM_2$ lead
to a good $\chi^2$ while, a priori, $\theta_{12}$ could vary, in say
$[-1,1]$. Suppose, in addition, that this does not happen for
$\calM_1$, namely that for $\theta_{11}$ in its natural range of
variation, the fit is always ``reasonable''. How do we take into
account this wasted parameter space for model $\calM_2$ in our
assessment of the respective performance of the two models?

\par

In order to answer these questions, one recalls that if
$\calL_2(\theta_{21},\theta_{22})\equiv p(D\vert
\theta_{21},\theta_{22}, \calM_2)$ is the likelihood of model
$\calM_2$ ($D$ represents the data, here we have of course CMB data in
mind), then the probability of the parameters
$\theta_{21},\theta_{22},$ can be expressed as (the Bayes'
theorem)~\cite{Trotta:2008qt}
\begin{equation}
\label{eq:postprobapara}
p(\theta_{21},\theta_{22} \vert D,\calM_2)
=\frac{1}{\calE(D\vert \calM_2)}
\calL_2(\theta_{21},\theta_{22})
\pi(\theta_{21}\vert \calM_2)\pi(\theta_{22}\vert \calM_2),
\end{equation}
where $\pi$ represents the prior distributions and $\calE$ is a
normalization factor which depends on the data and the model. We would
like to calculate the probability $p(\calM_2\vert D)$ of model
$\calM_2$ and, therefore, we expect a similar equation to hold, namely
\begin{equation}
\label{eq:postprobamodel}
p(\calM_2\vert D)=\frac{1}{p(D)}p(D\vert \calM_2)\pi(\calM_2),
\end{equation}
where $p(D)$ is a normalization factor depending on the data only and
$\pi$ encodes our a priori information about model $\calM_2$. Clearly,
Eqs.~(\ref{eq:postprobapara}) and~(\ref{eq:postprobamodel}) have the
same structure since they represent two applications of the Bayes's
theorem. To make progress we need to know $p(D\vert \calM_2)$. But
this quantity is in fact easy to calculate since $\int
p(\theta_{21},\theta_{22} \vert D,\calM_2) {\rm d}\theta_{21}{\rm
  d}\theta_{22}=1$, Eq.~(\ref{eq:postprobapara}) leads to
\begin{eqnarray}
\calE(D\vert \calM_2)
&=&\int \calL_2(\theta_{21},\theta_{22})
\pi(\theta_{21}\vert \calM_2)\pi(\theta_{22}\vert \calM_2)
 {\rm d}\theta_{21}{\rm d}\theta_{22}
\nonumber \\ 
&=& \int p(D\vert \theta_{21},\theta_{22},\calM_2)
\pi(\theta_{21}\vert \calM_2)\pi(\theta_{22}\vert \calM_2)
 {\rm d}\theta_{21}{\rm d}\theta_{22}
\nonumber \\
&=& p(D\vert \calM_2).
\end{eqnarray}
Of course the previous considerations apply in general and the
quantity $\calE(D\vert \calM_i)$ is called the Bayesian evidence of
the model $\calM_i$ and its definition
reads~\cite{Trotta:2008qt}
\begin{equation}
p(D \vert \calM_i) \equiv \calE \left(D\vert \calM_i\right)=
\int {\rm d}\theta_{ij}\calL \left(\theta_{ij}\right)
\pi\left(\theta_{ij}\vert \calM_i\right).
\label{eq:defevidence}
\end{equation}
The Bayesian evidence is often normalized to a reference model
$\calM_\usssREF$ and one defines $\Bref{i}\equiv \calE(D\vert\calM_i)
/\calE(D\vert \calM_\usssREF)$. In that case, the posterior
probability of the model $\calM_i$ (for non-committal model priors)
can be re-expressed as
\begin{equation}
p\left(\calM_i\vert D\right)=\frac{\Bref{i}}{\sum _j\Bref{j}}\,.
\end{equation}
In the following, we will give $\Bref{i}$ since this quantity is in
one-to-one correspondence with the probability of the model
$\calM_i$. In particular, one sees that $p(\calM_i)>p(\calM_j)$,
namely model $\calM_i$ is better than $\calM_j$ (or more probable), if
$\calE(\calM_i)>\calE(\calM_j)$ or, equivalently, $\Bref{i}>\Bref{j}$.

\begin{figure}
\begin{center}
\includegraphics[width=12cm]{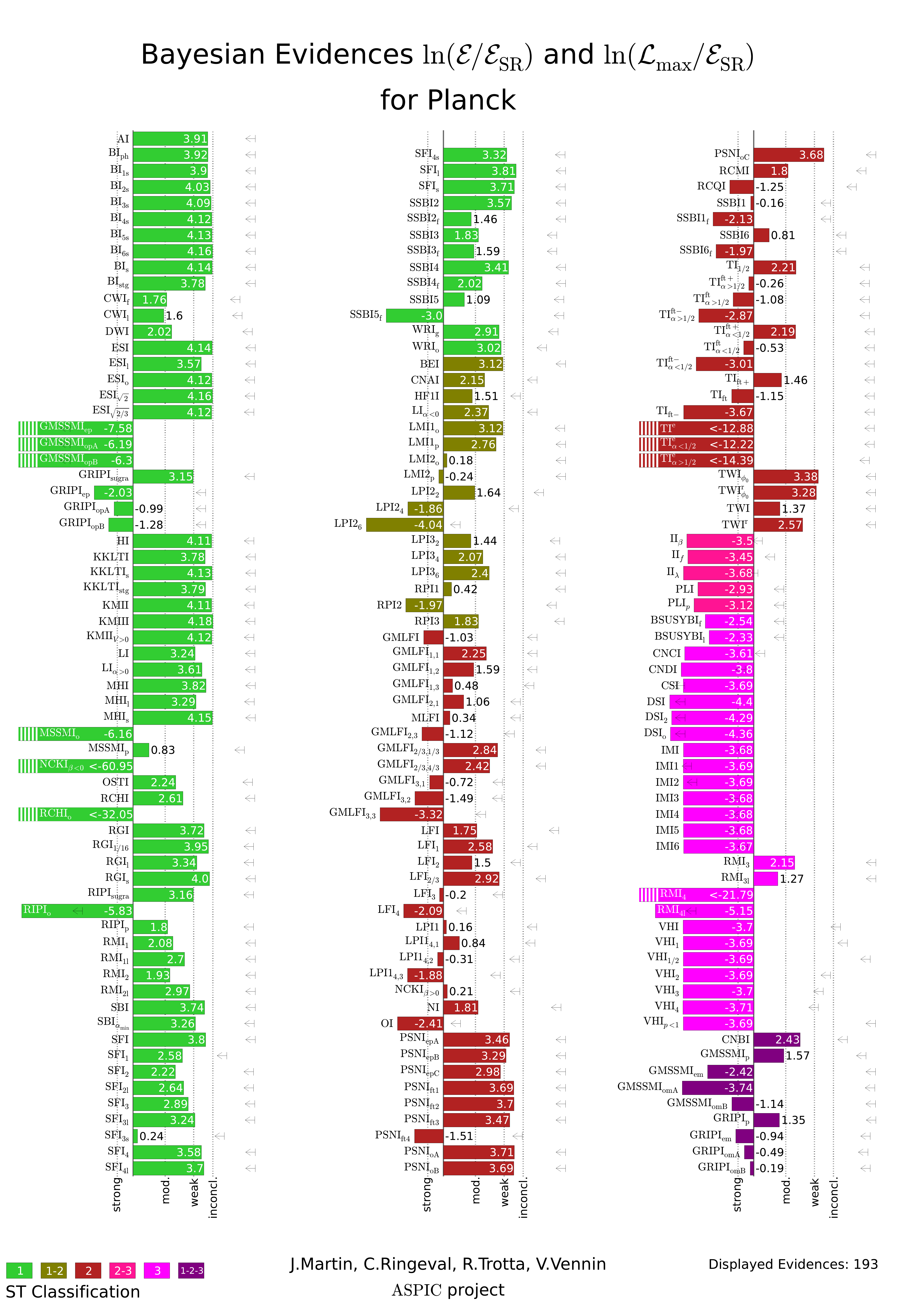}
\end{center}
\caption{$\ln \Bref{i}$ for the all single field slow-roll models with
  minimal kinetic terms. The reference model is taken to be the one
  where the priors are directly chosen on the Hubble flow
  parameters. Each model is represented by a bar, the length of which
  is directly proportional to $\ln \Bref{i}$ (the numerical value of
  $\ln \Bref{i}$ being indicated on the same line). A bar on the left
  means that $\ln \Bref{i}<0$ and a bar on the right that $\ln
  \Bref{i}>0$. The color code refers to the Schwarz-Cesaro Escalante
  classification~\cite{Schwarz:2001vv}. The vertical dotted black line
  indicates the Jeffreys' categories, see the text for more
  explanations. Figure taken from Ref.~\cite{Martin:2013nzq}.}
\label{fig:evidences}
\end{figure}

In order to see why computing the Bayesian evidence answers the
questions asked before and can give a fair estimate of the
performances of a model, let us consider the idealized following
situation. Let us assume that the likelihood function of model
$\calM_1$ has width $\delta \theta_{11}$ and that the prior is flat
and has width $\Delta \theta_{11}$. Since $\pi$ is normalized, we have
$\pi(\theta_{11})=1/\Delta \theta_{11}$. We also assume that the
likelihood is more informative than the prior, namely $\delta
\theta_{11}< \Delta \theta_{11}$. Then, Eq.~(\ref{eq:defevidence}) is 
approximately given by
\begin{equation}
  \calE(D\vert \calM_1)\simeq \calL_{1,{\rm max}}\frac{\delta \theta_{11}}
  {\Delta \theta_{11}}.
\end{equation}
A similar calculation for $\calM_2$ leads to
\begin{equation}
  \calE(D\vert \calM_2)\simeq \calL_{2,{\rm max}}\frac{\delta \theta_{21}}
  {\Delta \theta_{21}}\frac{\delta \theta_{22}}
  {\Delta \theta_{22}}.
\end{equation}
For simplicity, one can take the reference model to be model $\calM_1$
and, of course, one has $\Bref{1}=1$. For $\Bref{2}$, one finds
\begin{equation}
\Bref{2}=\frac{\calL_{2,{\rm max}}}{\calL_{1,{\rm max}}}
\frac{\Delta \theta_{11}}
  {\delta \theta_{11}}
\frac{\delta \theta_{21}}
  {\Delta \theta_{21}}\frac{\delta \theta_{22}}
  {\Delta \theta_{22}}.
\end{equation}
On this last equation, we see that deciding whether model $\calM_1$ is
better or worst than $\calM_2$ does not reduce to the comparison of
the likelihood function at the best fit, $\calL_{2,{\rm
    max}}/\calL_{1,{\rm max}}$ but that this ratio is corrected by a
factor which describes how much parameter space has been wasted. So
the best model is not the one which has the largest $\chi^2$ but the
one which achieves the best compromise between quality of the fit and
simplicity of the theoretical description.

\par

As explained before, here, we focus on single-field slow-roll
inflationary models (with minimal kinetic term) only. At this stage,
the strategy is clear: one must evaluate the Bayesian evidence of each
of these models in order to rank them according to their ability to
fit the data. This first requires to identify all models of this type
and this was recently done in {\EI}, see
Ref.~\cite{Martin:2014vha}. In this work, about $200$ models have been
identified. A model corresponds to a specific choice of potential and
of priors for its parameters. Two different models can have the same
potential but different priors. Each model is denoted by an acronym
according to the terminology introduced in Ref.~\cite{Martin:2013nzq}
and, in the present article, we just make use of this convention.  A
detailed justification of the priors chosen can also be found in that
reference. In Fig.~\ref{fig:evidences}, we have represented the
Bayesian evidence of the different models (being given the Planck
2013 data) by an horizontal bar the length of which is proportional to
$\ln \Bref{i}$, see also the caption of Fig.~\ref{fig:evidences} and
Refs.~\cite{Martin:2014vha,Martin:2013nzq,Ringeval:2013lea,
  Martin:2014lra}. In order to translate the numerical value of the
evidence into strength of belief, we introduce the Jeffrey's
scale~\cite{Trotta:2008qt}. If $\vert \ln \Bref{i}\vert <1$, then the
model is in the ``inconclusive zone'', if $1<\vert \ln \Bref{i}\vert
<2.5$, then it is in the ``weak evidence zone'', if $2.5<\vert \ln
\Bref{i}\vert <5$, then it is in the ``moderate evidence zone'' and,
finally, if $\vert \ln \Bref{i}\vert <5$, it is in the ``strong
evidence zone''. If the reference model is taken to be the best model,
then, by definition all $\ln \Bref{i}$ are negative. In that case, the
best models are those in the inconclusive zone and those in the
``strong evidence zone'' can be considered as ruled out.

\par

In Fig.~\ref{fig:evidences}, we see that the best Planck 2013 model is
$\kmiii$ and that $52$ models end up being in the inconclusive zone,
namely: $\kmiii$, $\esisqrtTWO$, $\biSIXs$, $\mhis$, $\bis$, $\esi$,
$\biFIVEs$, $\kkltis$, $\kmiivp$, $\biFOURs$, $\esio$,
$\esisqrtTWOTHREE$, $\kmii$, $\hi$, $\biTHREEs$, $\biTWOs$, $\rgis$,
$\rgiONEONESIX$, $\bi$, $\ai$, $\biONEs$, $\mhi$, $\sfil$, $\sfi$,
$\kkltistg$, $\bistg$, $\kklti$, $\sbi$, $\rgi$, $\sfis$, $\psnioA$,
$\sfiFOURl$, $\psniftTWO$, $\psnioB$, $\psniftONE$, $\psnioC$, $\lip$,
$\sfiFOUR$, $\esil$, $\ssbiTWO$, $\psniftTHREE$, $\psniepA$,
$\ssbiFOUR$, $\twiAONE$, $\rgil$, $\sfiFOURs$, $\mhil$, $\psniepB$,
$\twiATWO$, $\sbialphamin$, $\li$, $\sfiTHREEl$. They represent $\sim
26\%$ of the models analyzed. We also find that $21\%$ of the models
are in the ``weak evidence zone'', $17\%$ in the ``moderate evidence
zone'' and $34\%$ in the ``strong evidence zone''. Planck 2013 is
therefore able to rule out about one third of the inflationary
models. Model comparison with Planck $2015$ data cannot yet be done
since the scientific products are not delivered. However, given the
consistency of these two data sets, we do not expect very different
results.

\par

We have seen before that the ``winner'' is $\kmiii$ which is a string
inspired model with the following potential~\cite{Conlon:2005jm}
\begin{equation}
V(\phi)=M^4\left[1-\alpha\left(\frac{\phi}{\Mp}\right)^{4/3}
{\rm e}^{\beta (\phi/\Mp)^{4/3}}\right],
\end{equation}
where $\alpha $ and $\beta$ are two free parameters. Its heir apparent
is $\esisqrtTWO$ the potential of which is given
by~\cite{Stewart:1994ts,Dvali:1998pa,Cicoli:2008gp}
\begin{equation}
V(\phi)=M^4\left(1-{\rm e}^{-\sqrt{2}\phi/\Mp}\right).
\end{equation}
The fourteenth on the list is the Starobinsky
model~\cite{Starobinsky:1980te,Bezrukov:2007ep}, namely
\begin{equation}
V(\phi)=M^4\left(1-{\rm e}^{-\sqrt{2/3}\phi/\Mp}\right)^2.
\end{equation}
Actually, all these models being in the ``inconclusive zone'', the
difference between their Bayesian evidence is not significant. This
means that, for instance, one should view the Starobinsky model as as
good as $\kmiii$. In fact, the main common point between all these
scenarios is that they all possess a ``plateau-like'' potential,
meaning that $V_{\phi}(\phi)\rightarrow 0$ as $\phi \rightarrow
\infty$. We conclude that Planck 2013 has been able to constrain the
shape of the inflationary potential, a truly remarkable achievement
when one remembers that inflation can take place at $10^{15}\GeV$, and
certainly something impossible to do in an accelerator!

\par

We have seen before that, in order to explain the data, we do not need
to consider models more complicated than single field slow-roll
inflation with a minimal kinetic term. This does not mean that more
complicated models are ruled out (in the frequentist point of view) in
the sense that, with a carefully chosen set of parameters, they can
lead to good fits. However, from the previous considerations, we see
that those models must have a very ``bad'' Bayesian
evidence. Computing the evidence of those more complicated models is
certainly a difficult task (for instance, for models predicting a non
negligible level of Non-Gaussianities, one would need to take into
account the higher order correlation functions). But, in fact, we do
not need to carry out such a calculation which, at this stage appears
to be useless. Indeed we know in advance that they are much ``worse''
than single-field models because of their huge wasted parameter
space~\cite{Giannantonio:2014rva}. It is sufficient to know that they
all are in the ``strong evidence zone'' and, clearly, we are not much
interested in knowing the ranking in this Jeffreys category since the
models are ruled out (in the Bayesian sense) anyway.

\begin{figure}
\begin{center}
\includegraphics[width=\wdblefig]{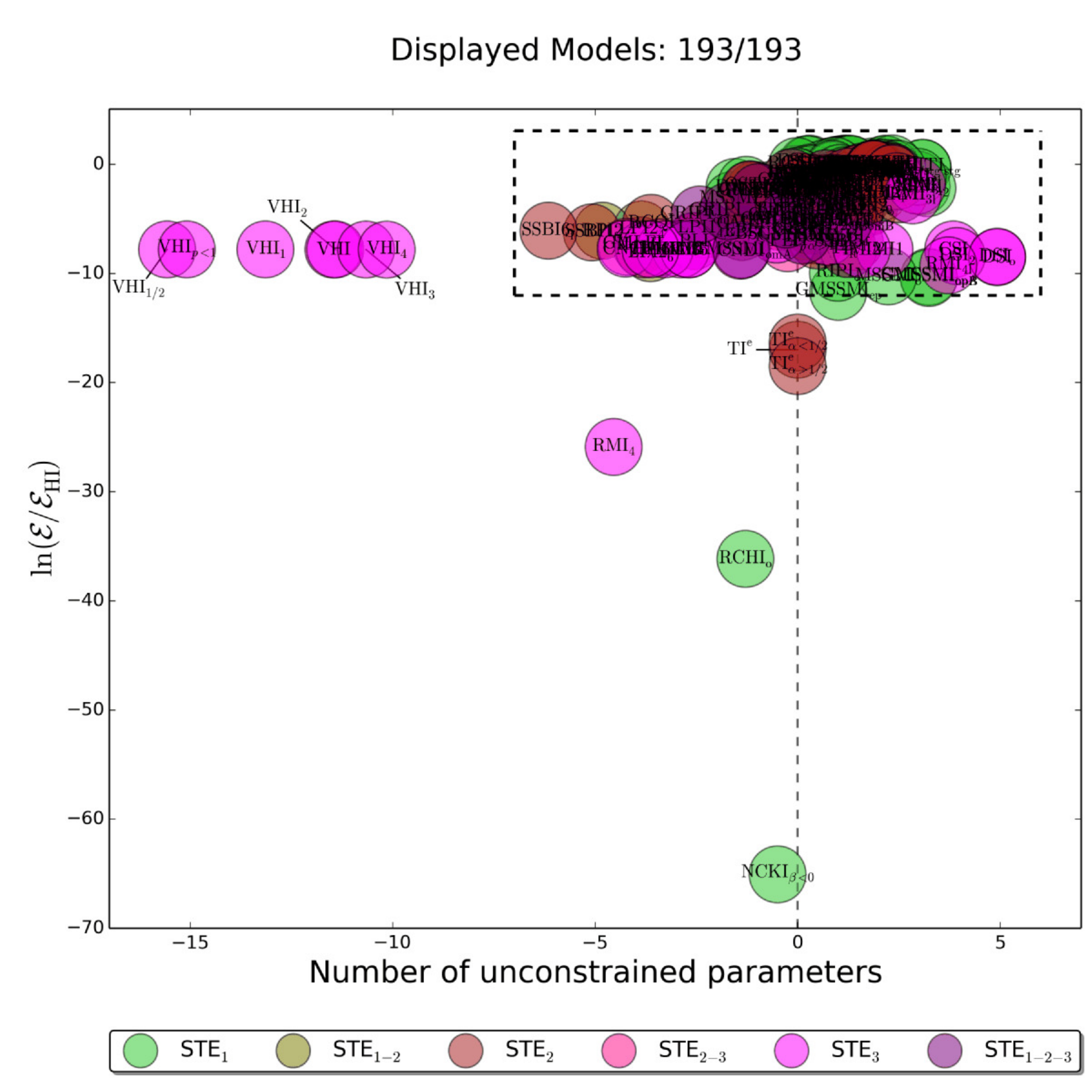}
\includegraphics[width=\wdblefig]{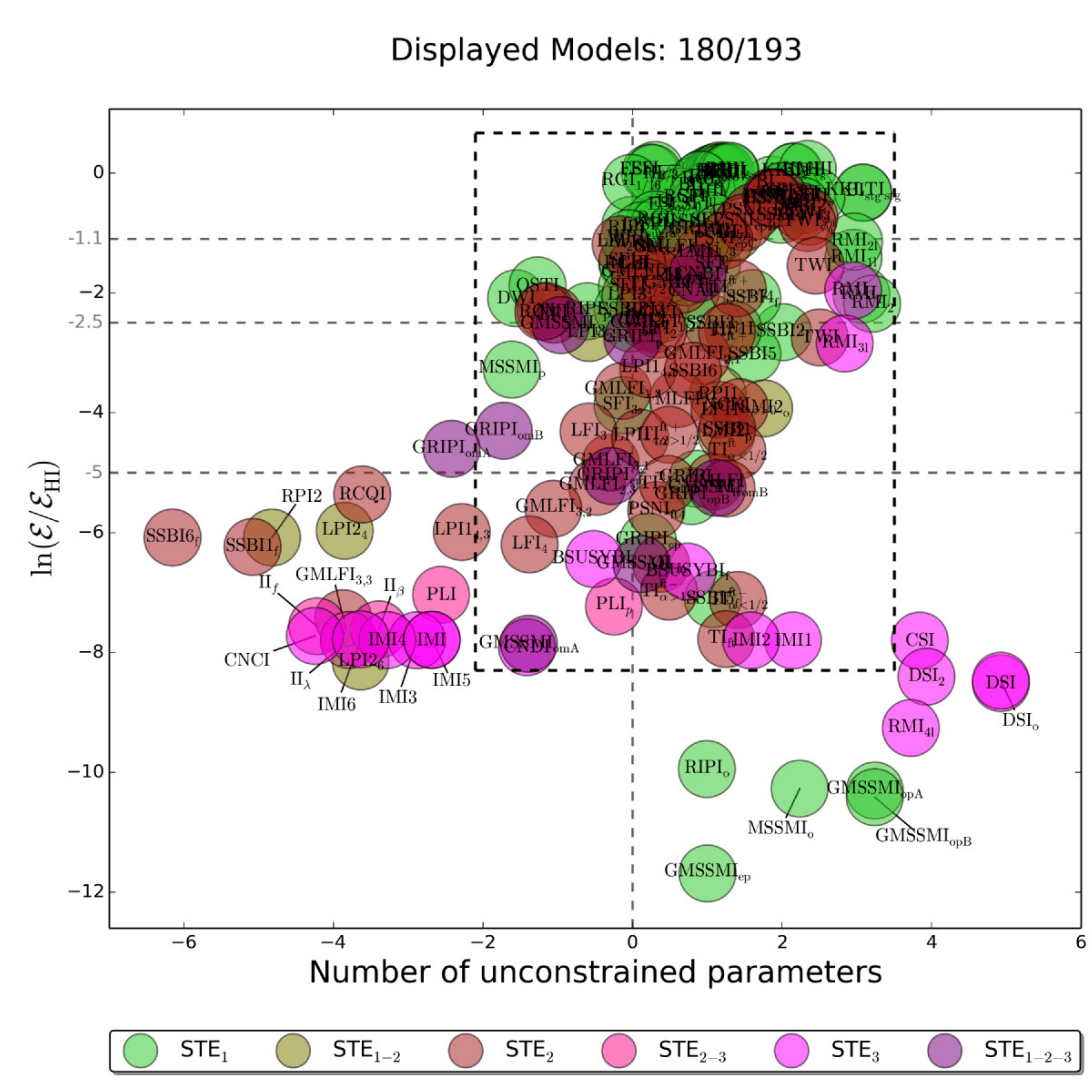}
\includegraphics[width=\wdblefig]{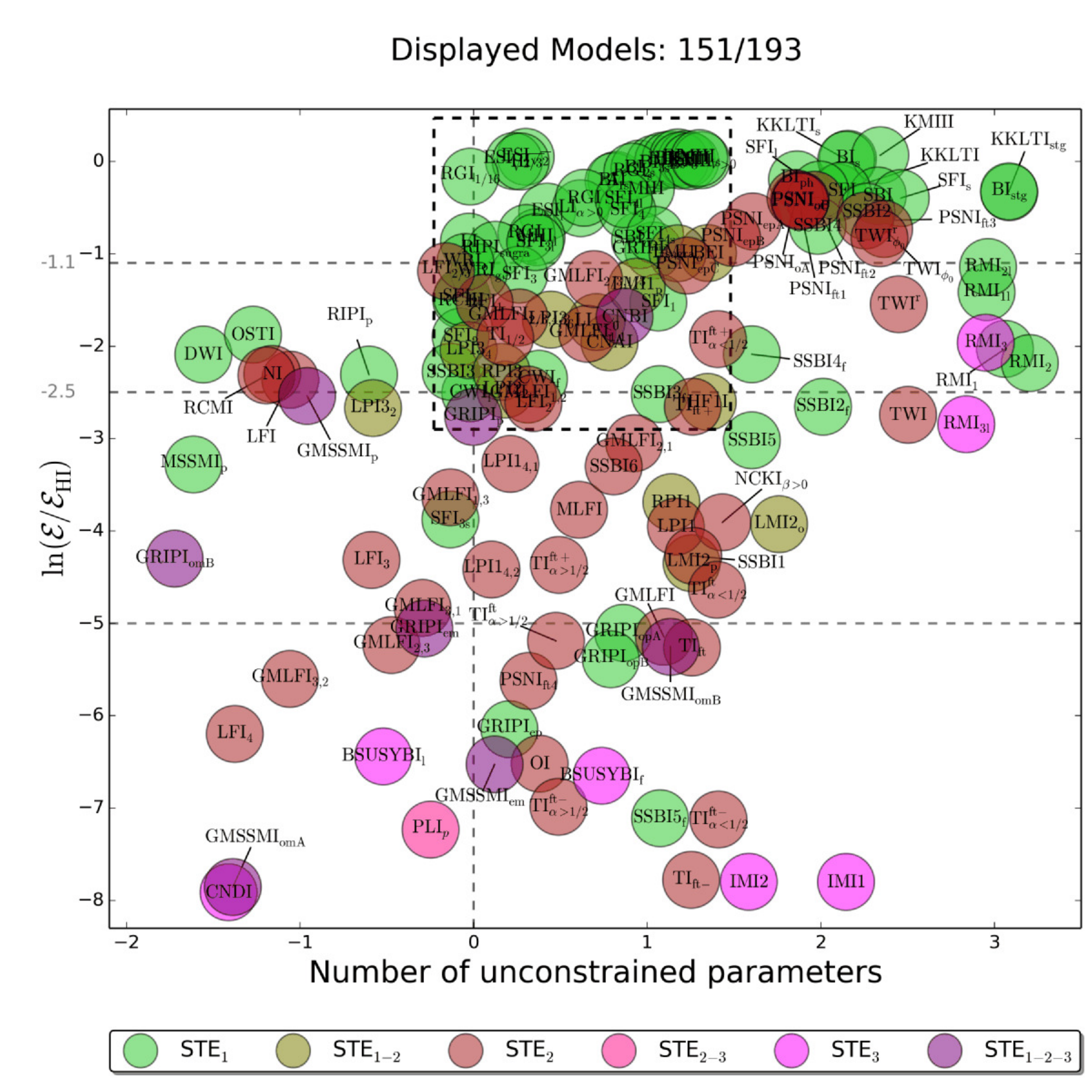}
\includegraphics[width=\wdblefig]{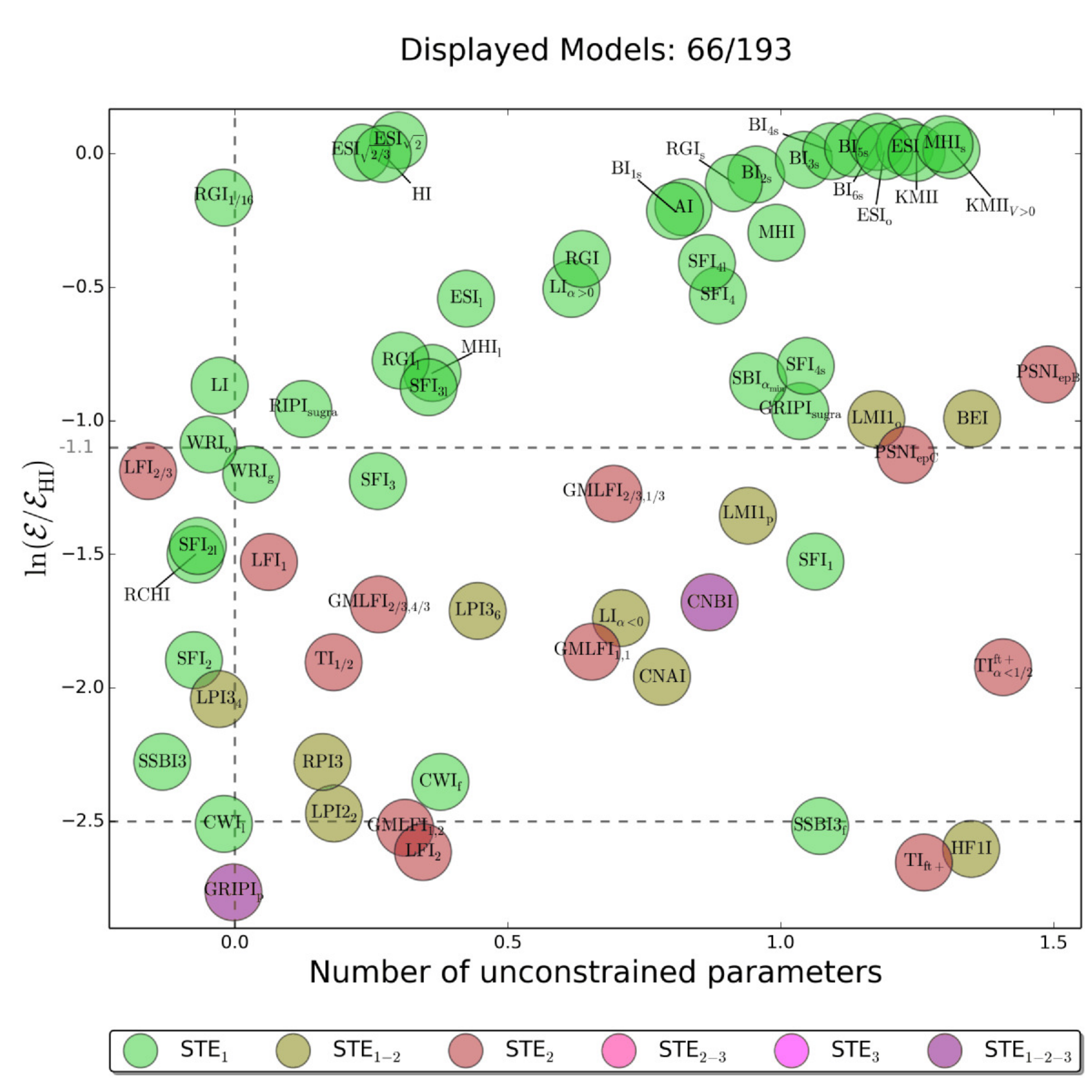}
\end{center}
\caption{Inflationary models in the space $\left(N_{\rm uc}, \ln
  \Bref{i} \right)$. Each model is represented by a circle (the radius
  of which has no meaning) with its acronym, taken from
  Ref.~\cite{Martin:2013nzq}, written inside. The four panels
  corresponds to successive zooms towards the best region (indicated
  by the dashed rectangles), the one with $0<N_{\rm uc}<1$, namely
  where all the parameters are constrained by the data and a large
  value of the evidence, namely the model achieves a good fit without
  wasting parameter space. Figures taken from
  Ref.~\cite{Martin:2013nzq}.}
\label{fig:compevid}
\end{figure}

\begin{figure}
\begin{center}
\includegraphics[width=10cm]{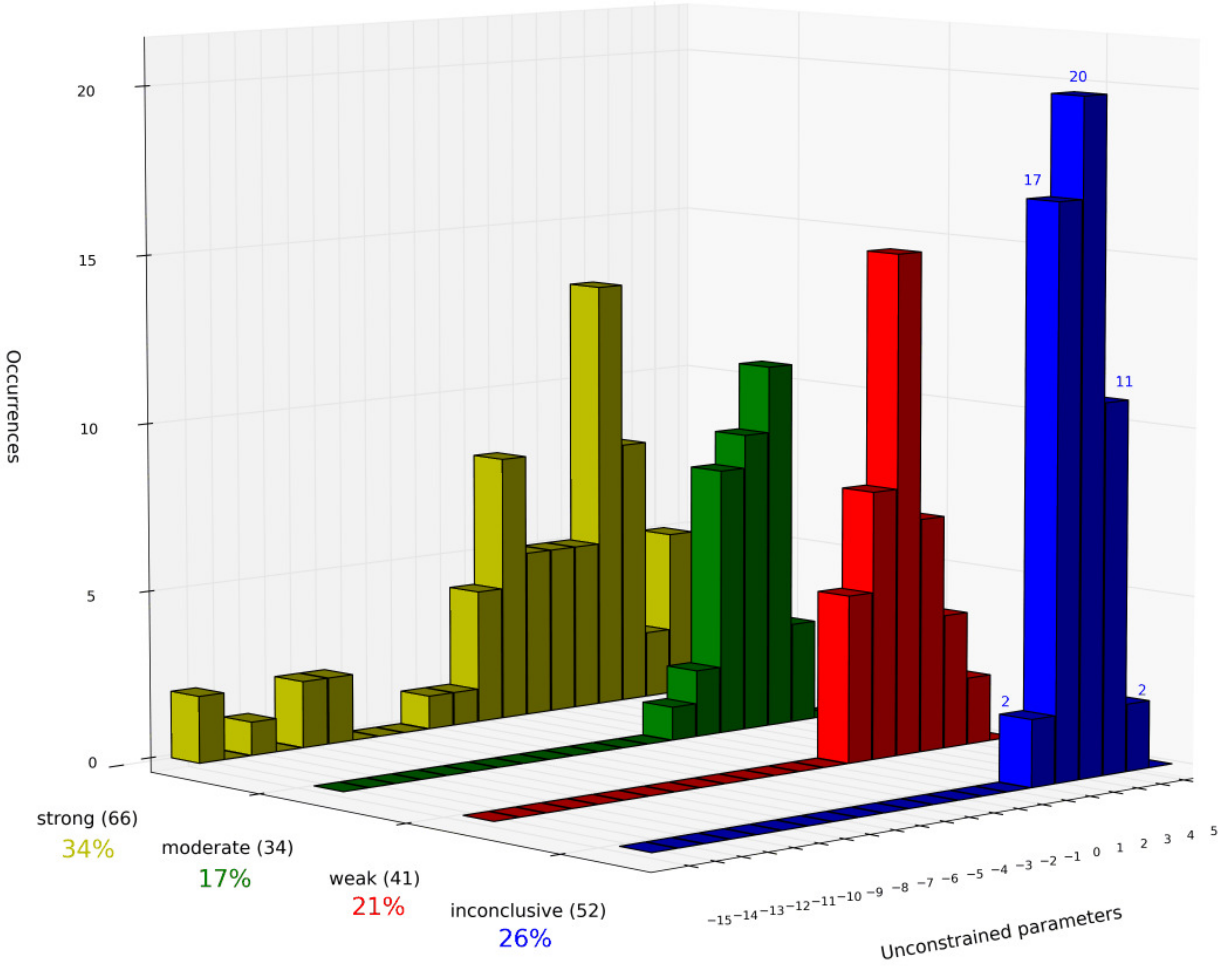}
\end{center}
\caption{Occurrences of inflationary models in the four Jeffreys
  categories for different values of the unconstrained number of
  parameters. Figure taken from Ref.~\cite{Martin:2013gra}.}
\label{fig:histo}
\end{figure}

We have seen how the Bayesian evidence allows us to rank the various
inflationary models. However, two models with a different number of
parameters can have the same evidence if the extra parameters are not
constrained by the data. This is certainly not a desirable property as
the model with less parameters is clearly simpler and, therefore,
should be favored. In order to break this degeneracy, we now introduce
the Bayesian complexity~\cite{Kunz:2006mc}. For a model $\calM_i$, it
is defined by~\cite{Kunz:2006mc}
\begin{equation}
\label{eq:defcomplex}
C_{\rm b}^i=\left \langle -2\log {\cal L}
\left(\theta_{ij}\right)\right \rangle +2\log {\cal L}
\left(\theta_{ij}^{\rm ML}\right),
\end{equation}
where $\langle \cdot \rangle$ means averaging over the posteriors and
$\theta_{ij}^{\rm ML}$ represents the maximum likelihood estimate of
the model's parameters. One can easily show, see for instance
Ref.~\cite{Martin:2013gra}, that the number of unconstrained
parameters, given a data set, can be expressed as
\begin{equation}
N_{\rm uc}^i=N_{\rm param}^i-C_{\rm b}^i,
\end{equation}
where $N_{\rm param}^i$ represents the number of free parameters of
model $\calM_i$. We see that this gives us a new criterion to
discriminate the various models since a model such that $0<N_{\rm
  uc}^i<1$ ought to be preferred. The Bayesian complexities (given the
Planck 2013 data) of all the {\EI} models have been computed in
Ref.~\cite{Martin:2013nzq}. In Fig.~\ref{fig:compevid}, we have
represented these scenarios in the space $\left(N_{\rm uc}, \ln
\Bref{i}\right)$. It can be noticed that, among the models in the
Planck 2013 inconclusive zone, those with a minimal number of
unconstrained parameters are: $\esisqrtTWO$, $\esisqrtTWOTHREE$,
$\hi$, $\biTWOs$, $\rgis$, $\ai$, $\biONEs$, $\mhi$, $\rgi$,
$\sfiFOURl$, $\lip$, $\sfiFOUR$, $\esil$, $\rgil$, $\mhil$,
$\sbialphamin$ and $\sfiTHREEl$. The number of preferred models is now
$17$, that is to say $\sim 9\%$ of the total number of models analyzed
here. Of course, as already remarked, these models are all of the
plateau shape. The distribution of models in the four Jeffreys
categories versus the number of unconstrained parameters is summarized
in Fig.~\ref{fig:histo}.

\subsection{Reheating}
\label{subsec:reheat}

\begin{figure}
\begin{center}
\includegraphics[width=10cm]{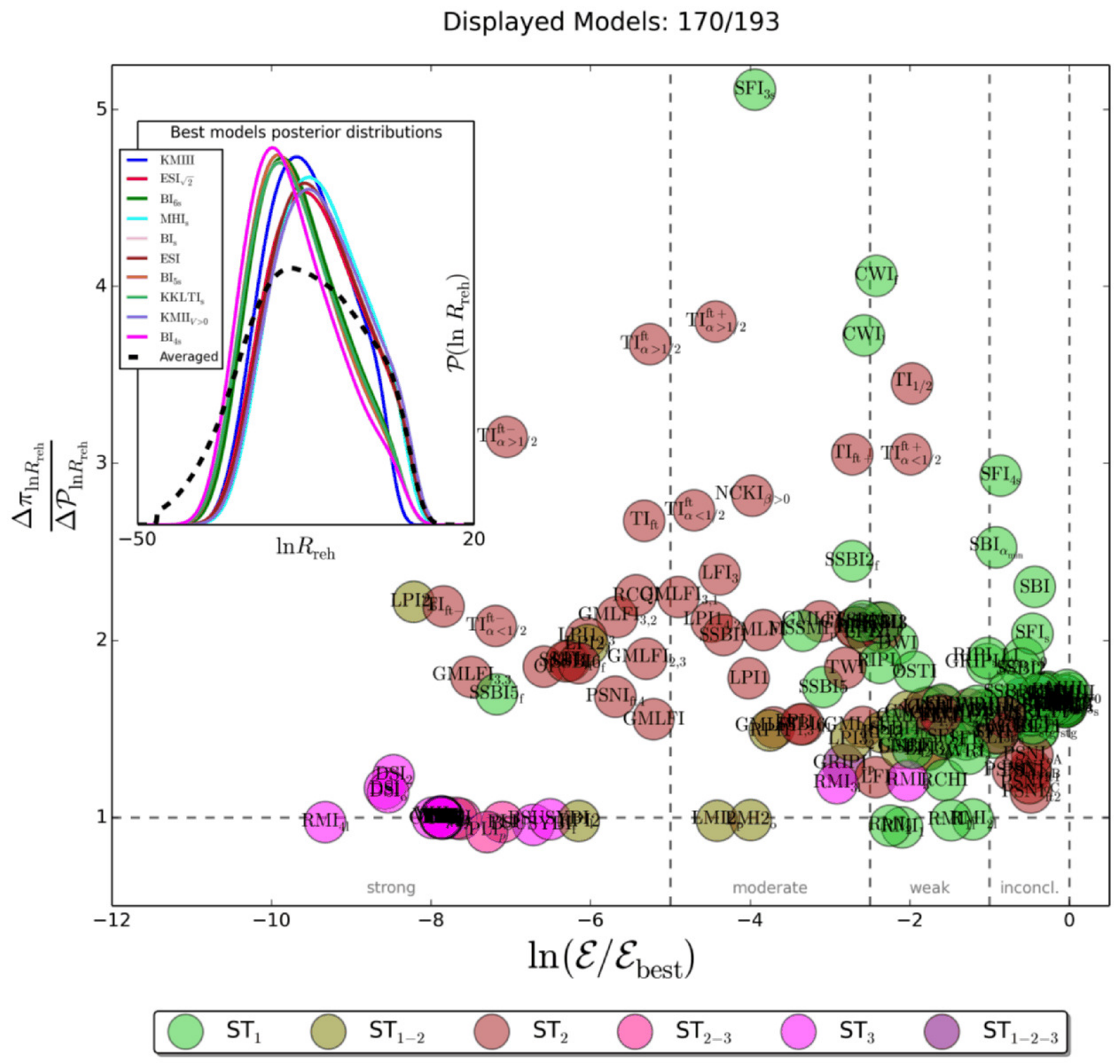}
\end{center}
\caption{The quantity $\Delta \pi_{\ln \Rreh}/\Delta \calP_{\ln
    \Rreh}$, quantifying how much the reheating is constrained, versus
  the Bayesian evidence for {\EI} models (each model is represented by
  a circle the size of which has no meaning). The inset shows the
  posterior distribution of the reheating parameter for the ten best
  Planck 2013 models. Figure taken from Ref.~\cite{Martin:2014nya}.}
\label{fig:reheat}
\end{figure}

\begin{figure*}
\begin{center}
\includegraphics[width=\wdblefig]{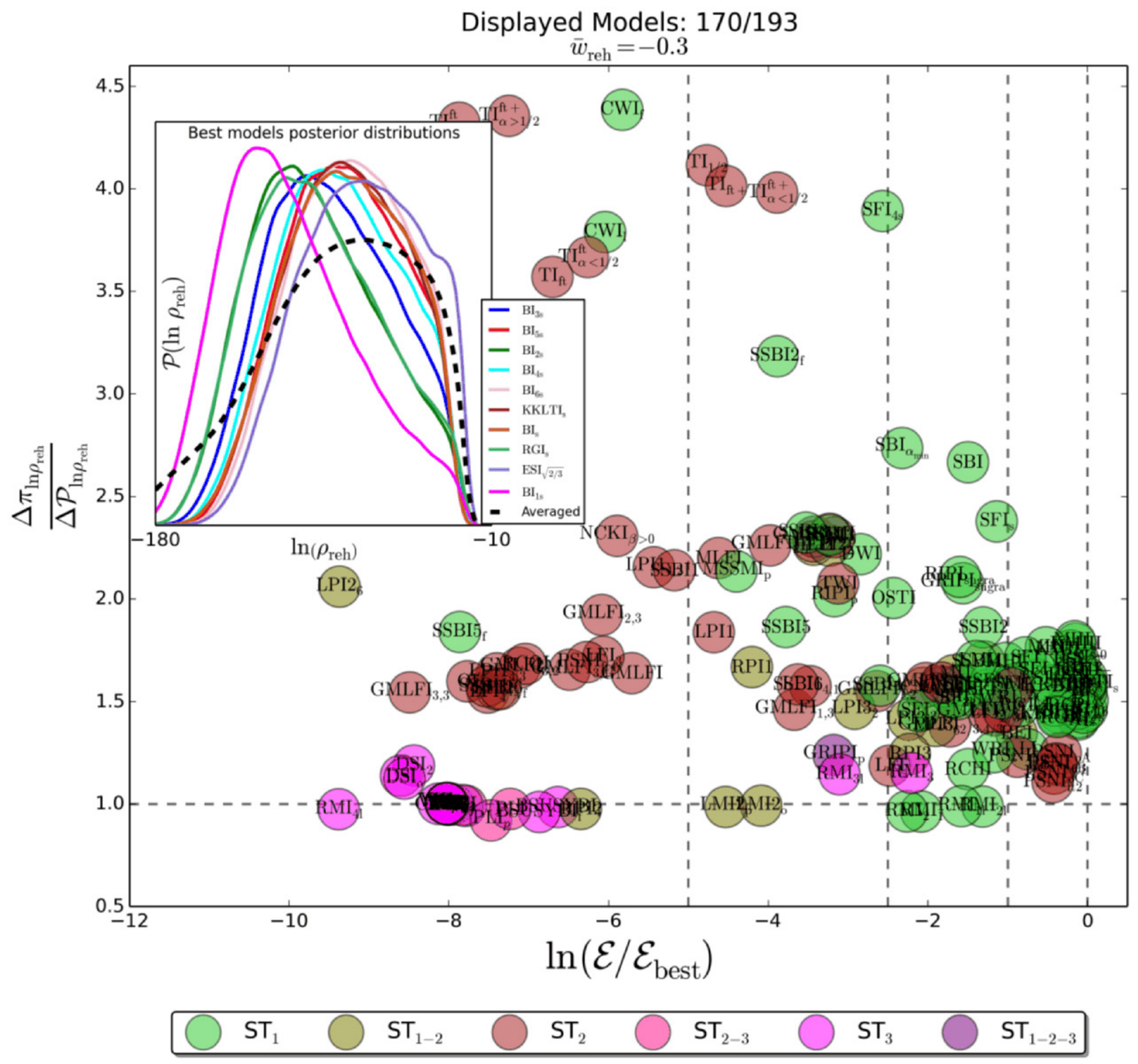}
\includegraphics[width=\wdblefig]{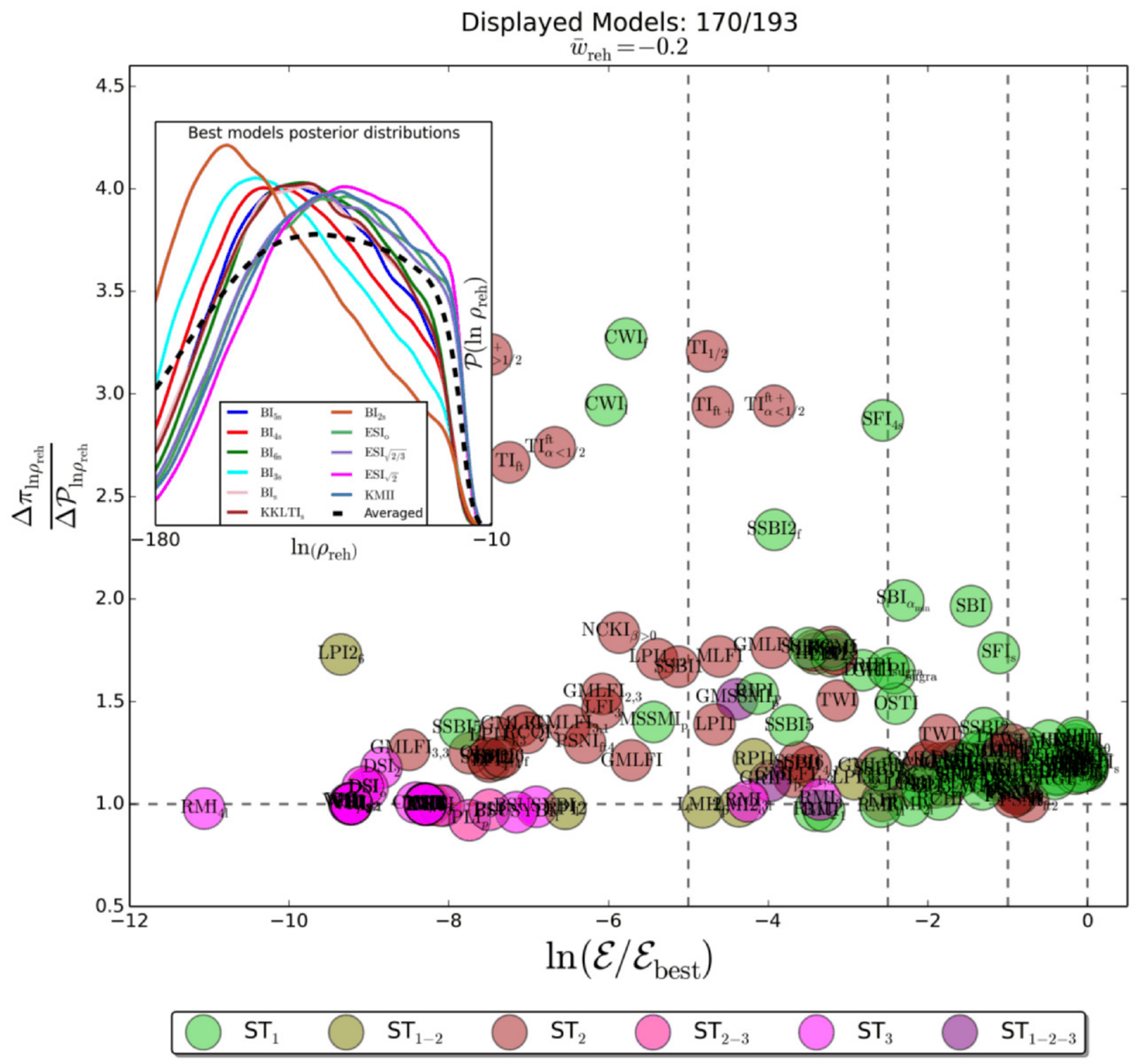}
\includegraphics[width=\wdblefig]{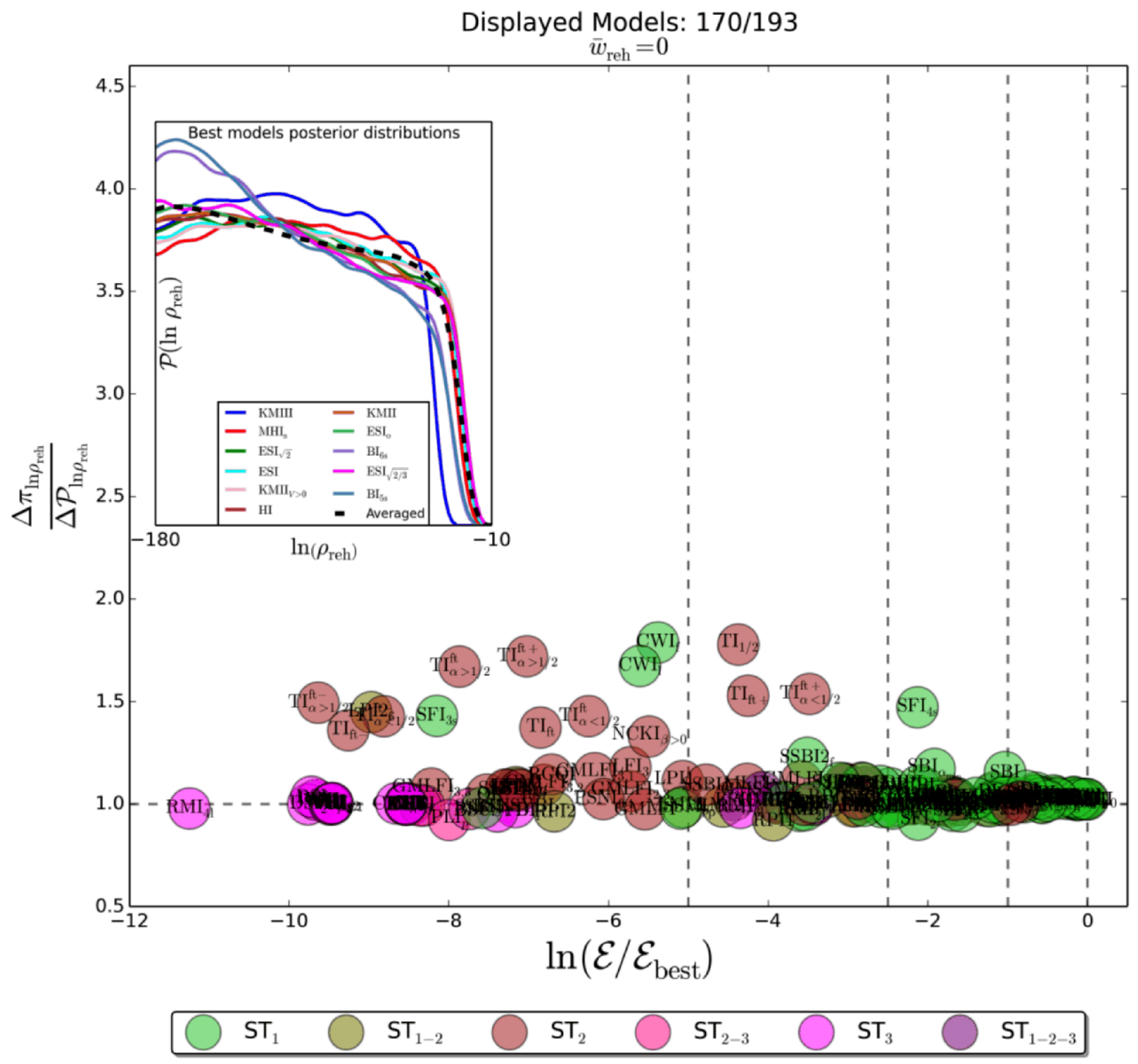}
\includegraphics[width=\wdblefig]{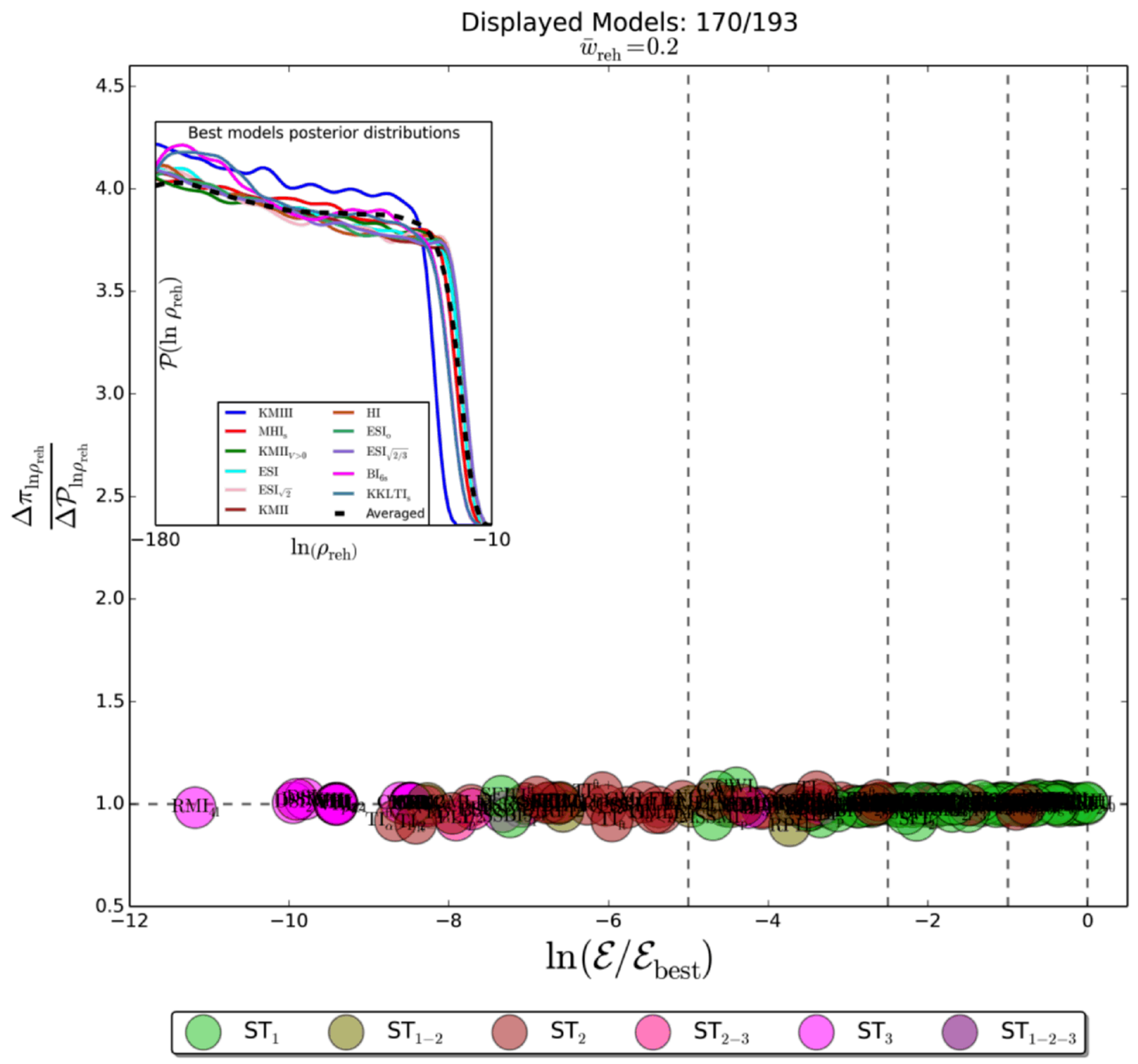}
\caption{Same as in Fig.~\ref{fig:reheat} but assuming the mean
  equation of state during reheating is known. The prior-to-posterior
  width for the reheating energy density $\ln(\rhoreh/\Mp^4)$ is
  represented assuming four values of the mean equation of state
  $\wrehbar$, namely $\wrehbar=-0.3$ (top left panel), $\wrehbar=-0.2$
  (top right panel), $\wrehbar=0$ (bottom left panel) and
  $\wrehbar=0.2$ (bottom right panel). Figures taken from
  Ref.~\cite{Martin:2014nya}.}
\label{fig:wfix}
\end{center}
\end{figure*}

We now describe the constraints on reheating that can be inferred from
the Planck 2013 data. This question was recently studied in
Ref.~\cite{Martin:2014nya}. In Sec.~\ref{subsec:reheatparam}, we have
seen that, as far as CMB data are concerned, reheating can be entirely
described by the parameter $\Rrad$, see Eq.~(\ref{eq:defRrad}) or,
equivalently, $\Rreh$, see Eq.~(\ref{eq:defR}). For each inflationary
model, Ref.~\cite{Martin:2014nya} has calculated the posterior
distribution of the parameter $\ln \Rreh $. In order to estimate how
much reheating is constrained, it is convenient to introduce the ratio
$\Delta \pi_{\ln \Rreh}/\Delta \calP_{\ln \Rreh}$. In this formula,
$\Delta \pi_{\ln \Rreh}$ is the standard width of the prior while
$\Delta \calP_{\ln \Rreh}$ is the standard width of the posterior
distribution. Therefore if $\Delta \pi_{\ln \Rreh}/\Delta \calP_{\ln
  \Rreh}=1$, the posterior is as wide as the prior and reheating is
not constrained at all. If, however, $\Delta \pi_{\ln \Rreh}/\Delta
\calP_{\ln \Rreh}>1$, then the posterior distribution is more peaked
than the prior and there is information gain. Clearly, the larger the
ratio $\Delta \pi_{\ln \Rreh}/\Delta \calP_{\ln \Rreh}$, the more
peaked the posterior.

\par

The prior on $\ln \Rreh$ has to be chosen carefully and must be
justified by physical considerations. Clearly, the energy density at
the end of reheating must be smaller than that at the end of inflation
and larger than at the BBN time where $\rho_{\rm
  nuc}=\left(10\MeV\right)^4$. Therefore, we require $\rho_{\rm
  nuc}<\rho_{\rm reh}<\rho_{\rm end}$. For the mean equation of state,
we take $-1/3<\wrehbar<1$ since, by definition, reheating is a non
accelerated phase of expansion. As a consequence, one can show that
this leads to
\begin{equation}
\ln \left(\frac{\rho_{\rm nuc}^{1/4}}{\Mp}\right)
<\ln \Rreh< \ln \left(\frac{\rho_{\rm nuc}^{1/4}}{\Mp}\right)
+\frac43 \ln \left(\frac{\rho_{\rm end}^{1/4}}{\Mp}\right).
\end{equation}
The order of magnitude of $\Rreh$ being unknown, we choose a Jeffreys
prior in the above range. Notice that this differs from what was done
in the Planck 2013 paper~\cite{Ade:2013uln}. Indeed, in that work,
specific reheating scenarios were considered such as instantaneous
reheating or ``restrictive reheating'' where, apparently without a
strong justification, the reheating energy density is fixed to
$10^{9}\GeV$. Moreover, it seems that a prior on the quantity $\Delta
N_*$ was chosen which is clearly awkward since it does not necessarily
guarantee that the two physical conditions on $\rho_{\rm reh}$ and
$\wrehbar$ discussed previously are valid\footnote{An additional
  problem comes from the description of Ref.~\cite{Martin:2010kz} by
  Ref.~\cite{Ade:2013uln}. Indeed it is claimed in this last paper
  that, for large field models where $V(\phi)\sim \phi^n$,
  Ref.~\cite{Martin:2010kz} considered only scenarios of reheating for
  which $\wrehbar=(n-2)/(n+2)$, a wrong claim as can be checked
  directly by reading Ref.~\cite{Martin:2010kz}.}. In the Planck
$2015$ paper~\cite{Ade:2015oja}, it seems that this weird approach has
been given up. The new method now seems closer to what is done in the
present article. Notice, however, that, if the prior on the reheating
energy density appears reasonable, only specific values of $\wrehbar$
are considered which is, of course, not the most general case. Let us
also remark that Ref.~\cite{Ade:2015oja} introduces an equation of
state parameter during reheating, denoted $w_{\rm int}$, called the
``effective equation of state'' but without defining it precisely. In
particular, it is difficult to know if it is equal to the parameter
introduced in Eq.~(\ref{eq:wrehbar}), which is the correct parameter
that ought to be used and was introduced for the first time in
Ref.~\cite{Martin:2010kz}.

\par

In Fig.~\ref{fig:reheat}, we have represented each {\EI} model in the
space $\left(\Delta \pi_{\ln \Rreh}/\Delta \calP_{\ln \Rreh}, \ln
\Bref{i} \right)$: good models are on the right and models for which
reheating is constrained are on the top. The horizontal dashed line
$\Delta \pi_{\ln \Rreh}/\Delta \calP_{\ln \Rreh}=1$ locates the models
for which reheating is not constrained. In order to globally assess the
value of the constraints, we can define the following quantity
\begin{equation}
\label{eq:defratio}
\left \langle \frac{\Delta \pi_{\ln \Rreh}}{\Delta \calP_{\ln \Rreh}}
\right \rangle\equiv 
\frac{1}{\sum_j\calE_j}
\displaystyle \sum_i
\calE_i
\left(\frac{\Delta \pi_{\ln \Rreh}}{\Delta \calP_{\ln \Rreh}}
\right)_i,
\end{equation}
which is the mean value of $\Delta \pi_{\ln \Rreh}/\Delta
\calP_{\ln\Rreh}$ weighted by the Bayesian evidence, {\ie} the mean
value in the space of models. This is a fair estimate since disfavored
models will not contribute a lot to this quantity due to their small
evidence. Numerically, the Planck 2013 data are such that
\begin{equation}
\label{eq:consR}
\left \langle \frac{\Delta
  \pi_{\ln\Rreh}}{\Delta \calP_{\ln\Rreh}} \right \rangle \simeq 1.66
\end{equation}
which, therefore, indicates that reheating is indeed constrained.

\par

It is also interesting to assume that the mean equation of state is
known. In that case, the parameter $\Rreh$ only depends on the energy
density at the end of reheating or, equivalently, on the reheating
temperature. In Fig.~\ref{fig:wfix}, we have represented the similar
quantities as in Fig.~\ref{fig:reheat} for four different values of
$\wrehbar$, namely $\wrehbar=-0.3, -0.2, 0, 0.2$. The most striking
feature of this plot is that, for positive values of $\wrehbar$, the
models tend to cluster around the horizontal line $\Delta \pi_{\ln
  \Rreh}/\Delta \calP_{\ln\Rreh}=1$. This indicates that, for those
values of the mean equation of state, reheating is not constrained. As
a consequence, the number obtained in Eq.~(\ref{eq:consR}) comes in
fact from a region in parameter space where $\wrehbar<0$. This
conclusion makes sense since, for $\wrehbar<0$, the dispersion of the
predictions in the $(r,\nS)$ space is much bigger than for positive
equation of state. More details can be found in
Ref.~\cite{Martin:2014nya}, in particular concrete bounds on the
reheating temperature for different models.

\par

Concluding, the reheating phase is already constrained by the Planck
$2013$ data. The precise values of the allowed reheating temperatures
depend on the model under consideration and on the mean equation of
state. If $\wrehbar>0$, the constraints are very mild. It is also
worth noticing that two identical models with two different reheating
histories can have different Bayesian evidence. This means that, given
the accuracy of the CMB measurements, reheating now needs to be
properly included in data analysis.

\section{Conclusion}
\label{sec:conclusion}

In this last section, we briefly summarize what we have learned about
inflation in the recent years. Inflation is a ``violent'' phenomenon
since it could occur at energies as high as the Grand Unified Theory
scale, \ie $\sim 10^{16}$ GeV. It is thus quite remarkable to be able
to say something about physics at such a high energy scale. The
picture that seems to emerge from the recent high accuracy
astrophysical measurements is that inflation is realized in its
simplest version, namely single-field slow-roll with a minimal kinetic
term. Additional features, such as the presence of several fields or
non-minimal kinetic term, which may appear as (natural) consequences
of embedding inflation in high energy physics, do not seem to be
relevant. If, indeed, inflation is really realized in its vanilla
version, an important challenge will be (is) to understand, from the
high energy point of view, why these extra ingredients are in fact not
present. Also, important questions such as the physical nature of the
inflaton field remains unanswered.

\par

The shape of the potential is also constrained and appears to be of
the ``plateau shape'', a typical example of this class of scenarios
being the Starobinsky model. Popular models such as monomial
potentials are now disfavored. 

\par

Interestingly enough, inflationary reheating is also constrained by
the Planck data. The constraints are model dependent and correspond to
an average reduction of the prior-to-posterior of about $40\%$.

\par

Given this situation, what should be done to increase our knowledge of
inflation? It is clear that in order to measure more precisely the
shape of the potential, one needs to constrain the values of the
Hubble flow parameters $\epsilon_n$. So far, we only have a good
measurement of the scalar spectral index which is a specific
combination of $\epsilon_1$ and $\epsilon_2$, namely
$\nS=1-2\epsilon_1-\epsilon_2$. To measure $\epsilon_1$ and
$\epsilon_2$ separately, one needs another observable. A more accurate
measurement of the scalar power spectrum cannot really do the job
since it involves an additional parameter, $\epsilon_3$, see
Eq.~(\ref{eq:alphast}). We are therefore left with either the
tensor-to-scalar ratio $r$, which is directly proportional to
$\epsilon_1$, see Eq.~(\ref{eq:defr}), or the bispectrum which depends
on $\epsilon_1$ and $\epsilon_2$ in a different combination than the
spectral index, see Eq.~(\ref{eq:finalBrsr}) and Eqs.~(\ref{eq:f1}),
(\ref{eq:f2}), (\ref{eq:f3}) and~(\ref{eq:f7}).

\par

Measuring primordial Non-Gaussianities has one great advantage: we
already know in advance where one should find the signal. If one dares
an analogy, it is like searching for the Higgs boson. We know that if
it is not found in a specific window, the consequences would be
drastic. However, the shortcomings is that the amplitude of the
signal, $\fnl\simeq 0.01$, is so small that it is not clear whether it
is technologically feasible. On the other hand, improving the limits
on Non-Gaussianities could be very rewarding. Many non-vanilla
scenarios predict $\fnl\simeq 1$ and reaching this limit could allow
us to rule out single field slow-roll models!

\par

Measuring the tensor-to-scalar ratio is the other possibility. It
requires to measure the tensor contribution which can be done through
a detection of $B$-mode CMB polarization. At the moment, there are
considerable efforts in this direction. The first claim of a detection
of primordial gravity waves was of course made by the BICEP2
team~\cite{Ade:2014xna}. The signal corresponds to a tensor-to-scalar
ratio of $r\sim 0.16$. However, as later shown by the Planck team and
already discussed before, the signal can probably be entirely
explained by dust emission~\cite{Ade:2014gaa}. Other ground based
experiments are currently operating in Antarctica such as BICEP3 \&
Keck (three channels: $100$ GHz, $150$ GHz and $200$ GHz, sky coverage
of $1-2\%$ and resolution of $30'$), SPTPol/SPT${}_3$G ($90$ GHz and
$150$ GHz, $6\%$, $1.2'$), in Chile such as Atacama B-mode Search
(ABS) ($145$ GHz, $2\%$, $30'$), Atacama Cosmology Telescope
(ACTPol)/AdvACT ($30$ GHz, $40$ GHz, $90$ GHz, $150$ GHz and $230$
GHz, $6\%$, $1.4'$), POLARBEAR/SIMONS ($90$ GHz, $150$ GHz and $220$
GHz, $6\%$, $3-5'$) and in the Canary islands such as QUIJOTE ($11-20$
GHz and $30$ GHz, $65\%$, $15'-55'$). Soon ($2016$) in Chile, the
experiment Cosmology Large Scale Surveyor (CLASS) ($40$ GHz, $90$ GHz
and $150$ GHz, $70\%$) will start taking data. There are also balloon
borne experiments such as EBEX ($150$ GHz, $250$ GHz and $410$ GHz,
$8\%$, $10'$) and SPIDER ($90$ GHz, $150$ GHz and $280$ GHz, $8\%$,
$30'-40'$) which are operating in Antarctica and Primordial Inflation
Polarization Explorer (PIPER) ($200$ GHz, $270$ GHz, $350$ GHz and
$600$ GHz, $70\%$, $10'-20'$) which will be starting in $2016$ in
Palestine in USA (Texas). The most efficient of these experiments will
reach a level corresponding to $r\sim 0.01$ in the following $3-5$
years. If one wants to go further, one needs space missions. Two
projects appear to be particularly promising: the Lite satellite for
the studies of $B$-modes polarization and Inflation from cosmic
background Radiation Detection (LiteBIRD)~\cite{Matsumura:2013aja}
selected as one of the prioritized projects in the master plan $2014$
by the Science Council of Japan and the Cosmic Origins Explorer
(COrE+)~\cite{2011arXiv1102.2181T} which is a proposal for European
Space Agency (ESA) M4 space mission. LiteBIRD has a polarization
sensitivity of $\sim 4.5 \mu\mbox{K}\times \mbox{arcmin}$, a
resolution of $\theta_{\rm fwhm}=38.5'$ and a sky coverage of
$70\%$. COrE+ can be ``light" with a sensitivity of $\sim 2.5
\mu\mbox{K}\times \mbox{arcmin}$ and a resolution of $\theta_{\rm
  fwhm}=6'$ or ``extended" with a sensitivity of $\sim 1.5
\mu\mbox{K}\times \mbox{arcmin}$ and a resolution of $\theta_{\rm
  fwhm}=4'$ (in both cases, the sky coverage is $70\%$). With these
space missions, one should be able to gain one order of magnitude on
$r$ and reach $r\sim 10^{-3}$ in the next decade, assuming no
delensing. With delensing, one might be able to probe even smaller
values of $r$.

\par

Using the above analogy, measuring $r$ is like searching for
super-symmetry. We do not know at which level it should show up (we do
not know the super-symmetry breaking scale) but it could be around the
corner and, hence, technologically realistic. In fact, a determination
of $r$ would immediately lead to the inflaton field excursion. An
excursion which is just Planckian corresponds to a tensor-to-scalar
ratio of $r\sim 10^{-3}$ that is to say precisely the limit reached by
future space missions. Therefore, given that $r=16\epsilon_1$, if
$\epsilon_1\gtrsim 10^{-4}$, then one should be able to measure it in
the next decade.

\par

Of course, a detection of primordial gravity waves would also impact
model comparison. It was recently shown in Ref.~\cite{Martin:2014rqa}
that this could allow us to rule out almost three-quarters of the
inflationary models compared to one-third for Planck $2013$.

\par

In conclusion, detecting $B$-mode CMB polarization and, hence,
primordial gravity waves, is probably the next challenge for
primordial Cosmology. An additional step would then be to check the
consistency relation, $r=-8\nT$, which would constitute the final
proof that vanilla inflation occurred in the early Universe. However,
if $r$ is very small, this measurement might be too difficult. In any
case, at the time of writing, detecting primordial gravity waves
appears to be the next frontier for inflation. Only time will tell
whether this is true or not.

\acknowledgements{It is a pleasure to thank J.~Fabris for inviting me
  to lecture at this school and for his hospitality. I am very
  grateful to all the participants for very interesting discussions. I
  also thank P.~Peter, C.~Ringeval L.~Sriramkumar and V.~Vennin for
  careful reading of the manuscript. I thank J.~Schwab for having
  given me the permission to reproduce Figs.~\ref{fig:mr},
  \ref{fig:mr_inder} and~\ref{fig:bbn}, that were originally made and
  published in Refs.~\cite{Schwab:2008ce}
  and~\cite{Rappaport:2007ct}.}

\bibliographystyle{apsrev4-1}
\bibliography{biblio}

\end{document}